\documentclass[twoside]{article}
\usepackage{graphicx}
\usepackage{url}

\newlength{\figurewidth}
\newcommand{\hide}[1]{}

\newcommand{\tabref}[1]{Table~\ref{#1}}
\newcommand{\secref}[1]{Section~\ref{#1}}

\renewcommand{\sectionmark}[1]{\markboth{\textsc{J. Leskovec et al.}}
{\textsc{The Dynamics of Viral Marketing}}}
\renewcommand{\subsectionmark}[1]{\markboth{\textsc{J. Leskovec et al.}}
{\textsc{The Dynamics of Viral Marketing}}}

\newcommand{\captionfonts}{\small}
\makeatletter
\long\def\@makecaption#1#2{%
  \vskip\abovecaptionskip
  \sbox\@tempboxa{{\captionfonts #1: #2}}%
  \ifdim \wd\@tempboxa >\hsize
    {\captionfonts #1: #2\par}
  \else
    \hbox to\hsize{\hfil\box\@tempboxa\hfil}%
  \fi
  \vskip\belowcaptionskip}
\makeatother

\begin{document}

\title{The Dynamics of Viral Marketing~\footnote{This work also appears
in: Leskovec, J., Adamic, L. A., and Huberman, B. A. 2007. The dynamics
of viral marketing. ACM Transactions on the Web, 1, 1 (May 2007).}}

\author{
  Jure Leskovec\\
  \textit{\normalsize Machine Learning Department, Carnegie Mellon University, Pittsburgh, PA}\\
  Lada A. Adamic\\
  \textit{\normalsize School of Information, University of Michigan, Ann Arbor, MI}\\
  Bernardo A. Huberman\\
  \textit{\normalsize HP Labs, Palo Alto, CA 94304}
}

\maketitle

\begin{abstract}
We present an analysis of a person-to-person recommendation network,
consisting of 4 million people who made 16 million recommendations on
half a million products. We observe the propagation of recommendations
and the cascade sizes, which we explain by a simple stochastic model. We
analyze how user behavior varies within user communities defined by a
recommendation network. Product purchases follow a 'long tail' where a
significant share of purchases belongs to rarely sold items. We
establish how the recommendation network grows over time and how
effective it is from the viewpoint of the sender and receiver of the
recommendations. While on average recommendations are not very effective
at inducing purchases and do not spread very far, we present a model
that successfully identifies communities, product and pricing categories
for which viral marketing seems to be very effective.
\end{abstract}

\nocite{jure06viral}

\section{Introduction}
With consumers showing increasing resistance to traditional forms of
advertising such as TV or newspaper ads, marketers have turned to
alternate strategies, including viral marketing. Viral marketing
exploits existing social networks by encouraging customers to share
product information with their friends. Previously, a few in depth
studies have shown that social networks affect the adoption of
individual innovations and products (for a review
see~\cite{rogers1995diffusion} or~\cite{strang98diffusion}). But until
recently it has been difficult to measure how influential
person-to-person recommendations actually are over a wide range of
products. Moreover, Subramani and Rajagopalan~\cite{subramani03sharing}
noted that ``there needs to be a greater understanding of the contexts
in which viral marketing strategy works and the characteristics of
products and services for which it is most effective. This is
particularly important because the inappropriate use of viral marketing
can be counterproductive by creating unfavorable attitudes towards
products. What is missing is an analysis of viral marketing that
highlights systematic patterns in the nature of knowledge-sharing and
persuasion by influencers and responses by recipients in online social
networks.''

Here we were able to in detail study the above mentioned problem. We
were able to directly measure and model the effectiveness of
recommendations by studying one online retailer's incentivised viral
marketing program. The website gave discounts to customers recommending
any of its products to others, and then tracked the resulting purchases
and additional recommendations.

Although word of mouth can be a powerful factor influencing purchasing
decisions, it can be tricky for advertisers to tap into. Some services
used by individuals to communicate are natural candidates for viral
marketing, because the product can be observed or advertised as part of
the communication. Email services such as Hotmail and Yahoo had very
fast adoption curves because every email sent through them contained an
advertisement for the service and because they were free. Hotmail spent
a mere \$50,000 on traditional marketing and still grew from zero to 12
million users in 18 months~\cite{jurvetson00viral}. The Hotmail user
base grew faster than any media company in history -- faster than CNN,
faster than AOL, even faster than Seinfeld's audience. By mid-2000,
Hotmail had over 66 million users with 270,000 new accounts being
established each day~\cite{wired98}. Google's Gmail also captured a
significant part of market share in spite of the fact that the
\textit{only} way to sign up for the service was through a referral.

Most products cannot be advertised in such a direct way. At the same
time the choice of products available to consumers has increased
manyfold thanks to online retailers who can supply a much wider variety
of products than traditional brick-and-mortar stores. Not only is the
variety of products larger, but one observes a `fat tail' phenomenon,
where a large fraction of purchases are of relatively obscure items. On
Amazon.com, somewhere between 20 to 40 percent of unit sales fall
outside of its top 100,000 ranked
products~\cite{brynjolfsson03consumer}. Rhapsody, a streaming-music
service, streams more tracks outside than inside its top 10,000
tunes~\cite{economist05longtail}. Some argue that the presence of the
long tail indicates that niche products with low sales are contributing
significantly to overall sales online.

We find that product purchases that result from recommendations are not
far from the usual 80-20 rule. The rule states that the top twenty
percent of the products account for 80 percent of the sales. In our case
the top 20\% of the products contribute to about half the sales.

Effectively advertising these niche products using traditional
advertising approaches is impractical. Therefore using more targeted
marketing approaches is advantageous both to the merchant and the
consumer, who would benefit from learning about new products.

The problem is partly addressed by the advent of online product and
merchant reviews, both at retail sites such as EBay and Amazon, and
specialized product comparison sites such as Epinions and CNET. Of
further help to the consumer are collaborative filtering recommendations
of the form ``people who bought $x$ also bought $y$''
feature~\cite{linden03amazon}. These refinements help consumers discover
new products and receive more accurate evaluations, but they cannot
completely substitute personalized recommendations that one receives
from a friend or relative. It is human nature to be more interested in
what a friend buys than what an anonymous person buys, to be more likely
to trust their opinion, and to be more influenced by their actions. As
one would expect our friends are also acquainted with our needs and
tastes, and can make appropriate recommendations. A Lucid Marketing
survey found that 68\% of individuals consulted friends and relatives
before purchasing home electronics -- more than the half who used search
engines to find product information~\cite{burke03lucid}.

In our study we are able to directly observe the effectiveness of person
to person word of mouth advertising for hundreds of thousands of
products for the first time.  We find that most recommendation chains do
not grow very large, often terminating with the initial purchase of a
product. However, occasionally a product will propagate through a very
active recommendation network. We propose a simple stochastic model that
seems to explain the propagation of recommendations.

Moreover, the characteristics of recommendation networks influence the
purchase patterns of their members. For example, individuals' likelihood
of purchasing a product initially increases as they receive additional
recommendations for it, but a saturation point is quickly reached.
Interestingly, as more recommendations are sent between the same two
individuals, the likelihood that they will be heeded decreases.

We find that communities (automatically found by graph theoretic
community finding algorithm) were usually centered around a product
group, such as books, music, or DVDs, but almost all of them shared
recommendations for all types of products. We also find patterns of
homophily, the tendency of like to associate with like, with communities
of customers recommending types of products reflecting their common
interests.

We propose models to identify products for which viral marketing is
effective: We find that the category and price of product plays a role,
with recommendations of expensive products of interest to small, well
connected communities resulting in a purchase more often. We also
observe patterns in the timing of recommendations and purchases
corresponding to times of day when people are likely to be shopping
online or reading email.

We report on these and other findings in the following sections. We
first survey the related work in section~\ref{sec:related}. We then
describe the characteristics of the incentivised recommendations program
and the dataset in section~\ref{sec:dataset}.
Section~\ref{sec:recNetStat} studies the temporal and static
characteristics of the recommendation network. We investigate the
propagation of recommendations and model the cascading behavior in
section~\ref{sec:propagation}. Next we concentrate on the various
aspects of the recommendation success from the viewpoint of the sender
and the recipient of the recommendation in section~\ref{sec:buyprob}.
The timing and the time lag between the recommendations and purchases is
studied in section~\ref{sec:lag}. We study network communities, product
characteristics and the purchasing behavior in
section~\ref{sec:communities}. Last, in section~\ref{sec:progReg} we
present a model that relates product characteristics and the surrounding
recommendation network to predict the product recommendation success. We
discuss the implications of our findings and conclude in
section~\ref{sec:conclusion}.

\section{Related work}
\label{sec:related}

Viral marketing can be thought of as a diffusion of information about
the product and its adoption over the network. Primarily in social
sciences there is a long history of the research on the influence of
social networks on innovation and product diffusion. However, such
studies have been typically limited to small networks and typically a
single product or service. For example, Brown and
Reingen~\cite{brown87wom} interviewed the families of students being
instructed by three piano teachers, in order to find out the network of
referrals. They found that strong ties, those between family or friends,
were more likely to be activated for information flow and were also more
influential than weak ties~\cite{granovetter73ties} between
acquaintances. Similar observations were also made by DeBruyn and Lilien
in~\cite{debruyn04} in the context of electronic referrals. They found
that characteristics of the social tie influenced recipients' behavior
but had different effects at different stages of decision making
process: tie strength facilitates awareness, perceptual affinity
triggers recipients' interest, and demographic similarity had a negative
influence on each stage of the decision-making process.

Social networks can be composed by using various information, i.e.
geographic similarity, age, similar interests and so on. Yang and
Allenby~\cite{yang03} showed that the geographically defined network of
consumers is more useful than the demographic network for explaining
consumer behavior in purchasing Japanese cars. A recent study by Hill et
al.~\cite{hill06viral} found that adding network information,
specifically whether a potential customer was already ``talking to" an
existing customer, was predictive of the chances of adoption of a new
phone service option. For the customers linked to a prior customer the
adoption rate of was 3--5 times greater than the baseline.

Factors that influence customers' willingness to actively share the
information with others via word of mouth have also been studied.
Frenzen and Nakamoto~\cite{frenzen93} surveyed a group of people and
found that the stronger the moral hazard presented by the information,
the stronger the ties must be to foster information propagation. Also,
the network structure and information characteristics interact when
individuals form decisions about transmitting information. Bowman and
Narayandas~\cite{bowman01} found that self-reported loyal customers were
more likely to talk to others about the products when they were
dissatisfied, but interestingly not more likely when they were
satisfied.

In the context of the internet word-of-mouth advertising is not
restricted to pairwise or small-group interactions between individuals.
Rather, customers can share their experiences and opinions regarding a
product with everyone. Quantitative marketing techniques have been
proposed~\cite{montgomery01marketing} to describe product information
flow online, and the rating of products and merchants has been shown to
effect the likelihood of an item being
bought~\cite{resnick01trust,chevalier04wordofmouth}. More sophisticated
online recommendation systems allow users to rate others' reviews, or
directly rate other reviewers to implicitly form a trusted reviewer
network that may have very little overlap with a person's actual social
circle. Richardson and Domingos~\cite{richardson02viral} used Epinions'
trusted reviewer network to construct an algorithm to maximize viral
marketing efficiency assuming that individuals' probability of
purchasing a product depends on the opinions on the trusted peers in
their network.  Kempe, Kleinberg and Tardos~\cite{kempe03maximizing}
have followed up on Richardson and Domingos' challenge of maximizing
viral information spread by evaluating several algorithms given various
models of adoption we discuss next.

Most of the previous research on the flow of information and influence
through the networks has been done in the context of epidemiology and
the spread of diseases over the network. See the works of
Bailey~\cite{Bailey1975Diseases} and Anderson and
May~\cite{anderson92infectious} for reviews of this area. The classical
disease propagation models are based on the stages of a disease in a
host: a person is first {\em susceptible} to a disease, then if she is
exposed to an infectious contact she can become {\em infected} and thus
{\em infectious}. After the disease ceases the person is {\em recovered}
or {\em removed}. Person is then {\em immune} for some period. The
immunity can also wear off and the person becomes again susceptible.
Thus SIR (susceptible -- infected -- recovered) models diseases where a
recovered person never again becomes susceptible, while SIRS (SIS,
susceptible -- infected -- (recovered) -- susceptible) models population
in which recovered host can become susceptible again. Given a network
and a set of infected nodes the {\em epidemic threshold} is studied,
i.e. conditions under which the disease will either dominate or die out.
In our case SIR model would correspond to the case where a set of
initially infected nodes corresponds to people that purchased a product
without first receiving the recommendations. A node can purchase a
product only once, and then tries to infect its neighbors with a
purchase by sending out the recommendations. SIS model corresponds to
less realistic case where a person can purchase a product multiple times
as a result of multiple recommendations. The problem with these type of
models is that they assume a known social network over which the
diseases (product recommendations) are spreading and usually a single
parameter which specifies the infectiousness of the disease. In our
context this would mean that the whole population is equally susceptible
to recommendations of a particular product.

There are numerous other models of influence spread in social networks.
One of the first and most influential diffusion models was proposed by
Bass~\cite{bass96}. The model of product diffusion predicts the number
of people who will adopt an innovation over time. It does not explicitly
account for the structure of the social network but it rather assumes
that the rate of adoption is a function of the current proportion of the
population who have already adopted (purchased a product in our case).
The diffusion equation models the cumulative proportion of adopters in
the population as a function of the intrinsic adoption rate, and a
measure of social contagion. The model describes an S-shaped curve,
where adoption is slow at first, takes off exponentially and flattens at
the end. It can effectively model word-of-mouth product diffusion at the
aggregate level, but not at the level of an individual person, which is
one of the topics we explore in this paper.

Diffusion models that try to model the process of adoption of an idea or
a product can generally be divided into two groups:
\begin{itemize}
\item {\em Threshold model}~\cite{Granovetter:1978} where each node
    in the network has a threshold $t \in [0,1]$, typically drawn
    from some probability distribution. We also assign {\em
    connection weights} $w_{u,v}$ on the edges of the network. A
    node adopts the behavior if a sum of the connection weights of
    its neighbors that already adopted the behavior (purchased a
    product in our case) is greater than the threshold: $t \le
    \sum_{\textrm{\scriptsize adopters}(u)} w_{u,v}$.

\item {\em Cascade model}~\cite{goldenberg01} where whenever a
    neighbor $v$ of node $u$ adopts, then node $u$ also adopts with
    probability $p_{u,v}$. In other words, every time a neighbor of
    $u$ purchases a product, there is a chance that $u$ will decide
    to purchase as well.
\end{itemize}

In the independent cascade model, Goldenberg et al.~\cite{goldenberg01}
simulated the spread of information on an artificially generated network
topology that consisted both of strong ties within groups of spatially
proximate nodes and weak ties between the groups. They found that weak
ties were important to the rate of information diffusion. Centola and
Macy~\cite{centola05} modeled product adoption on small world topologies
when a person's chance of adoption is dependent on having more than one
contact who had previously adopted. Wu and
Huberman~\cite{Wu04OpinionFormation} modeled opinion formation on
different network topologies, and found that if highly connected nodes
were seeded with a particular opinion, this would proportionally effect
the long term distribution of opinions in the network. Holme and
Newman~\cite{holme06opinion} introduced a model where individuals'
preferences are shaped by their social networks, but their choices of
whom to include in their social network are also influenced by their
preferences.

While these models address the question of how influence spreads in a
network, they are based on {\em assumed} rather than {\em measured}
influence effects. In contrast, our study tracks the actual diffusion of
recommendations through email, allowing us to quantify the importance of
factors such as the presence of highly connected individuals, or the
effect of receiving recommendations from multiple contacts. Compared to
previous empirical studies which tracked the adoption of a single
innovation or product, our data encompasses over half a million
different products, allowing us to model a product's suitability for
viral marketing in terms of both the properties of the network and the
product itself.

\section{The Recommendation Network}
\label{sec:dataset}

\subsection{Recommendation program and dataset description}
\label{sec:data}

Our analysis focuses on the recommendation referral program run by a
large retailer. The program rules were as follows. Each time a person
purchases a book, music, or a movie he or she is given the option of
sending emails recommending the item to friends. The first person to
purchase the same item through a referral link in the email gets a 10\%
discount. When this happens the sender of the recommendation receives a
10\% credit on their purchase.

The following information is recorded for each recommendation
\begin{enumerate}
  \item Sender Customer ID (shadowed)
  \item Receiver Customer ID (shadowed)
  \item Date of Sending
  \item Purchase flag (\emph{buy-bit})
  \item Purchase Date (error-prone due to asynchrony in the servers)
  \item Product identifier
  \item Price
\end{enumerate}

The recommendation dataset consists of 15,646,121 recommendations made
among 3,943,084 distinct users. The data was collected from June 5 2001
to May 16 2003. In total, 548,523 products were recommended, 99\% of
them belonging to 4 main product groups: Books, DVDs, Music and Videos.
In addition to recommendation data, we also crawled the retailer's
website to obtain product categories, reviews and ratings for all
products. Of the products in our data set, 5813 (1\%) were discontinued
(the retailer no longer provided any information about them).

Although the data gives us a detailed and accurate view of
recommendation dynamics, it does have its limitations. The only
indication of the success of a recommendation is the observation of the
recipient purchasing the product through the same vendor. We have no way
of knowing if the person had decided instead to purchase elsewhere,
borrow, or otherwise obtain the product. The delivery of the
recommendation is also somewhat different from one person simply telling
another about a product they enjoy, possibly in the context of a broader
discussion of similar products. The recommendation is received as a form
email including information about the discount program. Someone reading
the email might consider it spam, or at least deem it less important
than a recommendation given in the context of a conversation. The
recipient may also doubt whether the friend is recommending the product
because they think the recipient might enjoy it, or are simply trying to
get a discount for themselves. Finally, because the recommendation takes
place before the recommender receives the product, it might not be based
on a direct observation of the product. Nevertheless, we believe that
these recommendation networks are reflective of the nature of word of
mouth advertising, and give us key insights into the influence of social
networks on purchasing decisions.

\subsection{Identifying successful recommendations}

For each recommendation, the dataset includes information about the
recommended product, sender and received or the recommendation, and most
importantly, the success of recommendation. See section~\ref{sec:data}
for more details.

We represent this data set as a directed multi graph. The nodes
represent customers, and a directed edge contains all the information
about the recommendation. The edge $(i,j,p,t)$ indicates that $i$
recommended product $p$ to customer $j$ at time $t$. Note that as there
can be multiple recommendations of between the persons (even on the same
product) there can be multiple edges between two nodes.

The typical process generating edges in the recommendation network is as
follows: a node $i$ first buys a product $p$ at time $t$ and then it
recommends it to nodes $j_1, \dots, j_n$. The $j$ nodes can then buy the
product and further recommend it. The only way for a node to recommend a
product is to first buy it. Note that even if all nodes $j$ buy a
product, only the edge to the node $j_k$ that first made the purchase
(within a week after the recommendation) will be marked by a {\em
buy-bit}. Because the buy-bit is set only for the first person who acts
on a recommendation, we identify additional purchases by the presence of
outgoing recommendations for a person, since all recommendations must be
{\em preceded} by a purchase. We call this type of evidence of purchase
a {\em buy-edge}. Note that buy-edges provide only a lower bound on the
total number of purchases without discounts. It is possible for a
customer to not be the first to act on a recommendation and also to not
recommend the product to others. Unfortunately, this was not recorded in
the data set. We consider, however, the buy-bits and buy-edges as
proxies for the total number of purchases through recommendations.

As mentioned above the first buyer only gets a discount (the buy-bit is
turned on) if the purchase is made within one week of the
recommendation. In order to account for as many purchases as possible,
we consider all purchases where the recommendation preceded the purchase
(buy-edge) regardless of the time difference between the two events.

To avoid confusion we will refer to edges in a multi graph as
recommendations (or multi-edges) --- there can be more than one
recommendation between a pair of nodes. We will use the term edge (or
unique edge) to refer to edges in the usual sense, i.e. there is only
one edge between a pair of people. And, to get from recommendations to
edges we create an edge between a pair of people if they exchanged at
least one recommendation.

\section{The recommendation network}
\label{sec:recNetStat}

For each product group we took recommendations on all products from the
group and created a network. Table~\ref{tab:networkSizes} shows the
sizes of various product group recommendation networks with $p$ being
the total number of products in the product group, $n$ the total number
of nodes spanned by the group recommendation network, and $r$ the number
of recommendations (there can be multiple recommendations between two
nodes). Column $e$ shows the number of (unique) edges -- disregarding
multiple recommendations between the same source and recipient (i.e.,
number of pairs of people that exchanged at least one recommendation).

\begin{table}[tb]
\centering
\begin{tabular}{l||rrrrrr}\hline
  Group & $p$ & $n$ & $r$ & $e$ & $b_b$ & $b_e$ \\
  \hline
  Book  & 103,161 & 2,863,977 & 5,741,611 & 2,097,809  &65,344 & 17,769
  \\
  DVD   & 19,829  & 805,285   & 8,180,393 & 962,341  & 17,232 & 58,189
  \\
  Music & 393,598 & 794,148   & 1,443,847 & 585,738  & 7,837 & 2,739
  \\
  Video & 26,131  & 239,583   & 280,270   & 160,683   & 909 & 467 \\
  \hline
  Full network  & 542,719 & 3,943,084 & 15,646,121& 3,153,676 &
  91,322&  79,164 \\
  \hline
\end{tabular}
\caption{Product group recommendation statistics. $p$: number of
products, $n$: number of nodes, $r$: number of recommendations, $e$:
number of edges, $b_b$: number of buy bits, $b_e$: number of buy edges.
} \label{tab:networkSizes}
\end{table}

In terms of the number of different items, there are by far the most
music CDs, followed by books and videos. There is a surprisingly small
number of DVD titles. On the other hand, DVDs account for more half of
all recommendations in the dataset. The DVD network is also the most
dense, having about 10 recommendations per node, while books and music
have about 2 recommendations per node and videos have only a bit more
than 1 recommendation per node.

Music recommendations reached about the same number of people as DVDs
but used more than 5 times fewer recommendations to achieve the same
coverage of the nodes. Book recommendations reached by far the most
people -- 2.8 million. Notice that all networks have a very small number
of unique edges. For books, videos and music the number of unique edges
is smaller than the number of nodes -- this suggests that the networks
are highly disconnected~\cite{erdos60randgraph}.

Back to table~\ref{tab:networkSizes}: given the total number of
recommendations $r$ and purchases ($b_b$ + $b_e$) influenced by
recommendations we can estimate how many recommendations need to be
independently sent over the network to induce a new purchase. Using this
metric books have the most influential recommendations followed by DVDs
and music. For books one out of 69 recommendations resulted in a
purchase. For DVDs it increases to 108 recommendations per purchase and
further increases to 136 for music and 203 for video.

\begin{table}[tb]
\centering
\begin{tabular}{l||rrrrrrr}\hline
  Group & $n_c$ & $r_c$ & $e_c$ & $b_{bc}$ & $b_{ec}$ \\
  \hline
  Book  & 53,681 & 933,988   & 184,188 & 1,919 & 1,921 \\
  DVD   & 39,699 & 6,903,087 & 442,747 & 6,199 & 41,744 \\
  Music & 22,044 & 295,543   & 82,844 & 348 & 456 \\
  Video & 4,964  & 23,555    & 15,331 & 2 & 74 \\
  \hline
  Full network  & 100,460 & 8,283,753 & 521,803 & 8,468 & 44,195 \\
  \hline
\end{tabular}
\caption{Statistics for the largest connected component of each product
group. $n_c$: number of nodes in largest connected component, $r_c$:
number recommendations in the component, $e_c$: number of edges in the
component, $b_{bc}$: number of buy bits, $b_{ec}$: number of buy edges
in the largest connected component, and $b_{bc}$ and $b_{ec}$ are the
number of purchase through a buy-bit and a buy-edge, respectively.}
\label{tab:networkCcSizes}
\end{table}

Table~\ref{tab:networkCcSizes} gives more insight into the structure of
the largest connected component of each product group's recommendation
network. We performed the same measurements as in
table~\ref{tab:networkSizes} with the difference being that we did not
use the whole network but only its largest weakly connected component.
The table shows the number of nodes $n$, the number of recommendations
$r_c$, and the number of (unique) edges $e_c$ in the largest component.
The last two columns ($b_{bc}$ and $b_{ec}$) show the number of
purchases resulting in a discount (buy-bit, $b_{bc}$) and the number of
purchases through buy-edges ($b_{ec}$) in the largest connected
component.

First, notice that the largest connected components are very small. DVDs
have the largest - containing 4.9\% of the nodes, books have the
smallest at 1.78\%. One would also expect that the fraction of the
recommendations in the largest component would be proportional to its
size. We notice that this is not the case. For example, the largest
component in the full recommendation network contains 2.54\% of the
nodes and 52.9\% of all recommendations, which is the result of heavy
bias in DVD recommendations. Breaking this down by product categories we
see that for DVDs 84.3\% of the recommendations are in largest component
(which contains 4.9\% of all DVD nodes), vs. 16.3\% for book
recommendations (component size 1.79\%), 20.5\% for music
recommendations (component size 2.77\%), and 8.4\% for video
recommendations (component size 2.1\%). This shows that the dynamic in
the largest component is very much different from the rest of the
network. Especially for DVDs we can see that a very small fraction of
users generated most of the recommendations.

\subsection{Recommendation network over time}
\label{networkovertime}

\begin{figure}[t]
\begin{center}
  \includegraphics[width=0.6\textwidth]{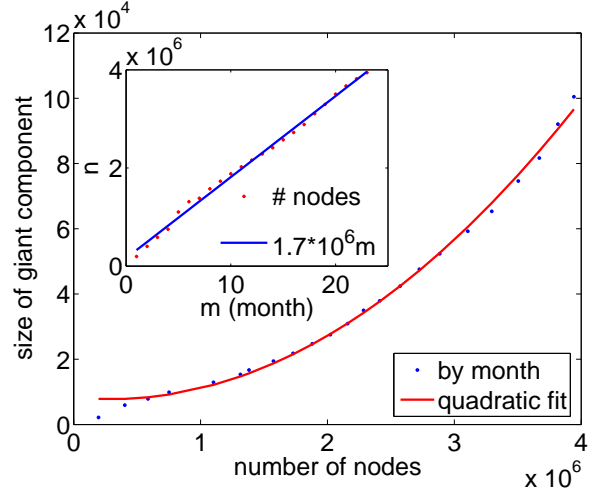}
\caption{(a) The size of the largest connected component of
customers over time. The inset shows the linear growth in the number
of customers $n$ over time.} \label{fig:recOverTm}
\end{center}
\end{figure}

The recommendations that occurred were exchanged over an existing
underlying social network. In the real world, it is estimated that any
two people on the globe are connected via a short chain of acquaintances
- popularly known as the small world
phenomenon~\cite{travers69smallworld}. We examined whether the edges
formed by aggregating recommendations over all products would similarly
yield a small world network, even though they represent only a small
fraction of a person's complete social network. We measured the growth
of the largest weakly connected component over time, shown in
Figure~\ref{fig:recOverTm}. Within the weakly connected component, any
node can be reached from any other node by traversing (undirected)
edges. For example, if $u$ recommended product $x$ to $v$, and $w$
recommended product $y$ to $v$, then $u$and $w$ are linked through one
intermediary and thus belong to the same weakly connected component.
Note that connected components do not necessarily correspond to
communities (clusters) which we often think of as densely linked parts
of the networks. Nodes belong to same component if they can reach each
other via an undirected path regardless of how densely they are linked.

Figure~\ref{fig:recOverTm} shows the size of the largest connected
component, as a fraction of the total network. The largest component is
very small over all time. Even though we compose the network using all
the recommendations in the dataset, the largest connected component
contains less than 2.5\% (100,420) of the nodes, and the second largest
component has only $600$ nodes. Still, some smaller communities,
numbering in the tens of thousands of purchasers of DVDs in categories
such as westerns, classics and Japanese animated films (anime), had
connected components spanning about 20\% of their members.

The insert in figure~\ref{fig:recOverTm} shows the growth of the
customer base over time. Surprisingly it was linear, adding on average
165,000 new users each month, which is an indication that the service
itself was not spreading epidemically. Further evidence of non-viral
spread is provided by the relatively high percentage (94\%) of users who
made their first recommendation without having previously received one.

\subsubsection{Growth of the largest connected component}

Next, we examine the growth of the largest connected component (LCC). In
figure~\ref{fig:recOverTm} we saw that the largest component seems to
grow quadratically over time, but at the end of the data collection
period is still very small, {\em i.e.} only 2.5\% of the nodes belong to
largest weakly connected component. Here we are not interested in how
fast the largest component grows over time but rather how big other
components are when they get merged into the largest component. Also,
since our graph is directed we are interested in determining whether
smaller components become attached to the largest component by a
recommendation sent from inside of the largest component. One can think
of these recommendations as being tentacles reaching out of largest
component to attach smaller components. The other possibility is that
the recommendation comes from a node outside the component to a member
of the largest component and thus the initiative to attach comes from
outside the largest component.

\begin{figure}[t]
\begin{center}
\begin{tabular}{ccc}
  \includegraphics[width=0.31\textwidth]{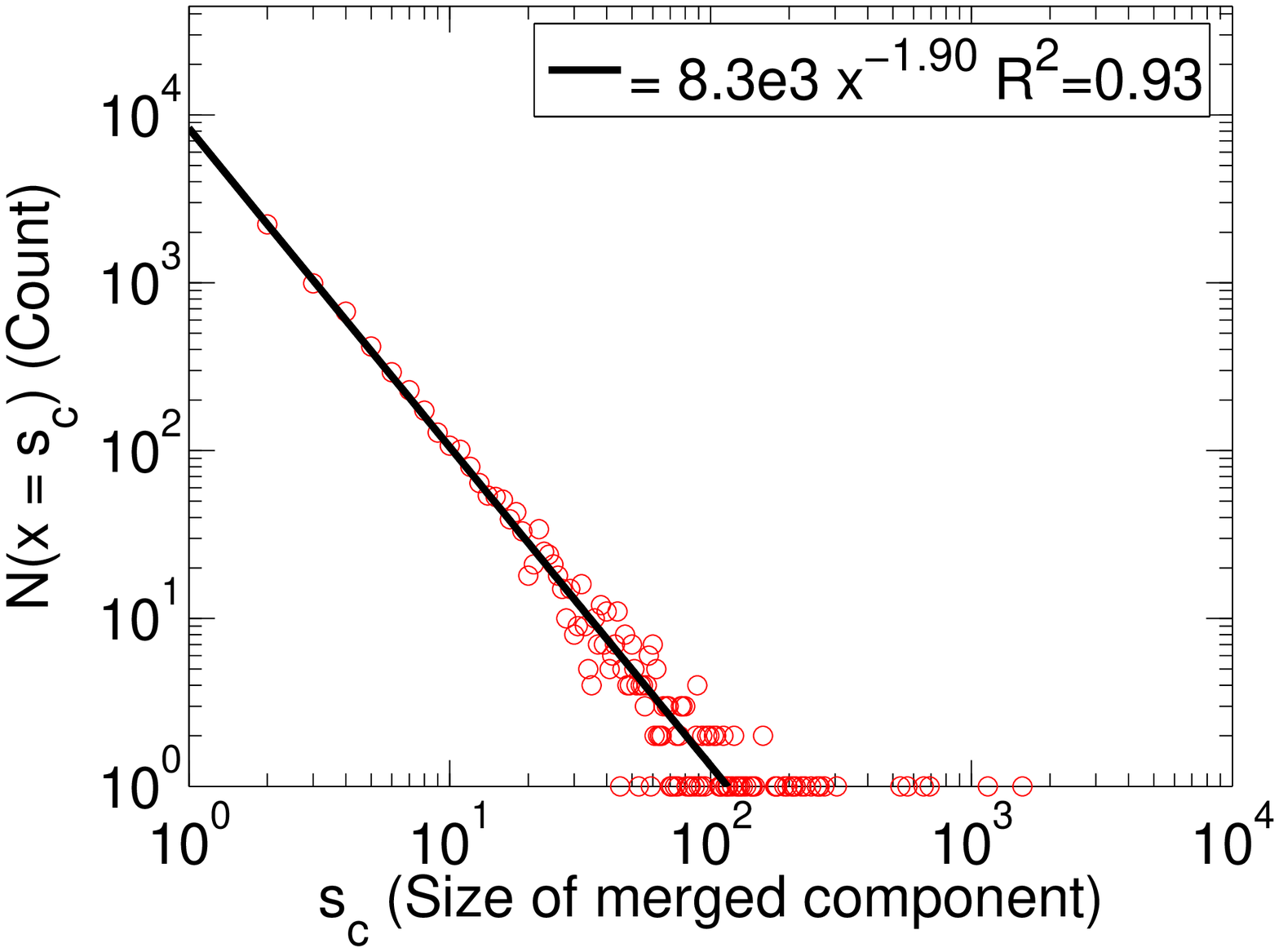} &
  \includegraphics[width=0.31\textwidth]{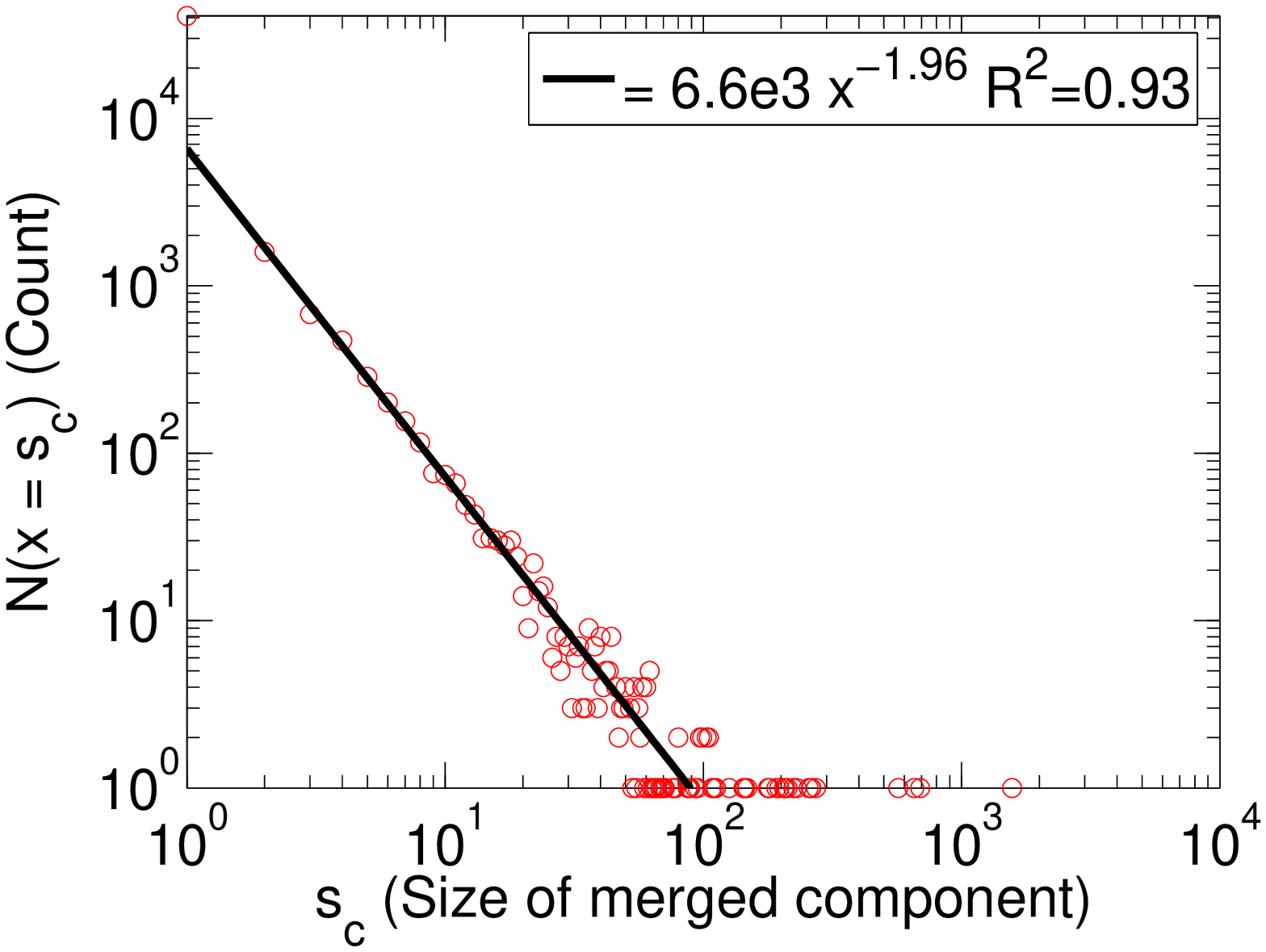} &
  \includegraphics[width=0.31\textwidth]{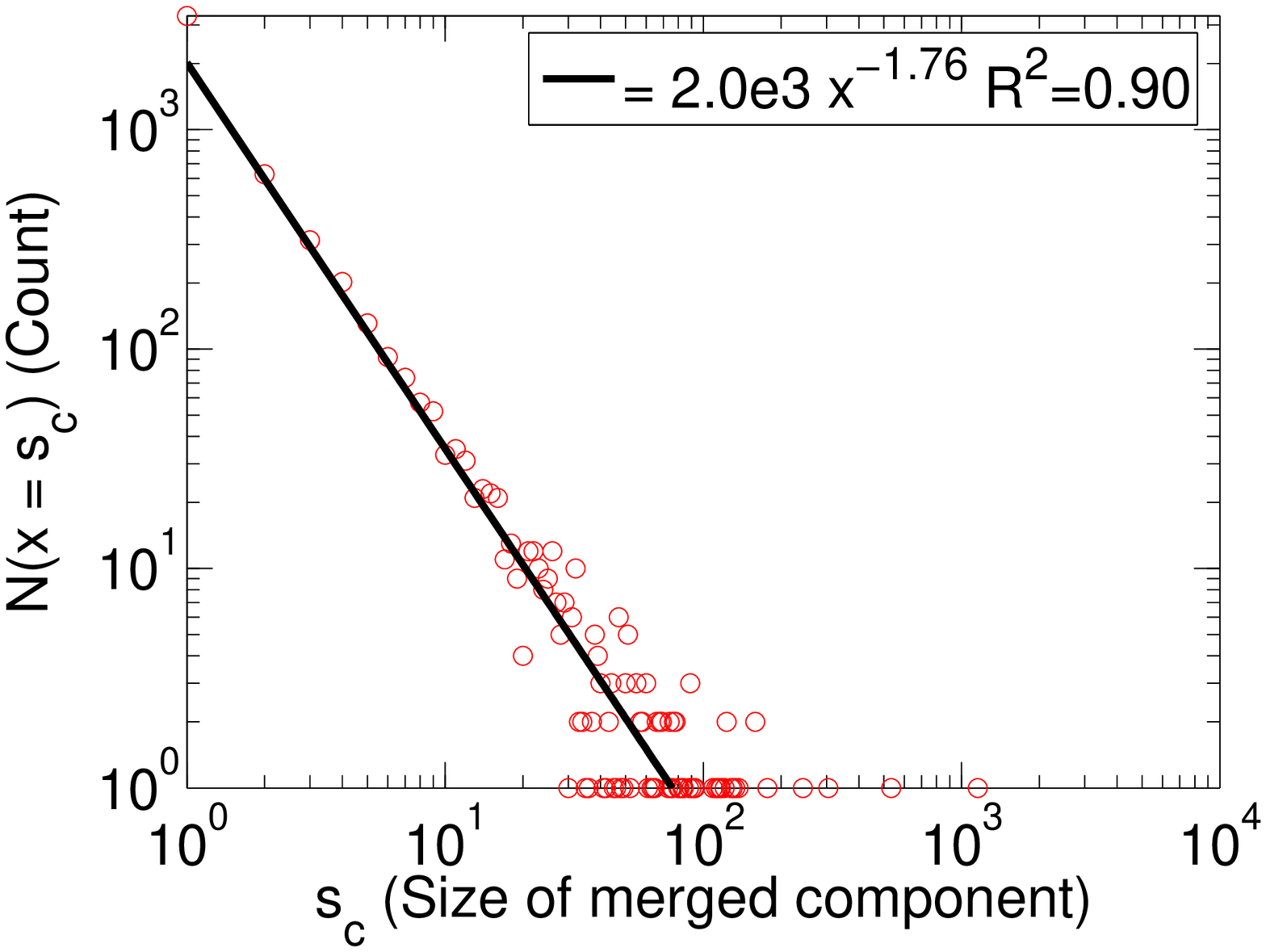} \\
  (a) LCC growth  & (b) Sender in LCC  & (c) Sender outside LCC
  \end{tabular}
\caption{Growth of the largest connected component (LCC). (a) the
distribution of sizes of components when they are merged into the
largest connected component. (b) same as (a), but restricted to
cases when a member of the LCC sends a recommendation to someone
outside the largest component. (c) a sender outside the largest
component sends a recommendation to a member of the component.}
\label{fig:ccGrowth}
\end{center}
\end{figure}

We look at whether the largest component grows gradually, adding nodes
one by one as the members send out more recommendations, or whether a
new recommendation might act as a bridge to a component consisting of
several nodes who are already linked by their previous recommendations.
To this end we measure the distribution of a component's size when it
gets merged to the largest weakly connected component.

We operate under the following setting. Recommendations are arriving
over time one by one creating edges between the nodes of the network. As
more edges are being added the size of largest connected component
grows. We keep track of the currently largest component, and measure how
big the separate components are when they get attached to the largest
component.

Figure~\ref{fig:ccGrowth}(a) shows the distribution of merged connected
component (CC) sizes. On the x-axis we plot the component size (number
of nodes $N$) and on the y-axis the number of components of size $N$
that were merged over time with the largest component. We see that a
majority of the time a single node (component of size 1) merged with the
currently largest component. On the other extreme is the case when a
component of $1,568$ nodes merged with the largest component.

Interestingly, out of all merged components, in 77\% of the cases the
source of the recommendation comes from inside the largest component,
while in the remaining 23\% of the cases it is the smaller component
that attaches itself to the largest one. Figure~\ref{fig:ccGrowth}(b)
shows the distribution of component sizes only for the case when the
sender of the recommendation was a member of the largest component, i.e.
the small component was attached from the largest component. Lastly,
Figure~\ref{fig:ccGrowth}(c) shows the distribution for the opposite
case when the sender of the recommendation was not a member of the
largest component, {\em i.e.} the small component attached itself to the
largest.

Also notice that in all cases the distribution of merged component sizes
follows a heavy-tailed distribution. We fit a power-law distribution and
note the power-law exponent of 1.90 (fig.~\ref{fig:ccGrowth}(a)) when
considering all merged components. Limiting the analysis to the cases
where the source of the edge that attached a small component to the
largest is in the largest component we obtain power-law exponent of 1.96
(fig.~\ref{fig:ccGrowth}(b)), and when the edge originated from the
small component to attached it to the largest, the power-law exponent is
1.76. This shows that even though in most cases the LCC absorbs the
small component, we see that components that attach themselves to the
LCC tend to be larger (smaller power-law exponent) than those attracted
by the LCC. This means that the component sometimes grows a bit before
it attaches itself to the largest component. Intuitively, an individual
node can get attached to the largest component simply by passively
receiving a recommendation. But if it is the outside node that sends a
recommendation to someone in the giant component, it is already an
active recommender and could therefore have recommended to several
others previously, thus forming a slightly bigger component that is then
merged.

From these experiments we see that the largest component is very active,
adding smaller components by generating new recommendations. Most of the
time these newly merged components are quite small, but occasionally
sizable components are attached.

\subsection{Preliminary observations and discussion}

Even with these simple counts and experiments we can already make a few
observations. It seems that some people got quite heavily involved in
the recommendation program, and that they tended to recommend a large
number of products to the same set of friends (since the number of
unique edges is so small as shown on table~\ref{tab:networkSizes}). This
means that people tend to buy more DVDs and also like to recommend them
to their friends, while they seem to be more conservative with books.
One possible reason is that a book is a bigger time investment than a
DVD: one usually needs several days to read a book, while a DVD can be
viewed in a single evening. Another factor may be how informed the
customer is about the product. DVDs, while fewer in number, are more
heavily advertised on TV, billboards, and movie theater previews.
Furthermore, it is possible that a customer has already watched a movie
and is adding the DVD to their collection. This could make them more
confident in sending recommendations before viewing the purchased DVD.

One external factor which may be affecting the recommendation patterns
for DVDs is the existence of referral websites (\url{www.dvdtalk.com}).
On these websites people, who want to buy a DVD and get a discount,
would ask for recommendations. This way there would be recommendations
made between people who don't really know each other but rather have an
economic incentive to cooperate.

In effect, the viral marketing program is altering, albeit briefly and
most likely unintentionally, the structure of the social network it is
spreading on. We were not able to find similar referral sharing sites
for books or CDs.

\section{Propagation of recommendations}
\label{sec:propagation}

\subsection{Forward recommendations}

Not all people who accept a recommendation by making a purchase also
decide to give recommendations. In estimating what fraction of people
that purchase also decide to recommend forward, we can only use the
nodes with purchases that resulted in a discount. Table~\ref{tab:fwdRec}
shows that only about a third of the people that purchase also recommend
the product forward. The ratio of forward recommendations is much higher
for DVDs than for other kinds of products. Videos also have a higher
ratio of forward recommendations, while books have the lowest. This
shows that people are most keen on recommending movies, possibly for the
above mentioned reasons, while more conservative when recommending books
and music.

\begin{figure}[t]
\begin{center}
\begin{tabular}{cc}
  \includegraphics[width=0.6\textwidth]{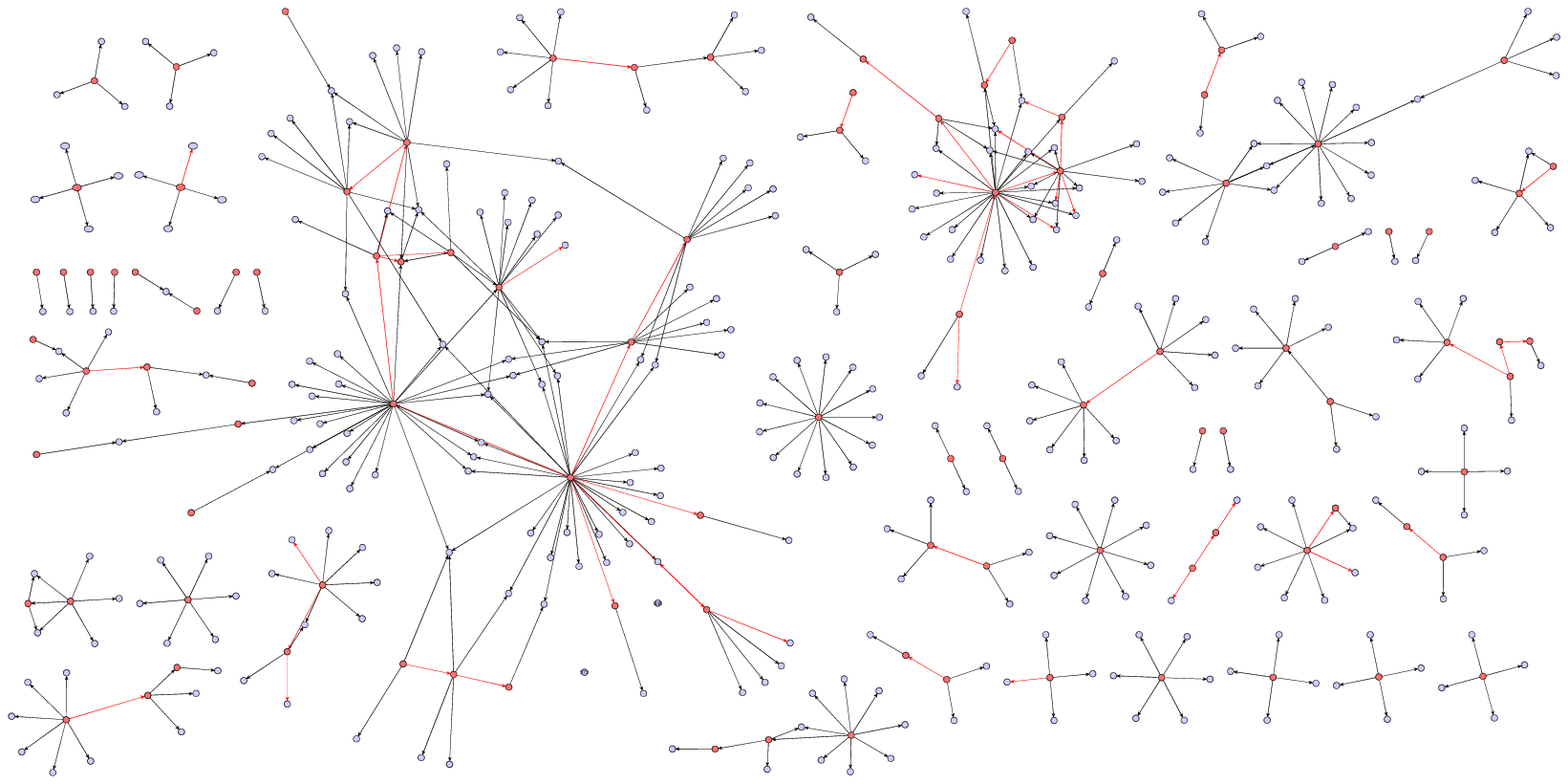}
  &
  \includegraphics[width=0.4\textwidth]{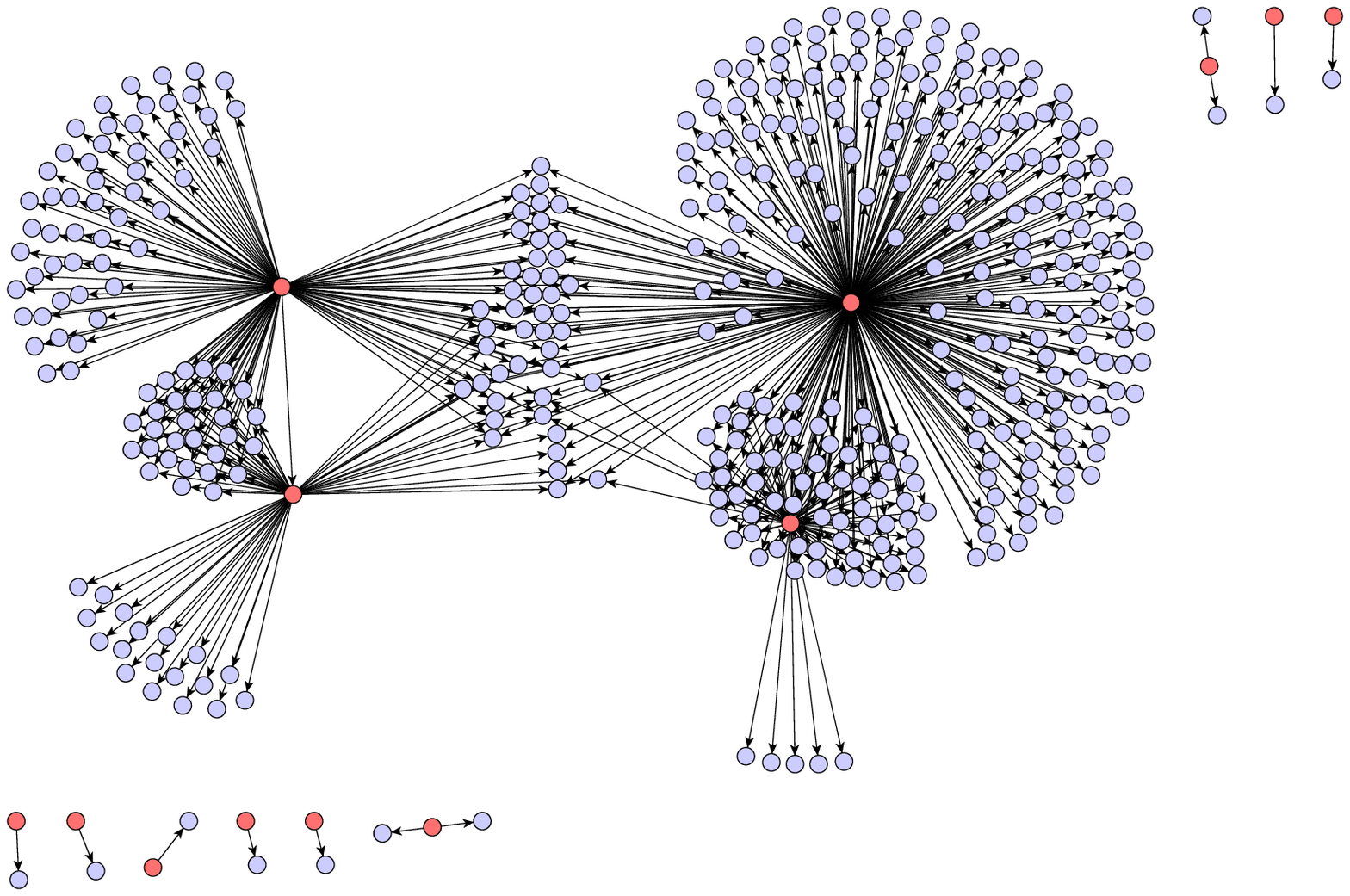} \\
  (a) Medical book & (b) Japanese graphic novel \\
\end{tabular}
\caption{Examples of two product recommendation networks: (a) First
aid study guide {\em First Aid for the USMLE Step}, (b) Japanese
graphic novel (manga) {\em Oh My Goddess!: Mara Strikes Back}.}
\label{fig:recNetPlot}
\end{center}
\end{figure}

\begin{table}[tb]
\centering
\begin{tabular}{l|rrr}\hline
  & \multicolumn{3}{c}{Number of nodes} \\
  Group & Purchases & Forward & Percent\\
  \hline
  Book  & 65,391 & 15,769 & 24.2 \\
  DVD   & 16,459 &  7,336 & 44.6 \\
  Music &  7,843 &  1,824 & 23.3 \\
  Video &    909 &    250 & 27.6 \\
  \hline
  Total & 90,602 &  25,179 & 27.8 \\
  \hline
\end{tabular}
\caption{Fraction of people that purchase and also recommend forward.
{\em Purchases}: number of nodes that purchased as a result of receiving
a recommendation. {\em Forward}: nodes that purchased and then also
recommended the product to others.} \label{tab:fwdRec}
\end{table}

\begin{figure}[t]
\begin{center}
\begin{tabular}{c}
  \includegraphics[width=0.60\textwidth]{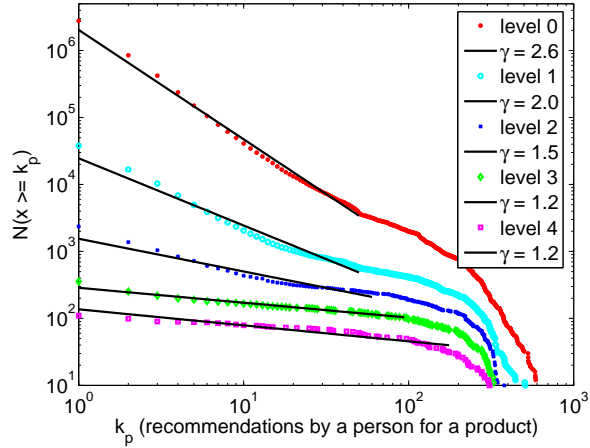}
  \end{tabular}
\caption{The number of recommendations sent by a user with each
curve representing a different depth of the user in the
recommendation chain. A power law exponent $\gamma$ is fitted to all
but the tail, which shows an exponential drop-off at around 100
recommendations sent). This drop-off is consistent across all depth
levels, and may reflect either a natural disinclination to send
recommendation to over a hundred people, or a technical issue that
might have made it more inconvenient to do so. The fitted
lines follow the order of the level number (i.e. top line
corresponds to level 0 and bottom to level 4).}
\label{fig:recByLev}
\end{center}
\end{figure}

Figure~\ref{fig:recByLev} shows the cumulative out-degree distribution,
that is the number of people who sent out at least $k_p$
recommendations, for a product. We fit a power-law to all but the tail
of the distribution. Also, notice the exponential decay in the tail of
the distribution which could be, among other reasons, attributed to the
finite time horizon of our dataset.

The figure~\ref{fig:recByLev} shows that the deeper an individual is in
the cascade, if they choose to make recommendations, they tend to
recommend to a greater number of people on average (the fitted line has
smaller slope $\gamma$, {\em i.e.} the distribution has higher
variance). This effect is probably due to only very heavily recommended
products producing large enough cascades to reach a certain depth. We
also observe, as is shown in Table~\ref{tab:levels}, that the
probability of an individual making a recommendation at all (which can
only occur if they make a purchase), declines after an initial increase
as one gets deeper into the cascade.

\begin{table}[tb]
\centering
\begin{tabular}{c|rr}\hline
level & prob. buy \& & average \\
 & forward & out-degree \\
\hline 0 & N/A & 1.99 \\
1 & 0.0069 & 5.34 \\
2 & 0.0149 & 24.43 \\
3 & 0.0115 & 72.79 \\
4 & 0.0082 & 111.75 \\
\end{tabular}
\caption{Statistics about individuals at different levels of the
cascade. } \label{tab:levels}
\end{table}

\subsection{Identifying cascades}

As customers continue forwarding recommendations, they contribute to the
formation of cascades. In order to identify cascades, i.e. the
``causal'' propagation of recommendations, we track {\em successful
recommendations} as they influence purchases and further
recommendations. We define a recommendation to be successful if it
reached a node before its {\em first} purchase. We consider only the
first purchase of an item, because there are many cases when a person
made multiple purchases of the same product, and in between those
purchases she may have received new recommendations. In this case one
cannot conclude that recommendations following the first purchase
influenced the later purchases.

Each cascade is a network consisting of customers (nodes) who purchased
the same product as a result of each other's recommendations (edges). We
delete \emph{late recommendations} --- all incoming recommendations that
happened after the first purchase of the product. This way we make the
network \emph{time increasing} or \emph{causal} --- for each node all
incoming edges (recommendations) occurred before all outgoing edges. Now
each connected component represents a time obeying propagation of
recommendations.

\begin{figure}
\begin{center}
\begin{tabular}{cc}
  \includegraphics[width=0.48\textwidth]{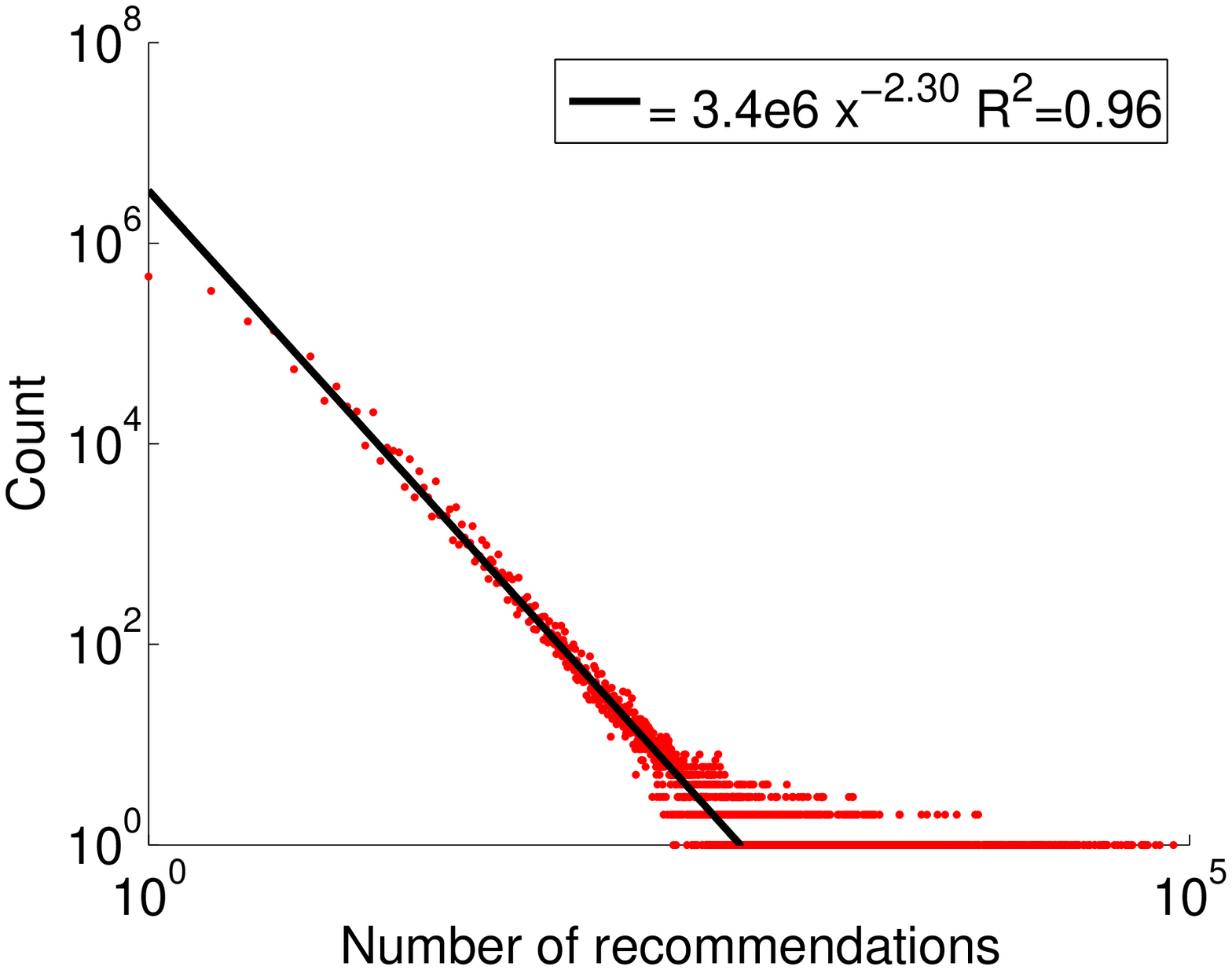} &
  \includegraphics[width=0.48\textwidth]{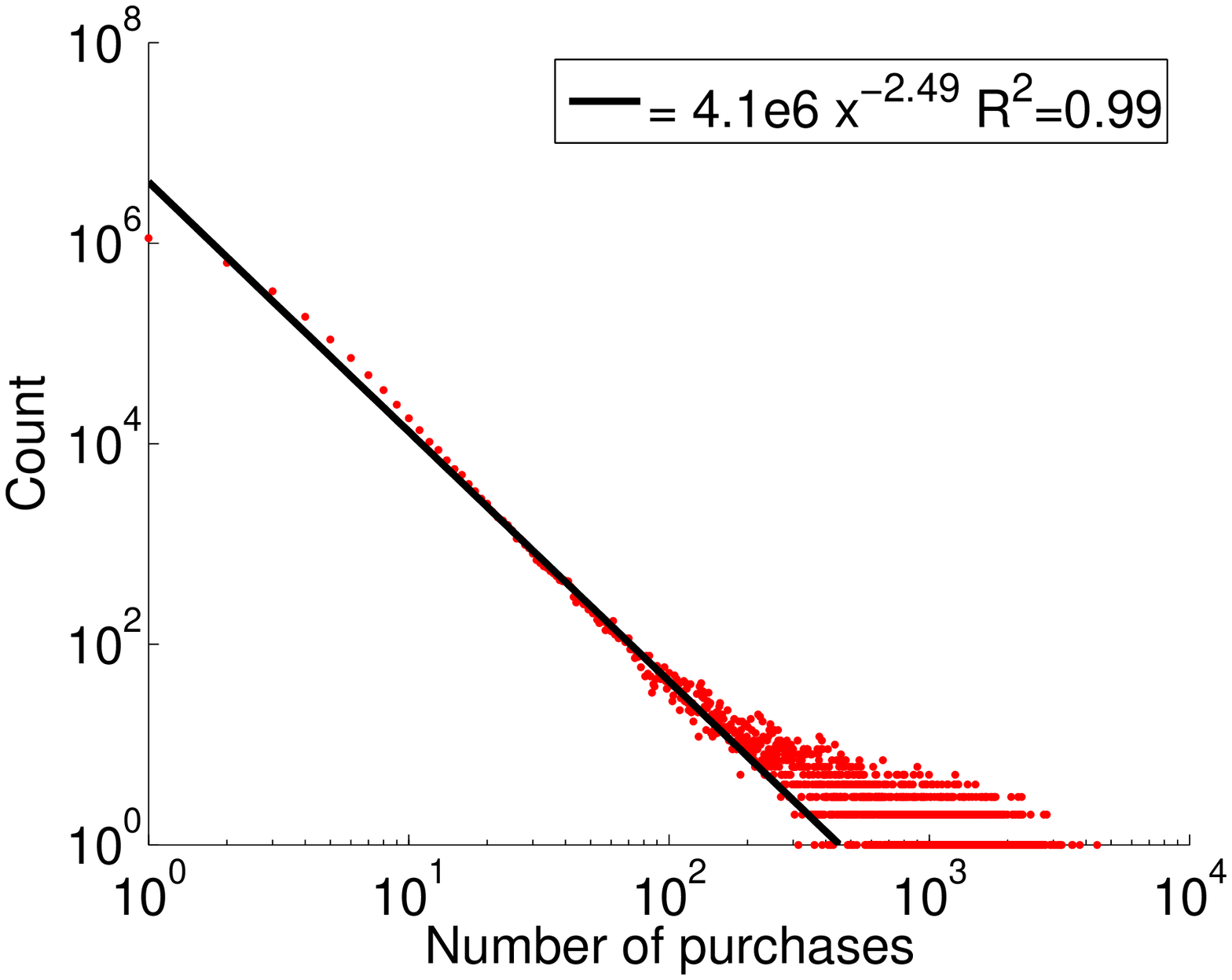} \\
  (a) Recommendations & (b) Purchases\\
\end{tabular}
\caption{Distribution of the number of recommendations and number of
purchases made by a customer.} \label{fig:recBuyDist}
\end{center}
\end{figure}

\begin{figure}
\begin{center}
\begin{tabular}{cc}
  \includegraphics[width=0.48\textwidth]{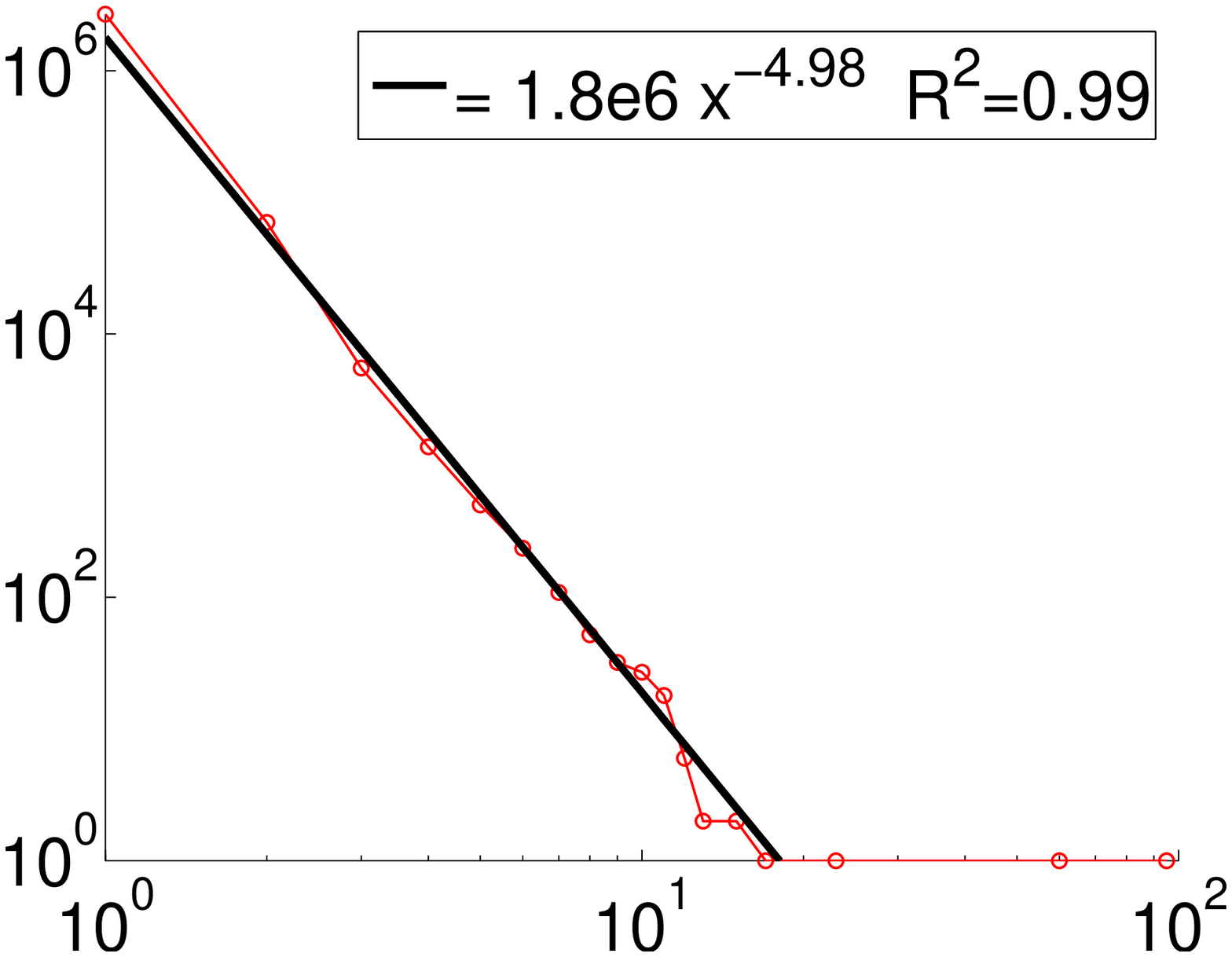} &
  \includegraphics[width=0.48\textwidth]{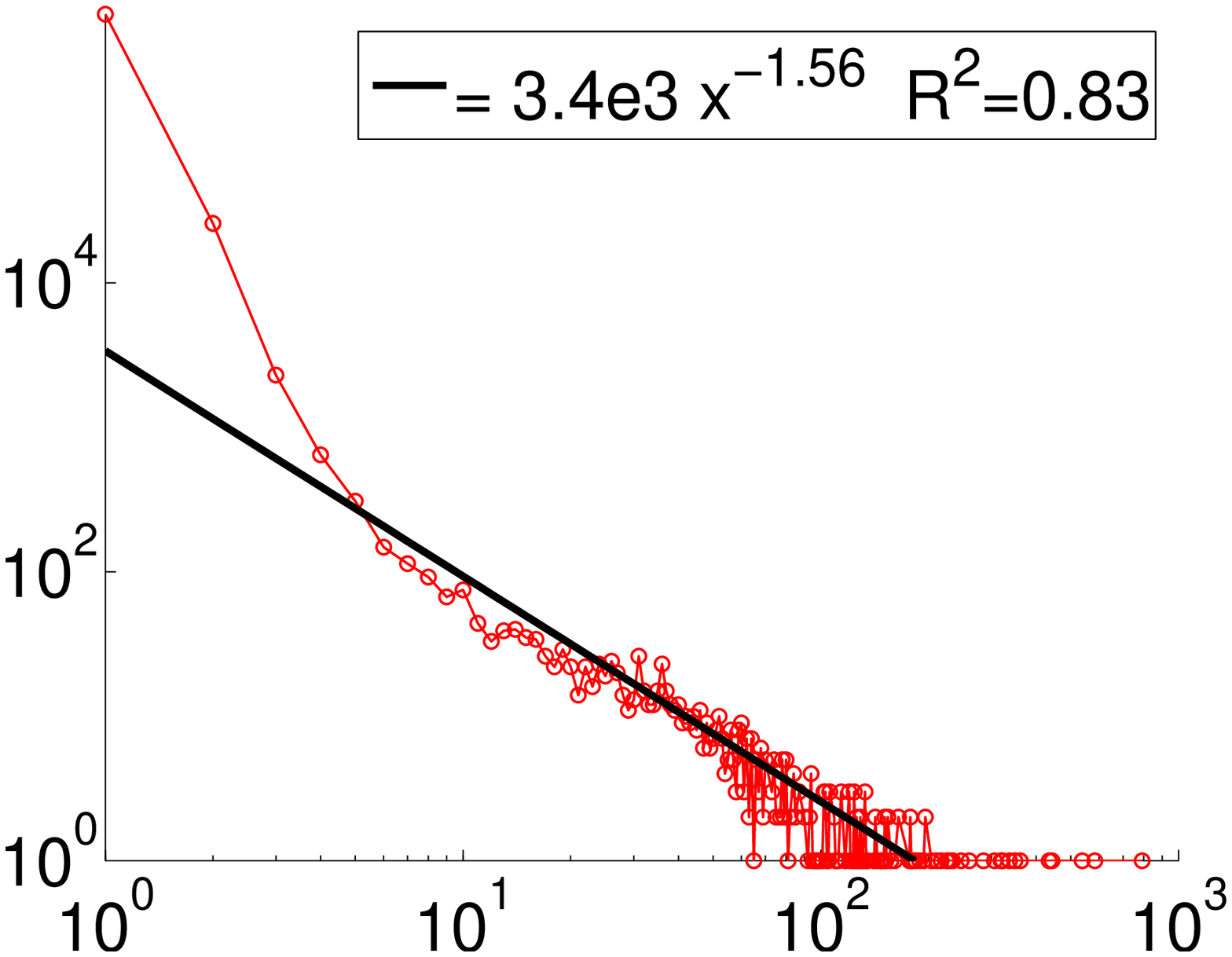} \\
  (a) Book & (b) DVD \\
  \includegraphics[width=0.48\textwidth]{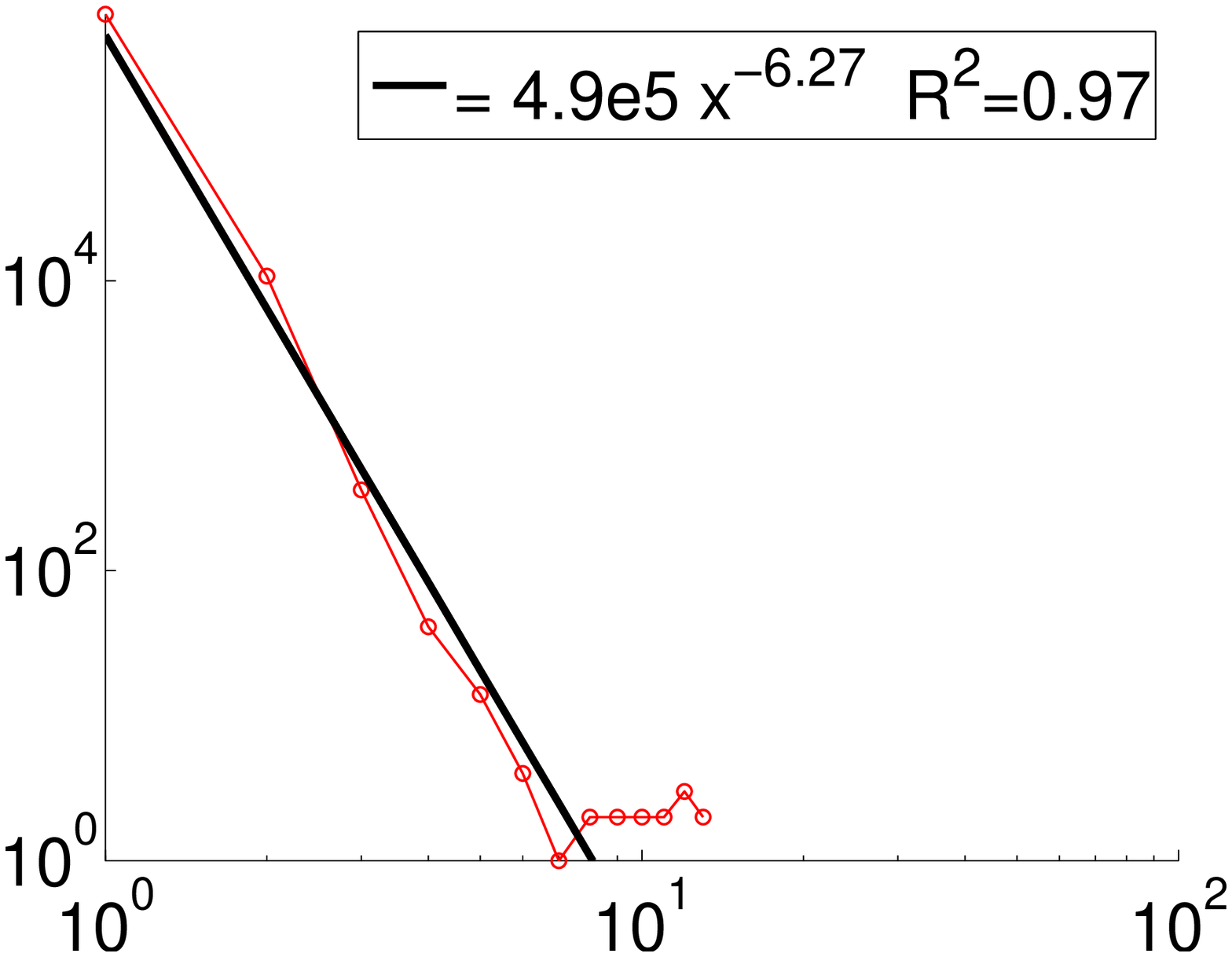} &
  \includegraphics[width=0.48\textwidth]{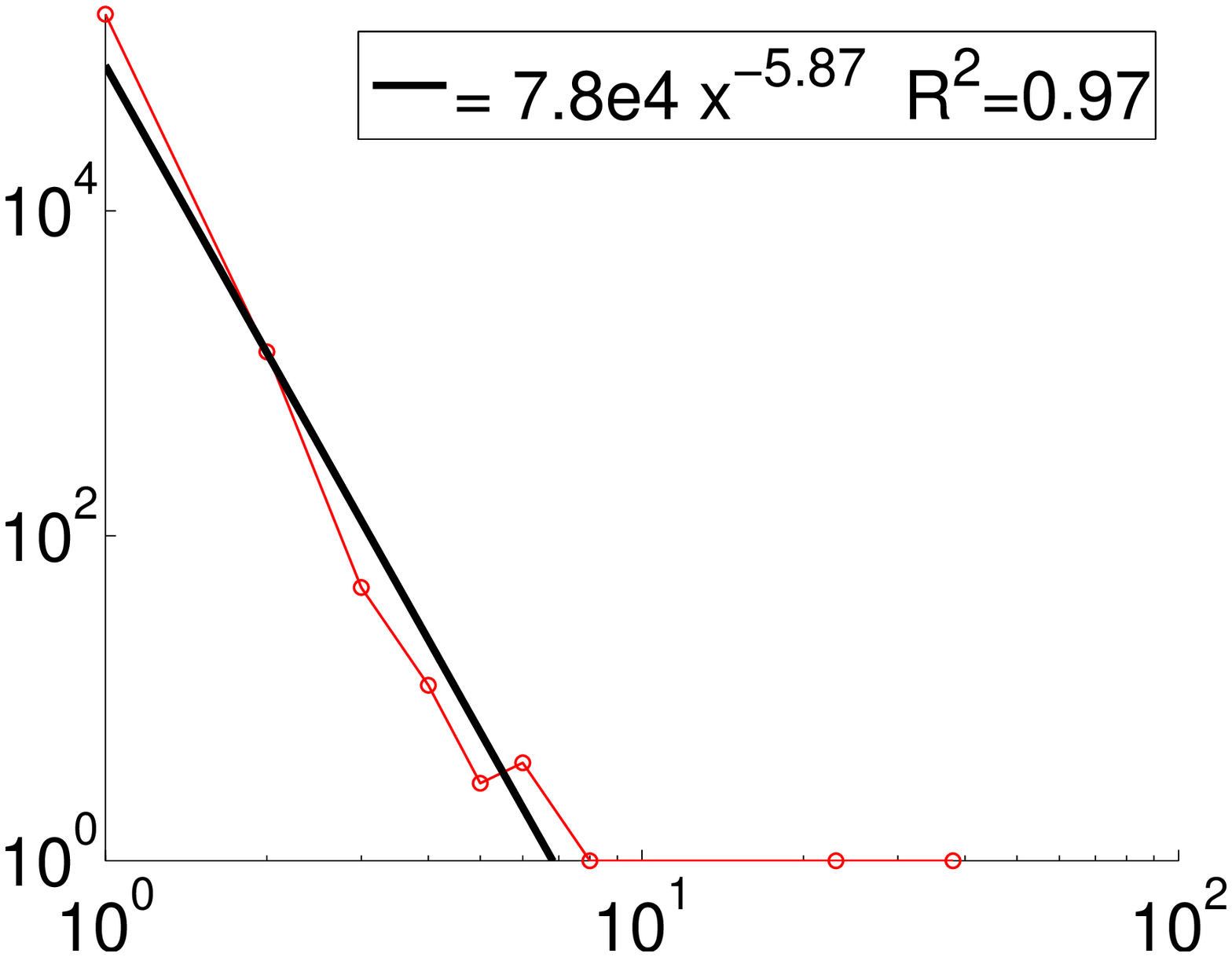} \\
  (c) Music & (d) Video\\
\end{tabular}
\caption{Size distribution of cascades (size of cascade vs. count).
Bold line presents a power-fit.} \label{fig:cascadeSzDist}
\end{center}
\end{figure}

Figure~\ref{fig:recNetPlot} shows two typical product recommendation
networks: (a) a medical study guide and (b) a Japanese graphic novel.
Throughout the dataset we observe very similar patters. Most product
recommendation networks consist of a large number of small disconnected
components where we do not observe cascades. Then there is usually a
small number of relatively small components with recommendations
successfully propagating. This observation is reflected in the heavy
tailed distribution of cascade sizes (see
figure~\ref{fig:cascadeSzDist}), having a power-law exponent close to 1
for DVDs in particular. We determined the power-law exponent by fitting
a line on log-log scales using the least squares method.

We also notice bursts of recommendations
(figure~\ref{fig:recNetPlot}(b)). Some nodes recommend to many friends,
forming a star like pattern. Figure~\ref{fig:recBuyDist} shows the
distribution of the recommendations and purchases made by a single node
in the recommendation network. Notice the power-law distributions and
long flat tails. The most active customer made 83,729 recommendations
and purchased 4,416 different items. Finally, we also sometimes observe
`collisions', where nodes receive recommendations from two or more
sources. A detailed enumeration and analysis of observed topological
cascade patterns for this dataset is made in~\cite{jurij05patterns}.

\begin{figure}[t]
\begin{center}
\begin{tabular}{ccc}
  \includegraphics[width=0.31\textwidth]{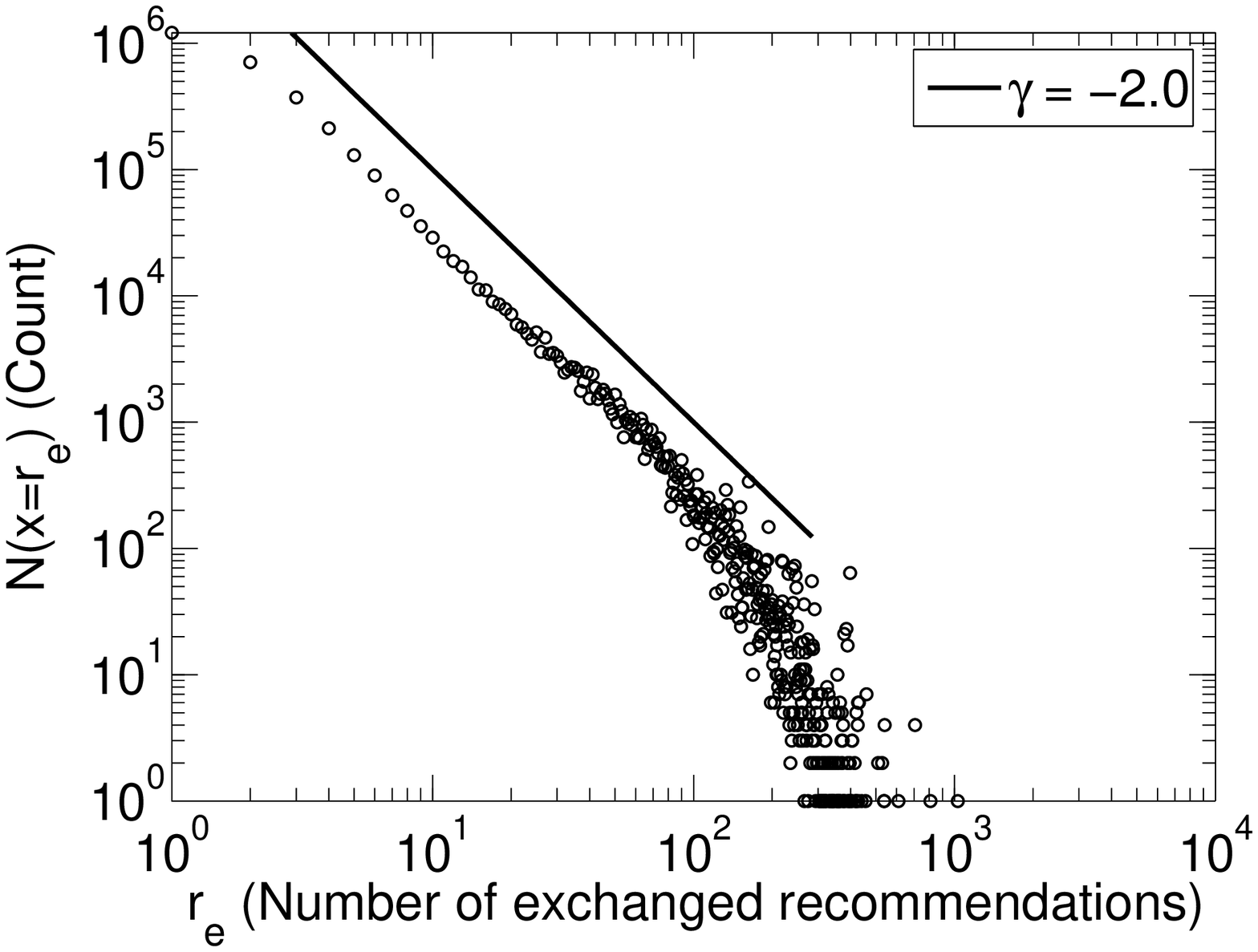} &
  \includegraphics[width=0.31\textwidth]{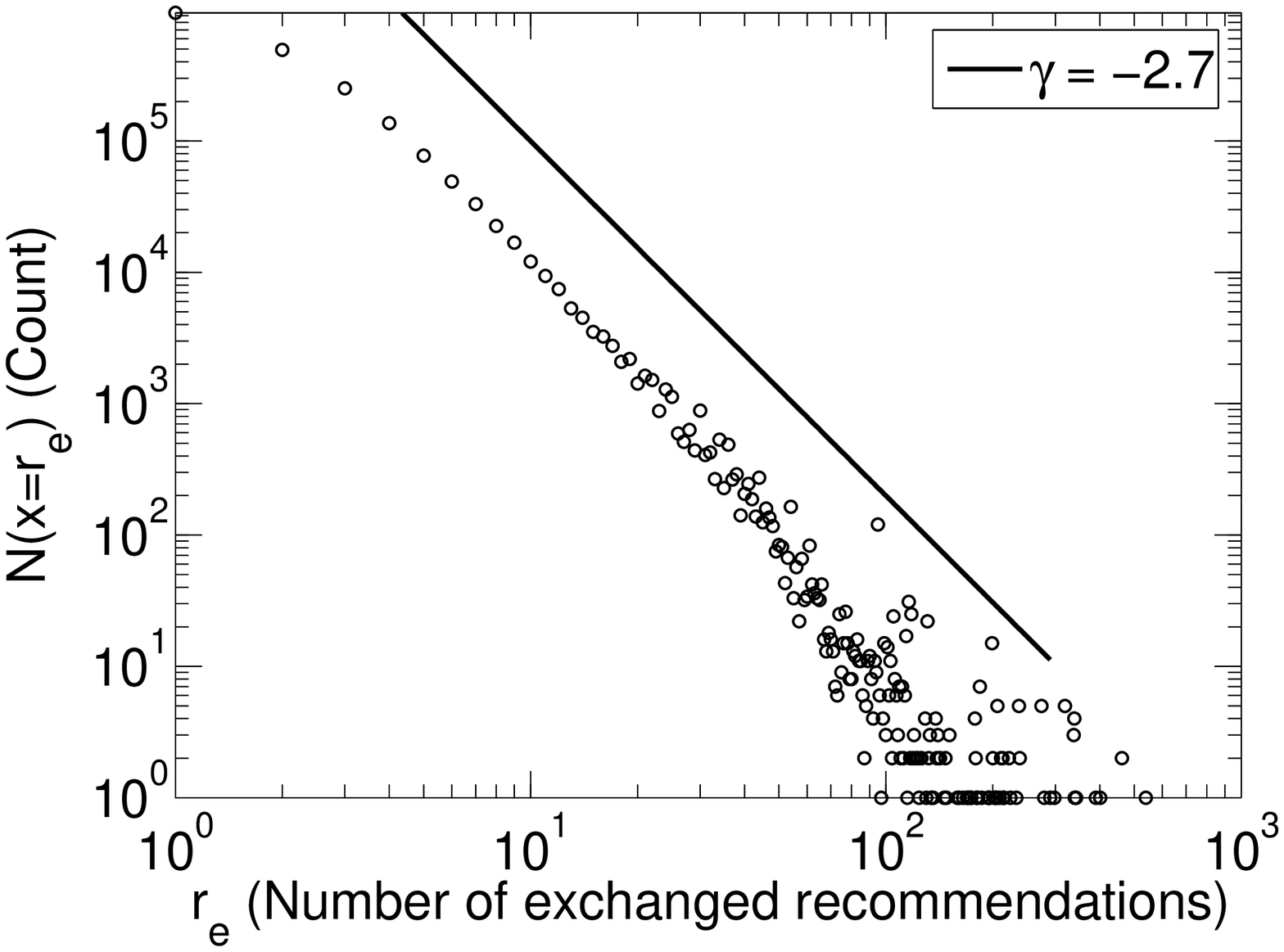} &
  \includegraphics[width=0.31\textwidth]{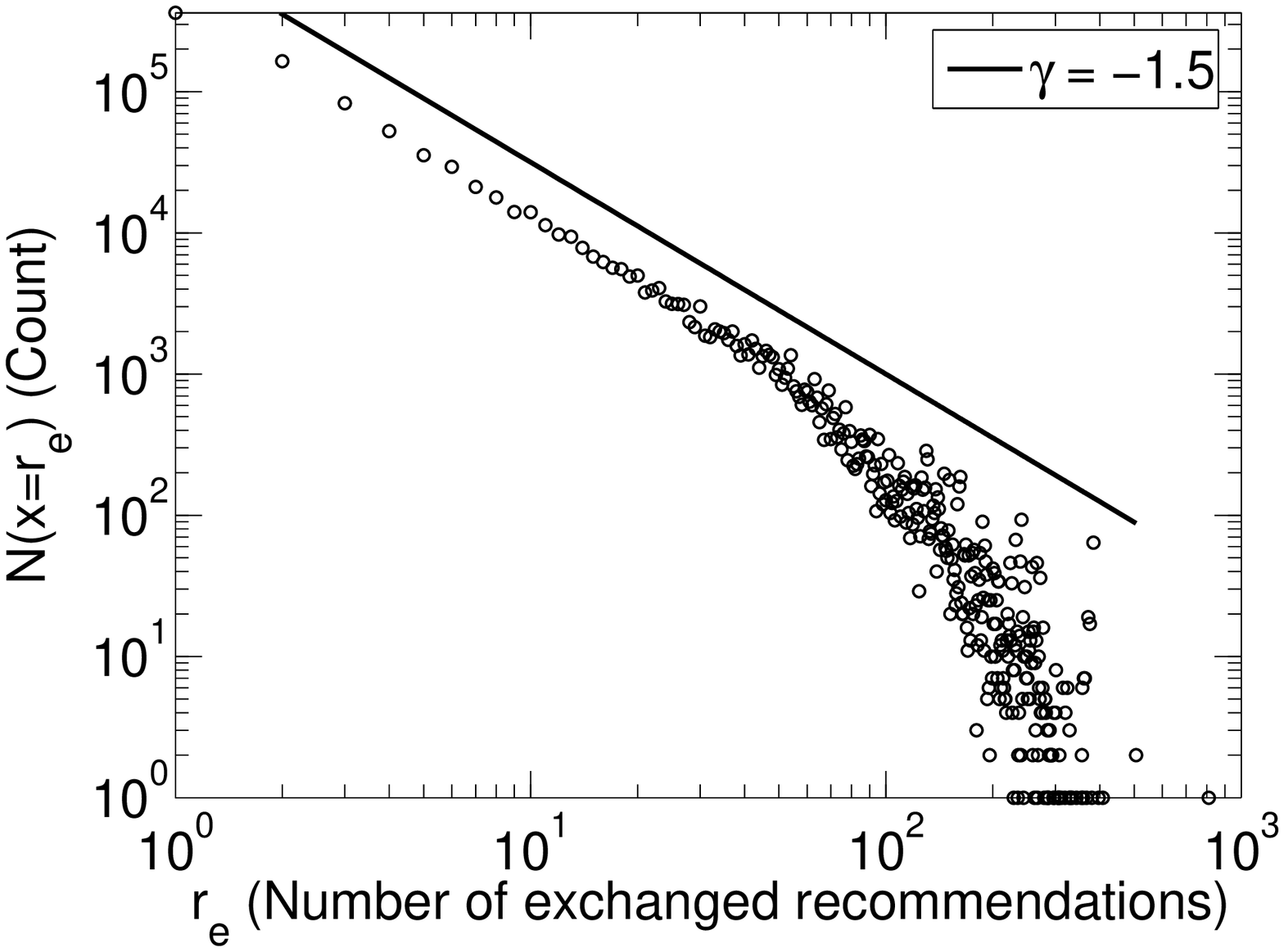} \\
  (a) All  & (b) Books  & (c) DVD
  \end{tabular}
\caption{Distribution of the number of exchanged
recommendations between pairs of people.} \label{fig:exchRecs}
\end{center}
\end{figure}

Last, we examine the number of exchanged recommendations between a pair
of people in figure~\ref{fig:exchRecs}. Overall, 39\% of pairs of people
exchanged just a single recommendation. This number decreases for DVDs
to 37\%, and increases for books to 45\%. The distribution of the number
of exchanged recommendations follows a heavy tailed distribution. To get
a better understanding of the distributions we show the power-law decay
lines. Notice that one gets much stronger decay exponent (distribution
has weaker tail) of -2.7 for books and a very shallow power-law exponent
of -1.5 for DVDs. This means that even a pair of people exchanges more
DVD than book recommendations.

\subsection{The recommendation propagation model}
\label{sec:propModel}

A simple model can help explain how the wide variance we observe in the
number of recommendations made by individuals can lead to power-laws in
cascade sizes (figure~\ref{fig:cascadeSzDist}). The model assumes that
each recipient of a recommendation will forward it to others if its
value exceeds an arbitrary threshold that the individual sets for
herself. Since exceeding this value is a probabilistic event, let's call
$p_t$ the probability that at time step $t$ the recommendation exceeds
the threshold. In that case the number of recommendations $N_{t+1}$ at
time $(t+1)$ is given in terms of the number of recommendations at an
earlier time by

\begin{equation}
  N_{t+1} = p_t N_t
\end{equation}

where the probability $p_t$ is defined over the unit interval.

Notice that, because of the probabilistic nature of the threshold being
exceeded, one can only compute the final distribution of recommendation
chain lengths, which we now proceed to do.

Subtracting from both sides of this equation the term $N_t$ and diving
by it we obtain

\begin{equation}
  \frac{N_{(t+1)}-N_{t}}{N_t} = p_t - 1
\end{equation}

Summing both sides from the initial time to some very large time $T$ and
assuming that for long times the numerator is smaller than the
denominator (a reasonable assumption) we get, up to a unit constant

\begin{equation}
  \frac{dN}{N} = \sum p_t
\end{equation}

The left hand integral is just $\log(N)$, and the right hand side is a
sum of random variables, which in the limit of a very large uncorrelated
number of recommendations is normally distributed (central limit
theorem).

This means that the logarithm of the number of messages is normally
distributed, or equivalently, that the number of messages passed is
log-normally distributed. In other words the probability density for $N$
is given by

\begin{equation}
  P(N) = \frac{1}{N\sqrt{2\pi\sigma^2}} \exp
    \frac{-(\log(N)-\mu)^{2}}{2\sigma^{2}}
\end{equation}

\noindent which, for large variances describes a behavior whereby the
typical number of recommendations is small (the mode of the
distribution) but there are unlikely events of large chains of
recommendations which are also observable.

Furthermore, for large variances, the lognormal distribution can behave
like a power law for a range of values. In order to see this, take the
logarithms on both sides of the equation (equivalent to a log-log plot)
and one obtains

\begin{equation}
  \log(P(N)) = -\log(N) -\log(\sqrt{2\pi\sigma^2}) -
  \frac{(\log{(N)-\mu)^{2}}}{2\sigma^{2}}
\end{equation}

So, for large $\sigma$, the last term of the right hand side goes to
zero, and since the second term is a constant one obtains a power law
behavior with exponent value of minus one. There are other models which
produce power-law distributions of cascade sizes, but we present ours
for its simplicity, since it does not depend on network
topology~\cite{Gruhl:2004a} or critical thresholds in the probability of
a recommendation being accepted~\cite{Watts:2002}.

\section{Success of Recommendations}
\label{sec:buyprob}

So far we only looked into the aggregate statistics of the
recommendation network. Next, we ask questions about the effectiveness
of recommendations in the recommendation network itself. First, we
analyze the probability of purchasing as one gets more and more
recommendations. Next, we measure recommendation effectiveness as two
people exchange more and more recommendations. Lastly, we observe the
recommendation network from the perspective of the sender of the
recommendation. Does a node that makes more recommendations also
influence more purchases?

\subsection{Probability of buying versus number of incoming
recommendations} \label{sec:errBar}

First, we examine how the probability of purchasing changes as one gets
more and more recommendations. One would expect that a person is more
likely to buy a product if she gets more recommendations. On the other
had one would also think that there is a saturation point -- if a person
hasn't bought a product after a number of recommendations, they are not
likely to change their minds after receiving even more of them. So, how
many recommendations are too many?

\begin{figure}
\begin{center}
\begin{tabular}{cc}
  \includegraphics[width=0.48\textwidth]{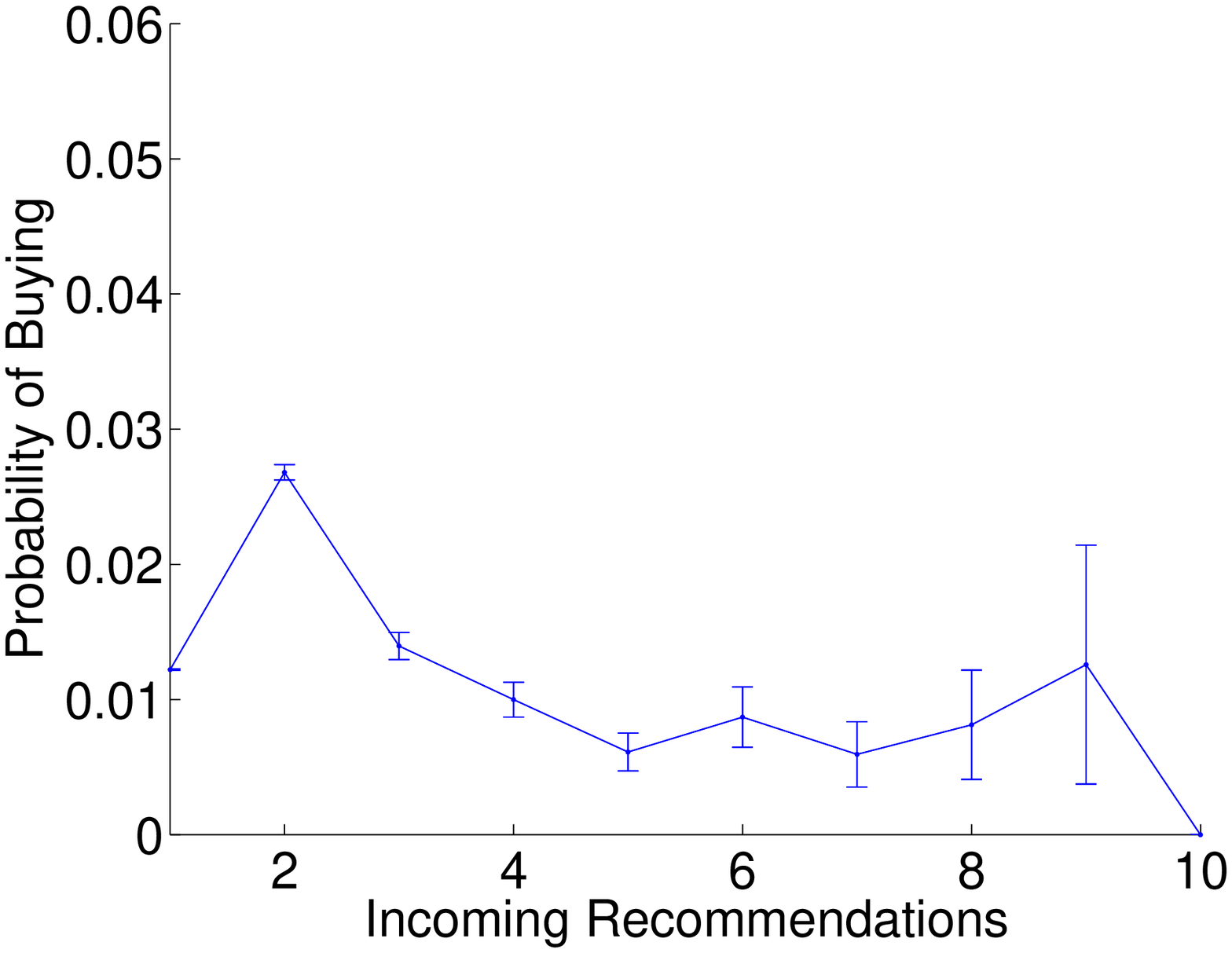} &
  \includegraphics[width=0.48\textwidth]{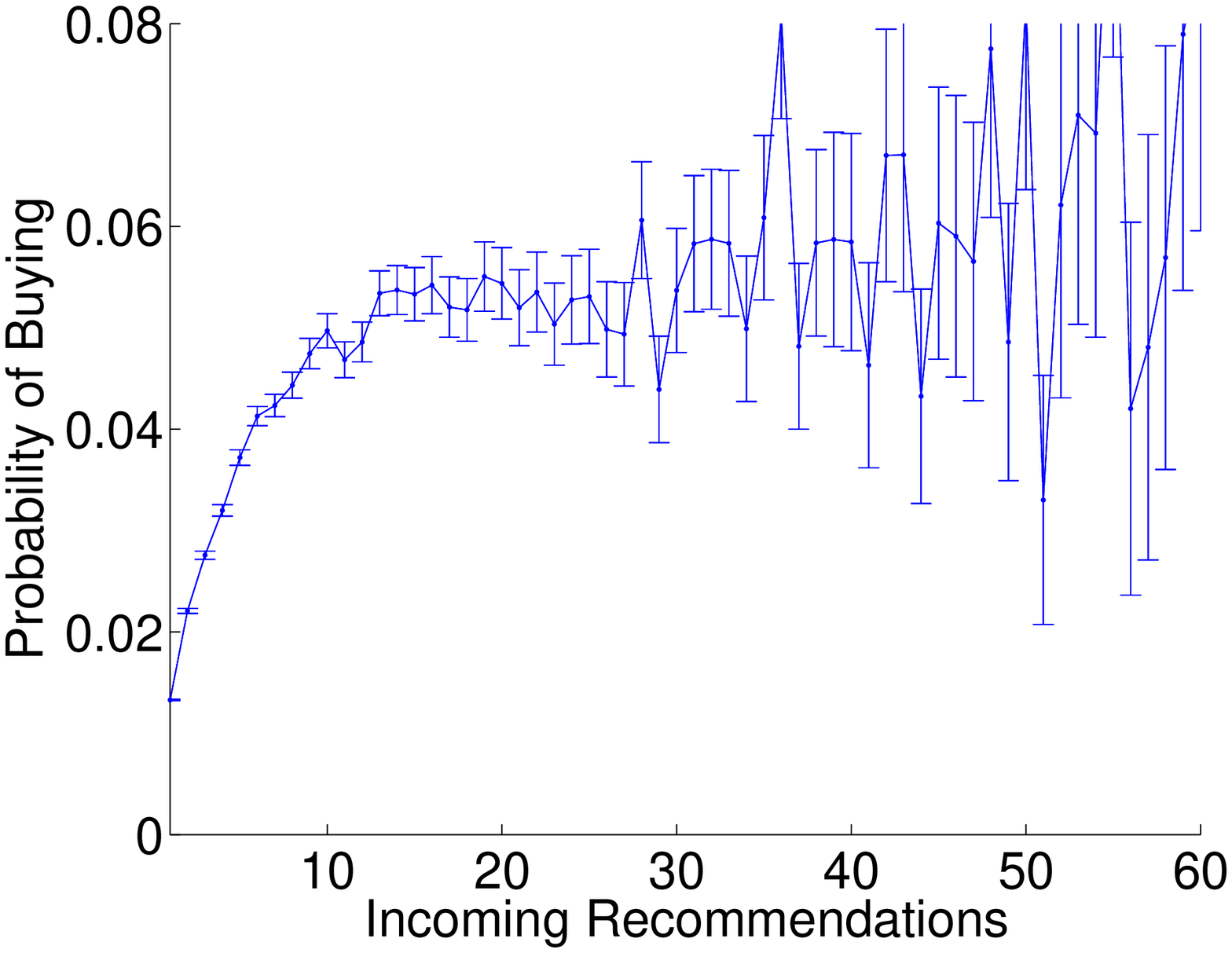} \\
  (a) Books & (b) DVD\\
  \includegraphics[width=0.48\textwidth]{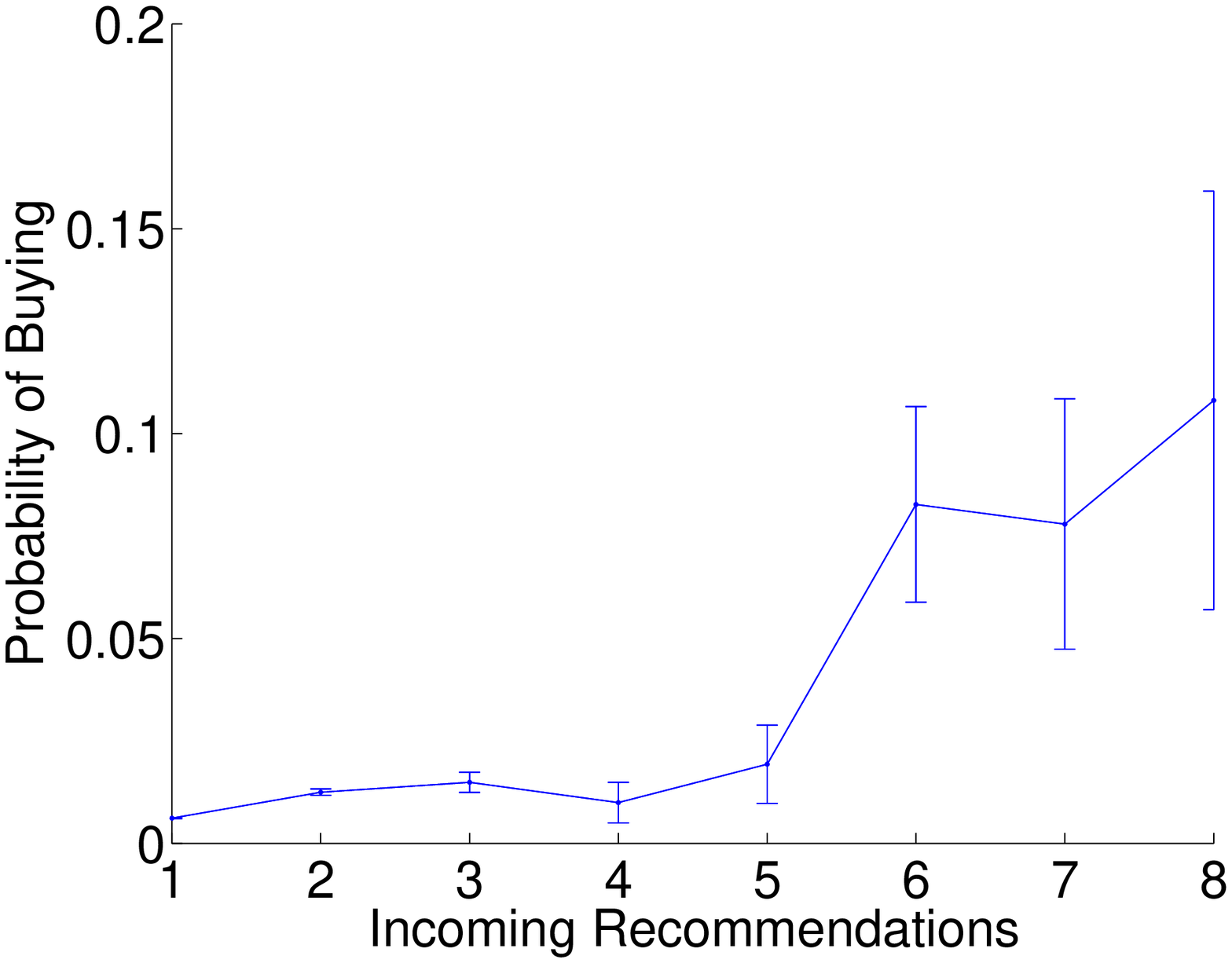} &
  \includegraphics[width=0.48\textwidth]{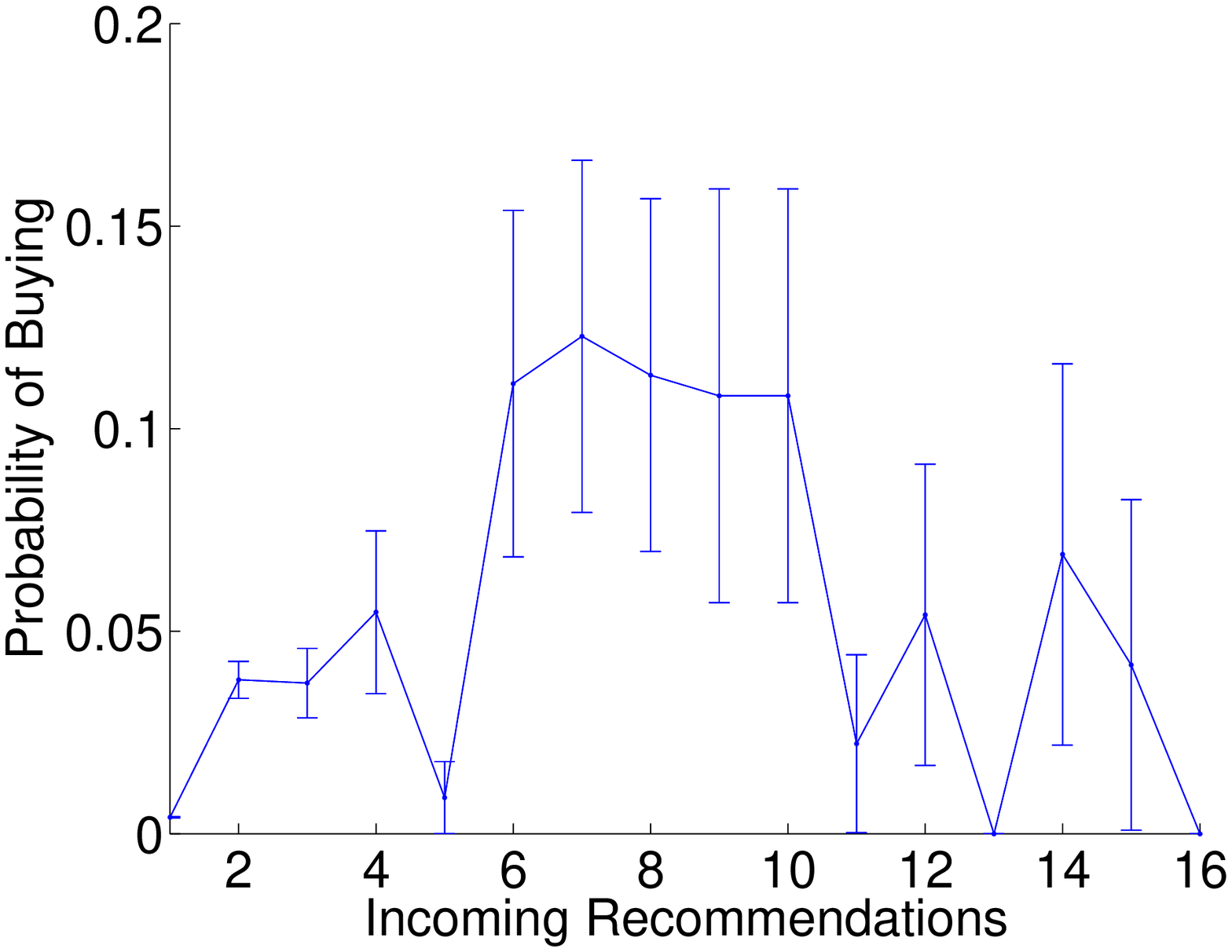} \\
  (c) Music & (d) Video \\
\end{tabular}
\caption{Probability of buying a book (DVD) given a number of
incoming recommendations.} \label{fig:inRecBuyProb}
\end{center}
\end{figure}

Figure~\ref{fig:inRecBuyProb} shows the probability of purchasing a
product as a function of the number of incoming recommendations on the
product. Because we exclude late recommendations, those that were
received after the purchase, an individual counts as having received
three recommendations only if they did not make a purchase after the
first two, and either purchased or did not receive further
recommendations after receiving the third one. As we move to higher
numbers of incoming recommendations, the number of observations drops
rapidly. For example, there were 5 million cases with 1 incoming
recommendation on a book, and only 58 cases where a person got 20
incoming recommendations on a particular book. The maximum was 30
incoming recommendations. For these reasons we cut-off the plot when the
number of observations becomes too small and the error bars too large.

We calculate the purchase probabilities and the standard errors of the
estimates which we use to plot the error bars in the following way. We
regard each point as a binomial random variable. Given the number of
observations n, let m be the number of successes, and k (k=n-m) the
number of failures. In our case, m is the number of people that first
purchased a product after receiving r recommendations on it, and k is
the number of people that received the total of r recommendations on a
product (till the end of the dataset) but did purchase it, then the
estimated probability of purchasing is $\hat{p} = m / n$ and the
standard error $s_{\hat{p}}$ of estimate $\hat{p}$ is $s_{\hat{p}} =
\sqrt{{p (1-p)}/{n}}$.

Figure~\ref{fig:inRecBuyProb}(a) shows that, overall, book
recommendations are rarely followed. Even more surprisingly, as more and
more recommendations are received, their success decreases. We observe a
peak in probability of buying at 2 incoming recommendations and then a
slow drop. This implies that if a person doesn't buy a book after the
first recommendation, but receives another, they are more likely to be
persuaded by the second recommendation. But thereafter, they are less
likely to respond to additional recommendations, possibly because they
perceive them as spam, are less susceptible to others' opinions, have a
strong opinion on the particular product, or have a different means of
accessing it.

For DVDs (figure~\ref{fig:inRecBuyProb}(b)) we observe a saturation
around 10 incoming recommendations. This means that with each additional
recommendation, a person is more and more likely to be persuaded - up to
a point. After a person gets 10 recommendations on a particular DVD,
their probability of buying does not increase anymore. The number of
observations is 2.5 million at 1 incoming recommendation and 100 at 60
incoming recommendations. The maximal number of received recommendations
is 172 (and that person did not buy), but someone purchased a DVD after
169 receiving recommendations. The different patterns between book and
DVD recommendations may be a result of the recommendation exchange
websites for DVDs. Someone receiving many DVD recommendations may have
signed up to receive them for a product they intended to purchase, and
hence a greater number of received recommendations corresponds to a
higher likelihood of purchase (up to a point).

\subsection{Success of subsequent recommendations}

Next, we analyze how the effectiveness of recommendations changes as one
received more and more recommendations from the same person. A large
number of exchanged recommendations can be a sign of trust and
influence, but a sender of too many recommendations can be perceived as
a spammer. A person who recommends only a few products will have her
friends' attention, but one who floods her friends with all sorts of
recommendations will start to loose her influence.

We measure the effectiveness of recommendations as a function of the
total number of previously received recommendations from a particular
node. We thus measure how spending changes over time, where time is
measured in the number of received recommendations.

We construct the experiment in the following way. For every
recommendation $r$ on some product $p$ between nodes $u$ and $v$, we
first determine how many recommendations node $u$ received from $v$
before getting $r$. Then we check whether $v$, the recipient of
recommendation, purchased $p$ after the recommendation $r$ arrived. If
so, we count the recommendation as successful since it influenced the
purchase. This way we can calculate the recommendation success rate as
more recommendations were exchanged. For the experiment we consider only
node pairs $(u,v)$, where there were at least a total of 10
recommendations sent from $u$ to $v$. We perform the experiment using
only recommendations from the same product group.

We decided to set a lower limit on the number of exchanged
recommendations so that we can measure how the effectiveness of
recommendations changes as the {\em same} two people exchange more and
more recommendations. Considering all pairs of people would heavily bias
our findings since most pairs exchange just a few or even just a single
recommendation. Using the data from figure~\ref{fig:exchRecs} we see
that 91\% of pairs of people that exchange at least 1 recommendation
exchange less than 10. For books this number increases to 96\%, and for
DVDs it is even smaller (81\%). In the DVD network there are 182
thousand pairs that exchanged more than 10 recommendations, and 70
thousand for the book network.

\begin{figure}[t]
\begin{center}
\begin{tabular}{cc}
  \includegraphics[width=0.48\textwidth]{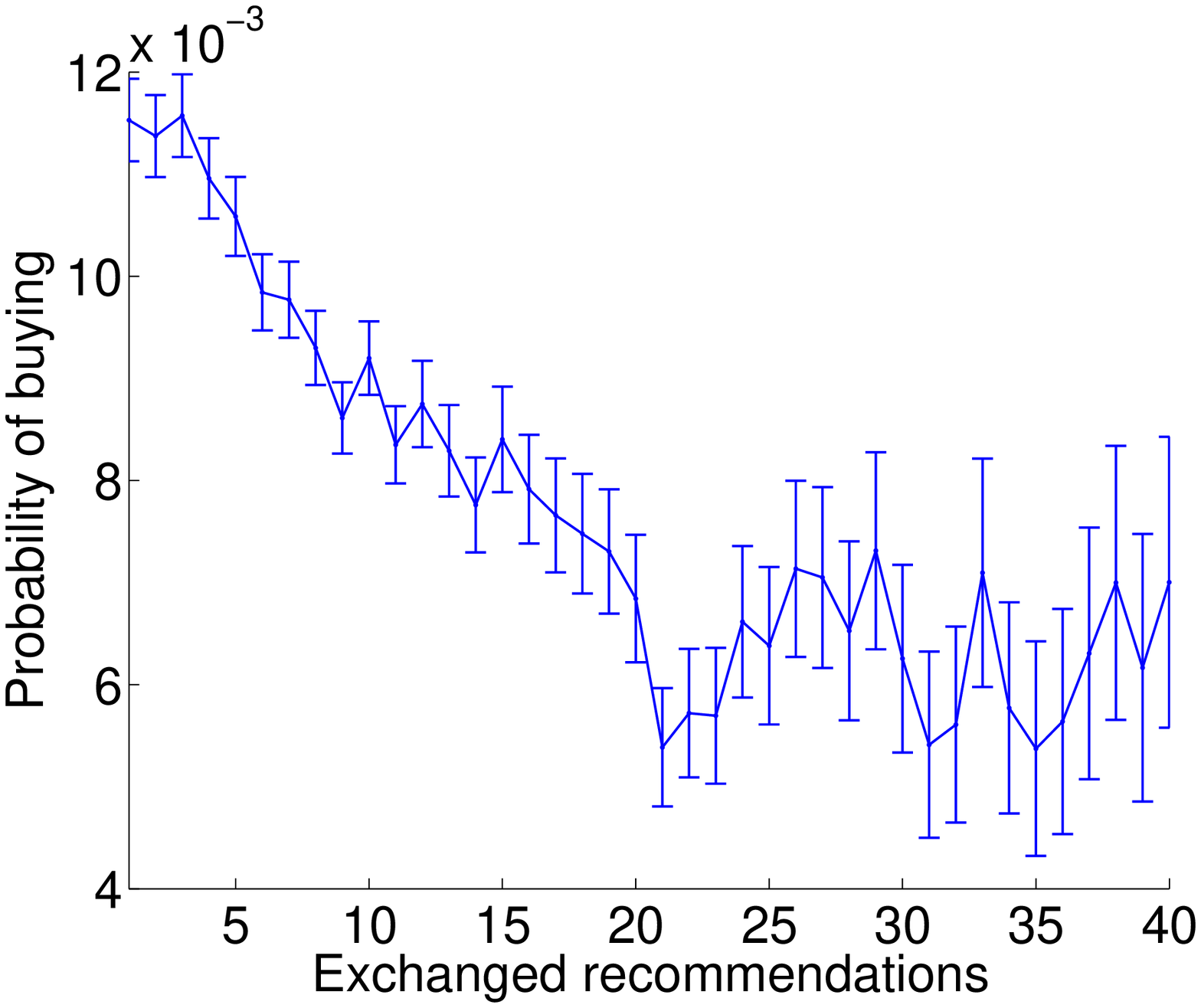} &
  \includegraphics[width=0.48\textwidth]{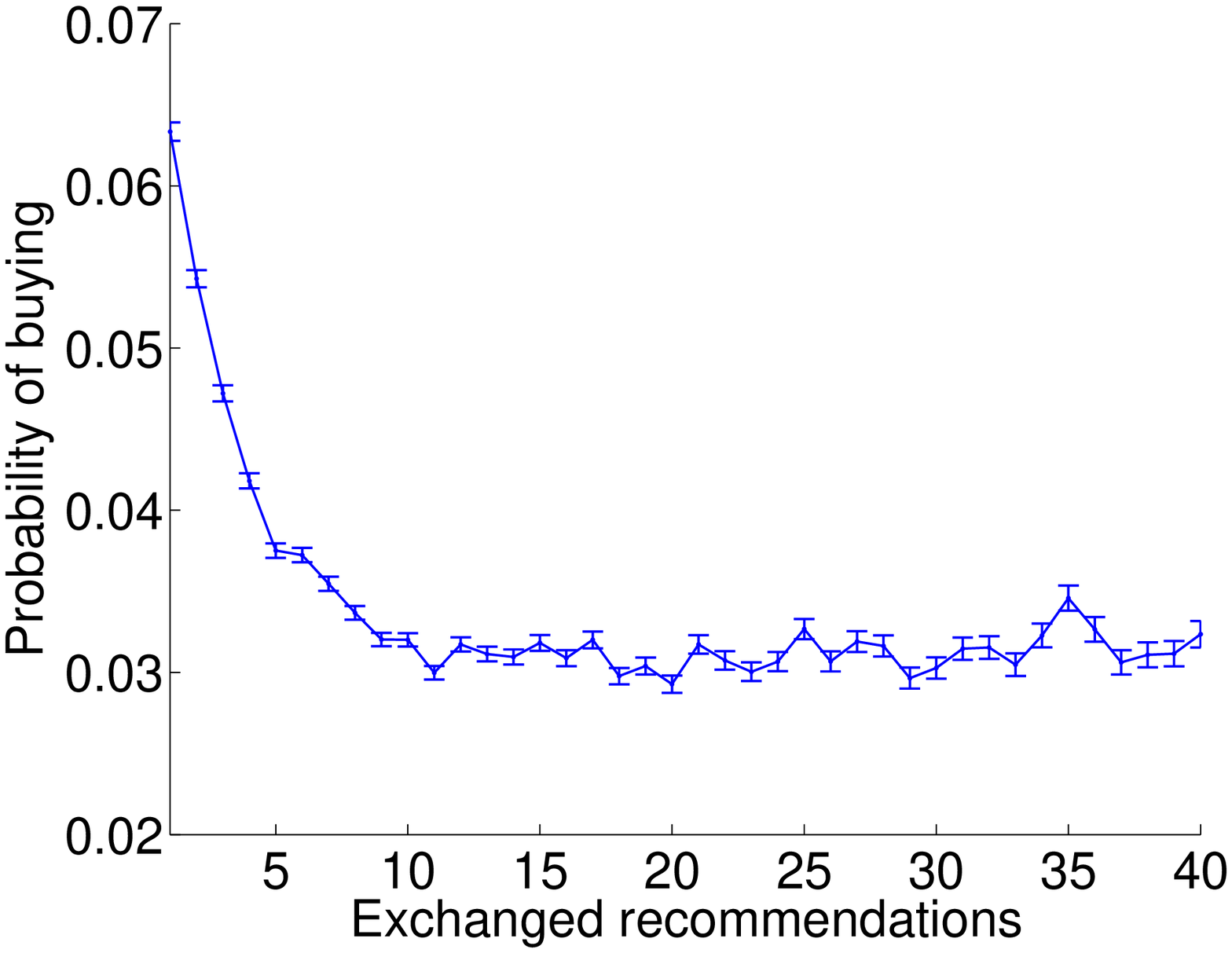} \\
  (a) Books & (b) DVD \\
\end{tabular}
\caption{The effectiveness of recommendations with the number of
received recommendations.} \label{fig:recEffect}
\end{center}
\end{figure}

Figure~\ref{fig:recEffect} shows the probability of buying as a function
of the total number of received recommendations from a particular person
up to that point. One can think of x-axis as measuring time where the
unit is the number of received recommendations from a particular person.

For books we observe that the effectiveness of recommendation remains
about constant up to 3 exchanged recommendations. As the number of
exchanged recommendations increases, the probability of buying starts to
decrease to about half of the original value and then levels off. For
DVDs we observe an immediate and consistent drop. We performed the
experiment also for video and music, but the number of observations was
too low and the measurements were noisy. This experiment shows that
recommendations start to lose effect after more than two or three are
passed between two people. Also, notice that the effectiveness of book
recommendations decays much more slowly than that of DVD
recommendations, flattening out at around 20 recommendations, compared
to around 10 DVD exchanged recommendations.

The result has important implications for viral marketing because
providing too much incentive for people to recommend to one another can
weaken the very social network links that the marketer is intending to
exploit.

\subsection{Success of outgoing recommendations}

In previous sections we examined the data from the viewpoint of the
receiver of the recommendation. Now we look from the viewpoint of the
sender. The two interesting questions are: how does the probability of
getting a 10\% credit change with the number of outgoing
recommendations; and given a number of outgoing recommendations, how
many purchases will they influence?

One would expect that recommendations would be the most effective when
recommended to the right subset of friends. If one is very selective and
recommends to too few friends, then the chances of success are slim. One
the other hand, recommending to everyone and spamming them with
recommendations may have limited returns as well.

\begin{figure}[t]
\begin{center}
\begin{tabular}{cccc}
  \includegraphics[width=0.22\textwidth]{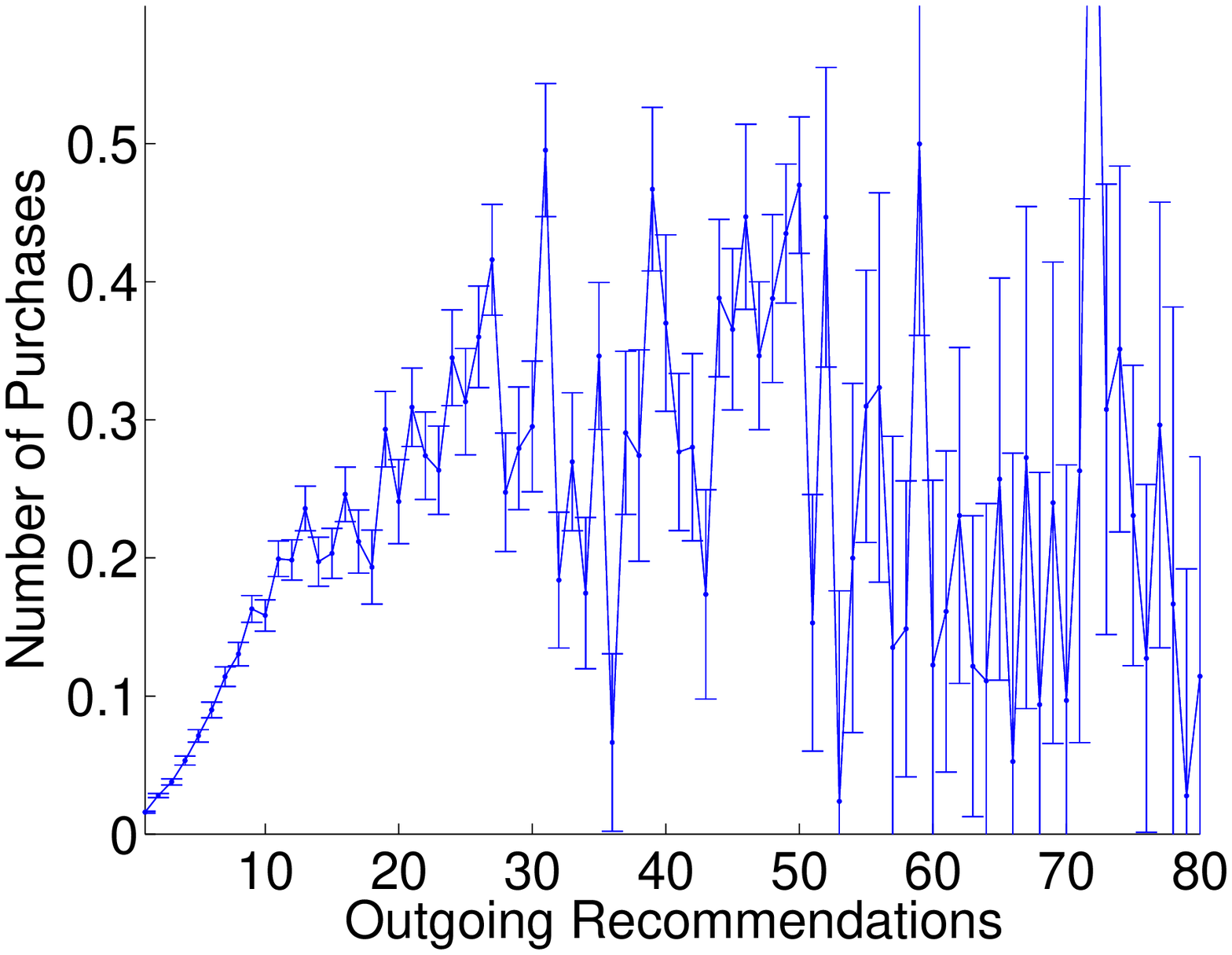} &
  \includegraphics[width=0.22\textwidth]{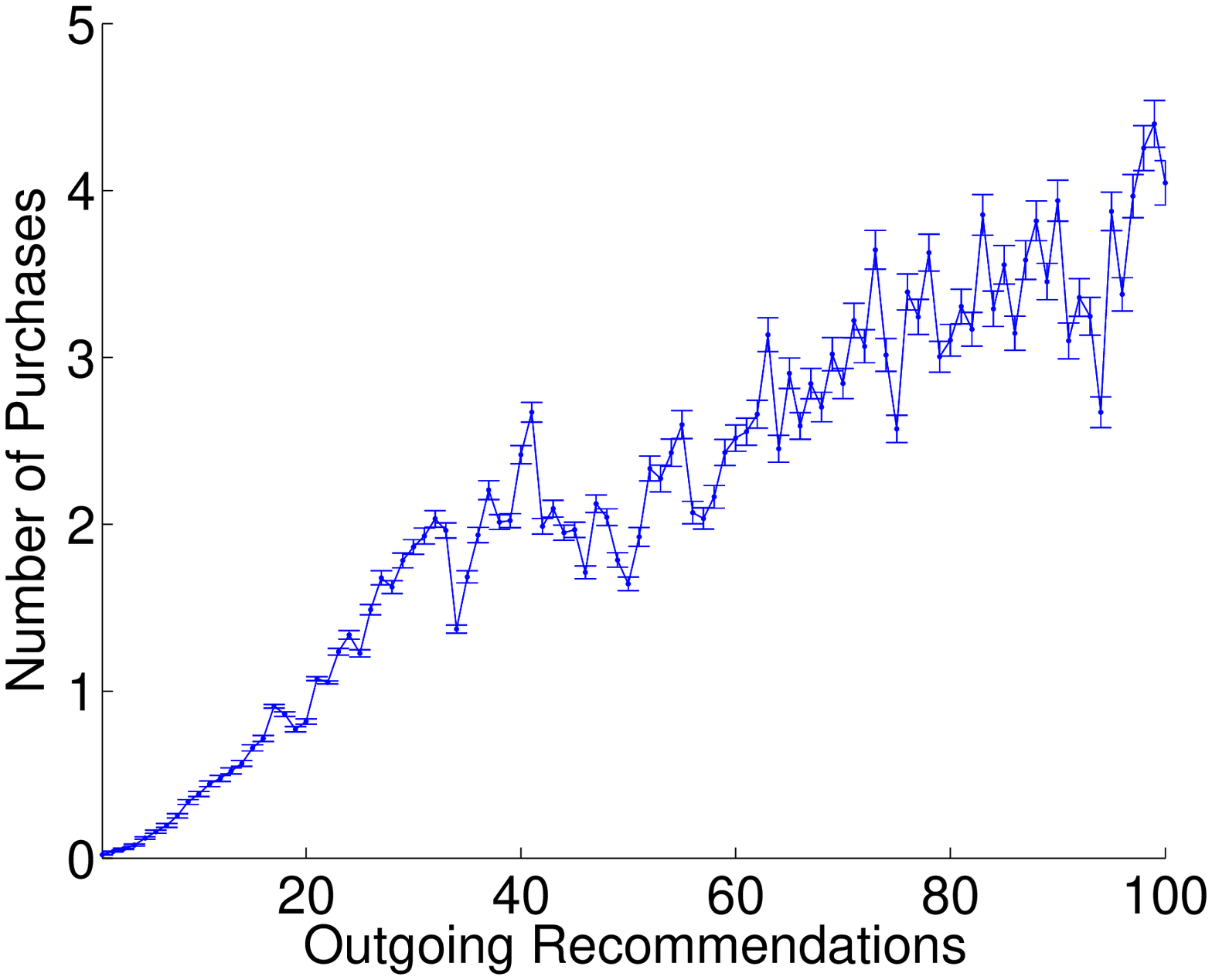} &
  \includegraphics[width=0.22\textwidth]{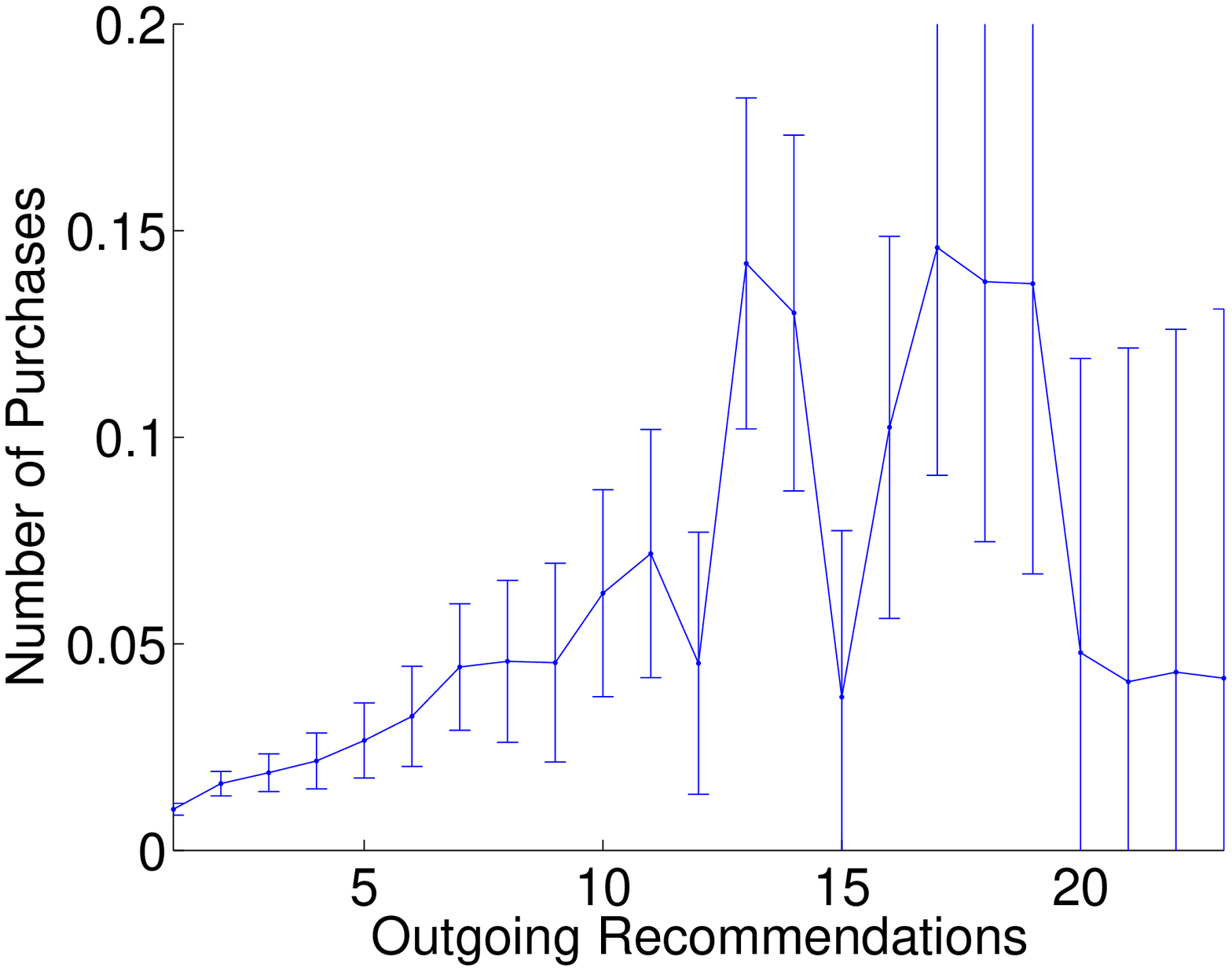} &
  \includegraphics[width=0.22\textwidth]{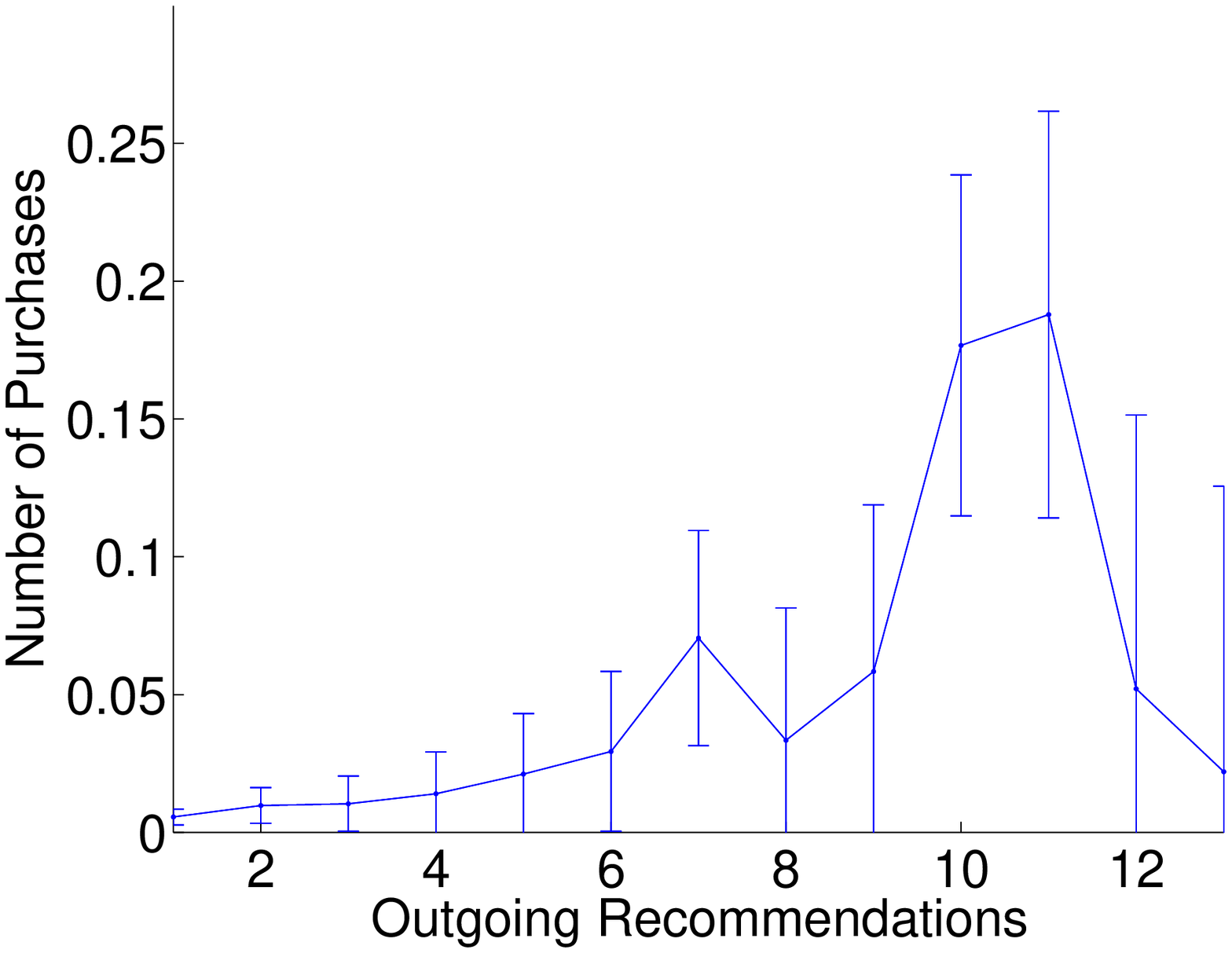} \\
  \includegraphics[width=0.22\textwidth]{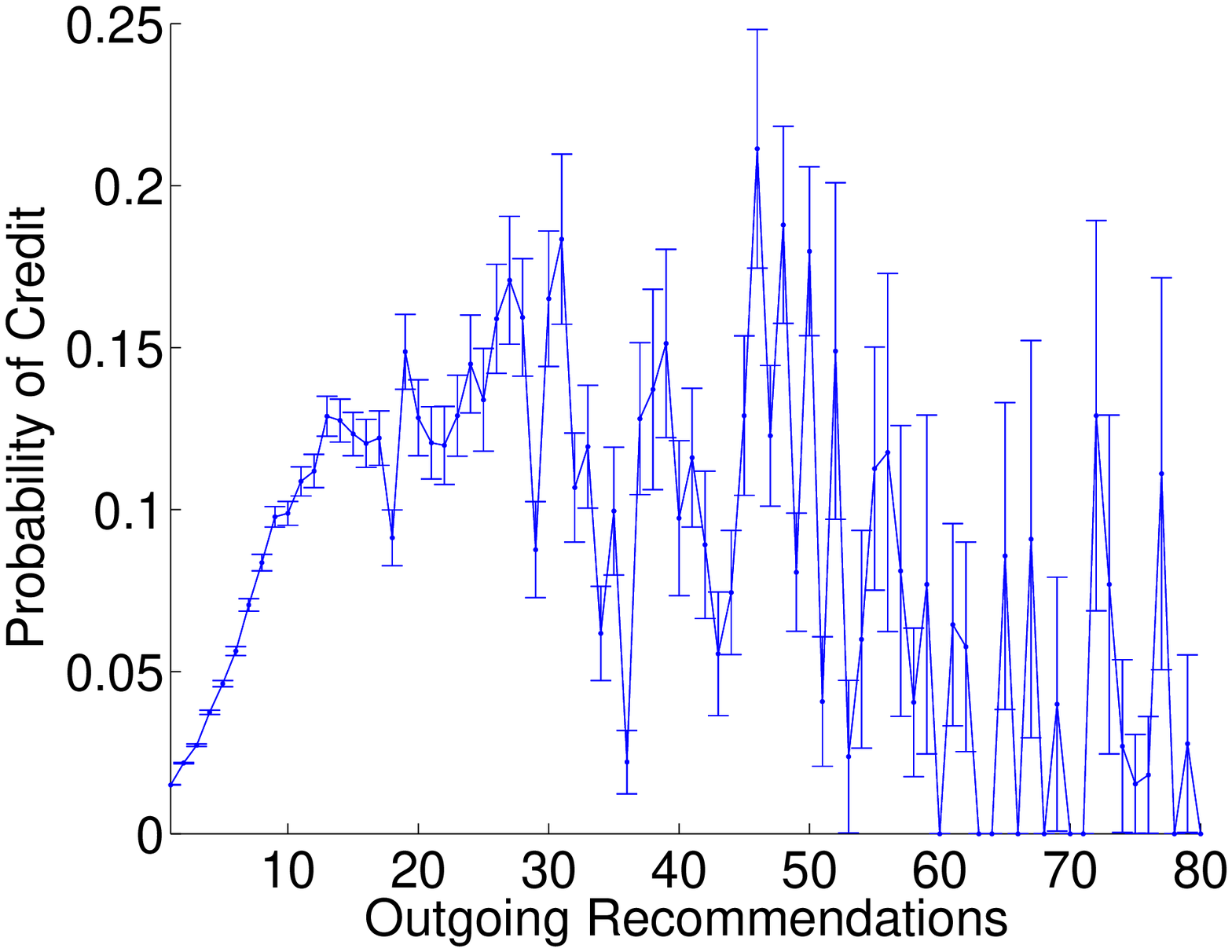} &
  \includegraphics[width=0.22\textwidth]{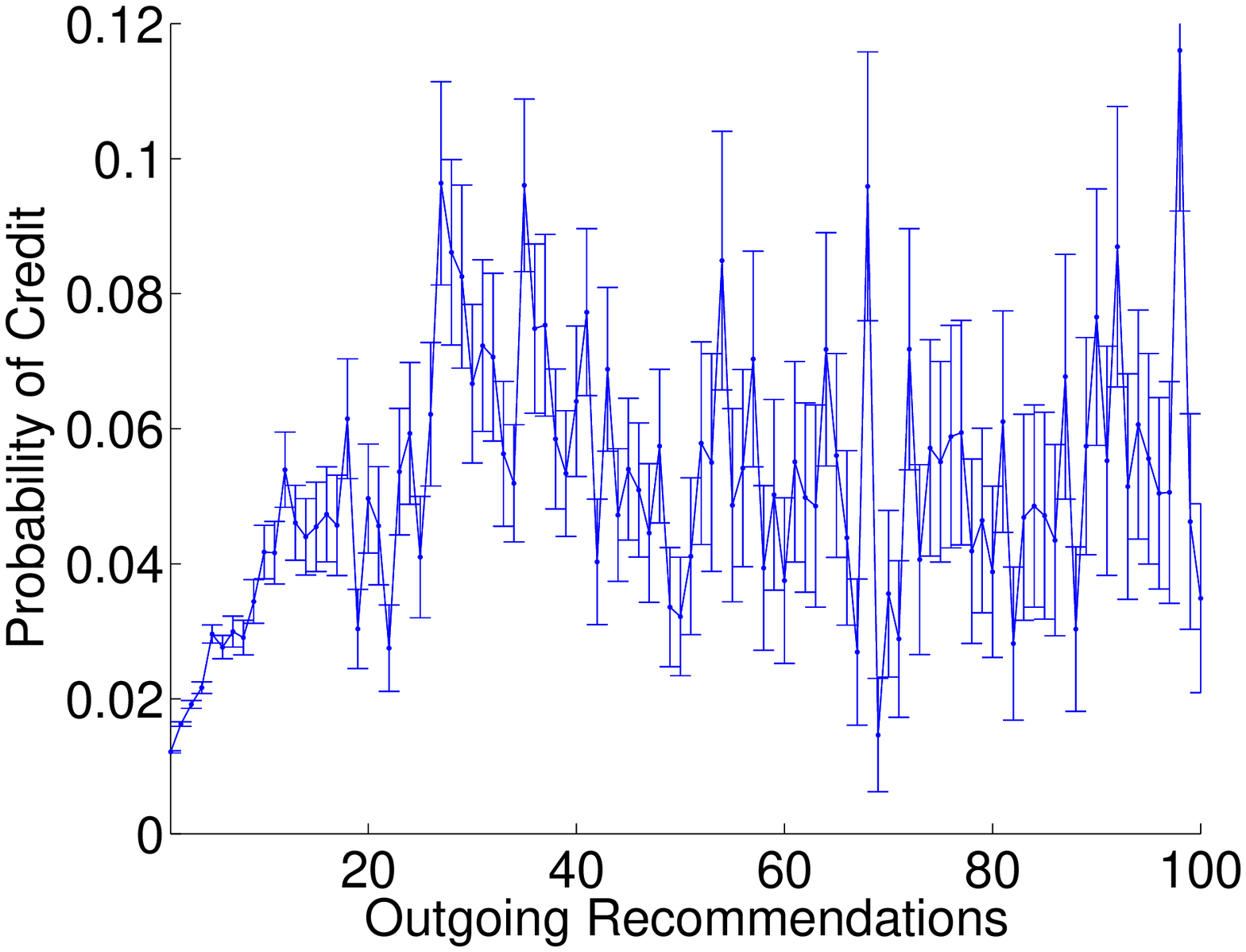} &
  \includegraphics[width=0.22\textwidth]{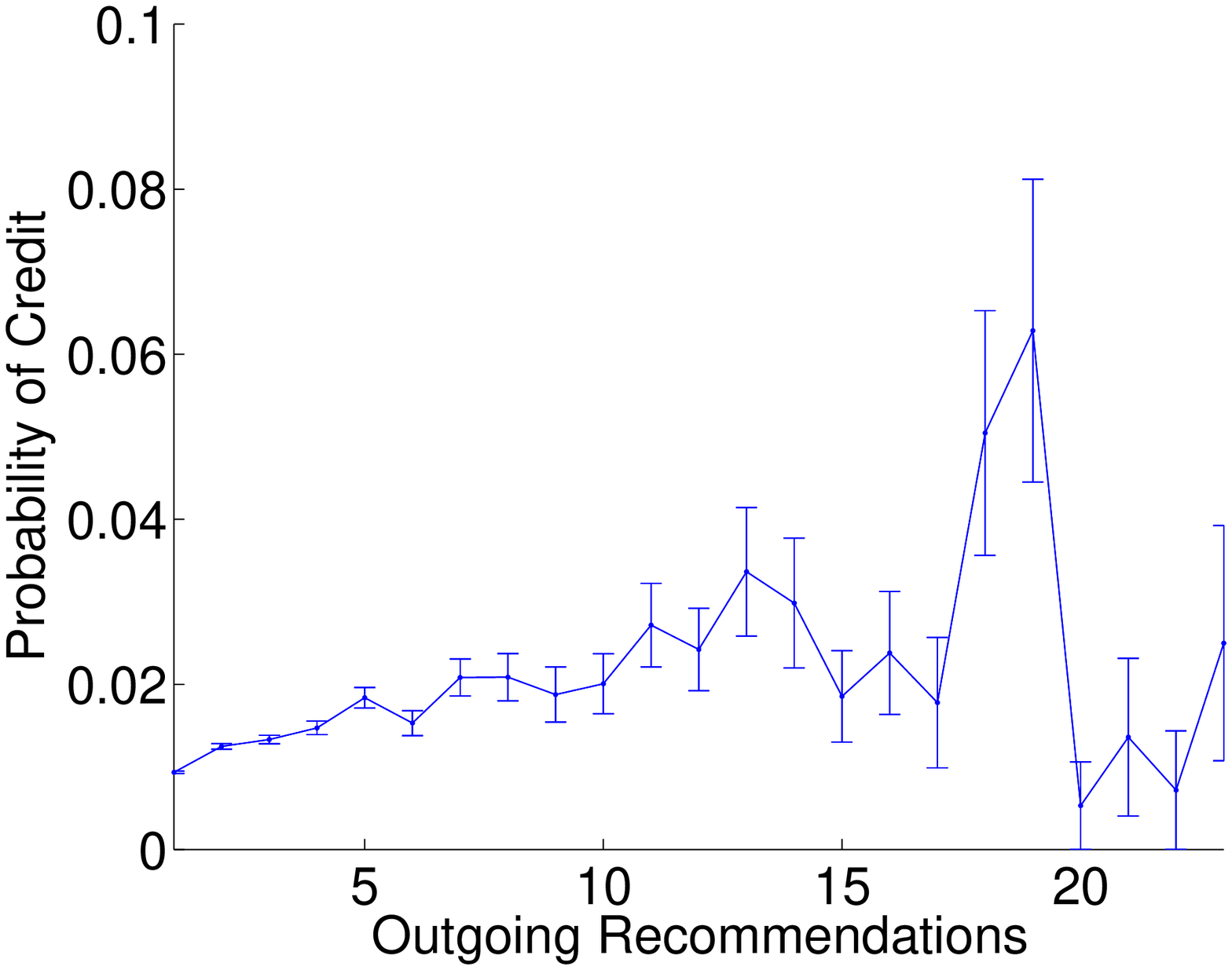} &
  \includegraphics[width=0.22\textwidth]{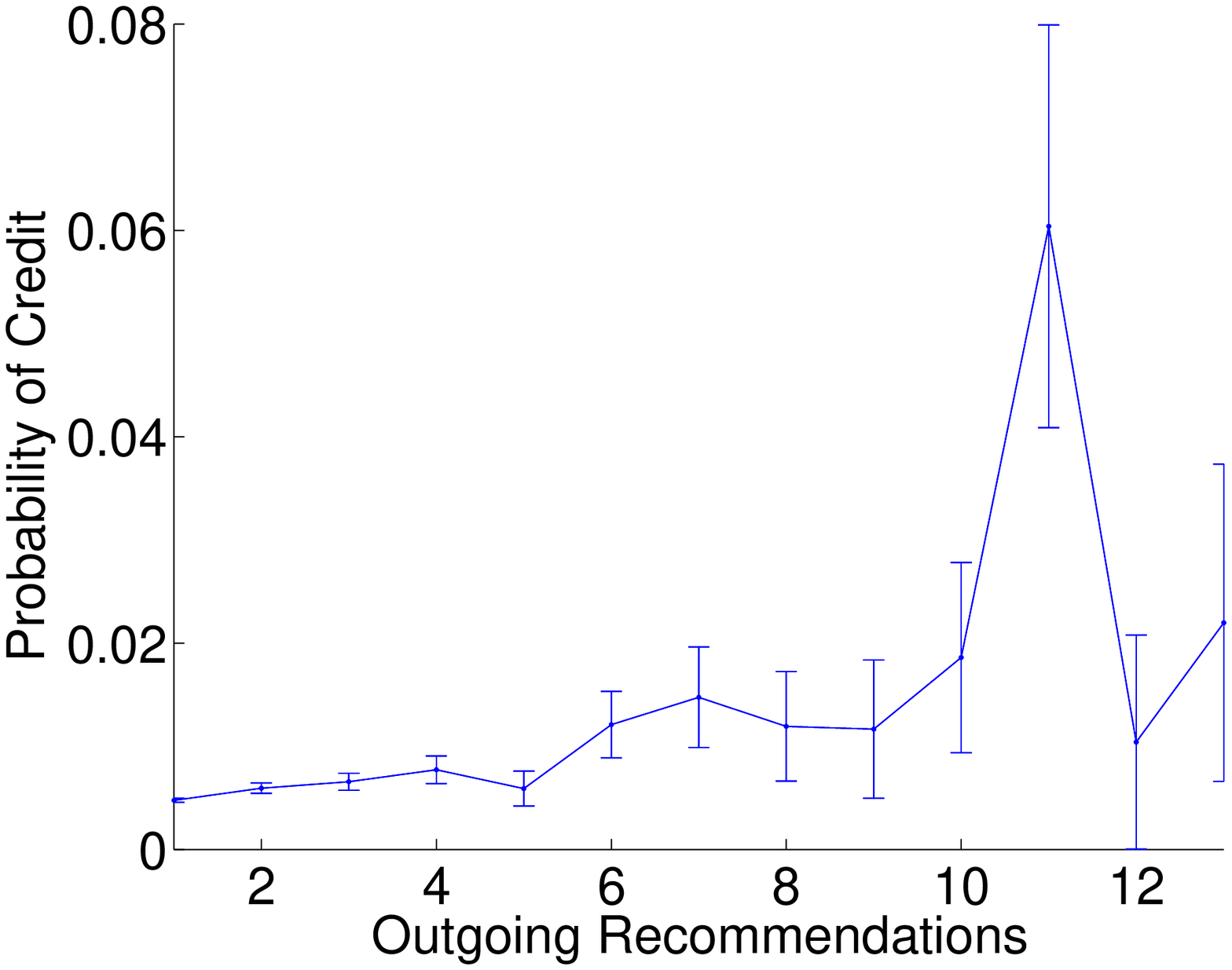} \\
    (a) Books & (b) DVD & (c) Music & (d) Video \\
\end{tabular}
\caption{Top row: Number of resulting purchases given a number of
outgoing recommendations. Bottom row: Probability of getting a
credit given a number of outgoing recommendations.}
\label{fig:outRecBuyProb}
\end{center}
\end{figure}

The top row of figure~\ref{fig:outRecBuyProb} shows how the average
number of purchases changes with the number of outgoing recommendations.
For books, music, and videos the number of purchases soon saturates: it
grows fast up to around 10 outgoing recommendations and then the trend
either slows or starts to drop. DVDs exhibit different behavior, with
the expected number of purchases increasing throughout.

These results are even more interesting since the receiver of the
recommendation does not know how many other people also received the
recommendation. Thus the plots of figure~\ref{fig:outRecBuyProb} show
that there are interesting dependencies between the product
characteristics and the recommender that manifest through the number of
recommendations sent. It could be the case that widely recommended
products are not suitable for viral marketing (we find something similar
in section~\ref{sec:regression}), or that the recommender did not put
too much thought into who to send the recommendation to, or simply that
people soon start to ignore mass recommenders.

Plotting the probability of getting a 10\% credit as a function of the
number of outgoing recommendations, as in the bottom row of
figure~\ref{fig:outRecBuyProb}, we see that the success of DVD
recommendations saturates as well, while books, videos and music have
qualitatively similar trends. The difference in the curves for DVD
recommendations points to the presence of collisions in the dense DVD
network, which has 10 recommendations per node and around 400 per
product --- an order of magnitude more than other product groups. This
means that many different individuals are recommending to the same
person, and after that person makes a purchase, even though all of them
made a `successful recommendation' by our definition, only one of them
receives a credit.

\subsection{Probability of buying given the total number of incoming
recommendations}

The collisions of recommendations are a dominant feature of the DVD
recommendation network. Book recommendations have the highest chance of
getting a credit, but DVD recommendations cause the most purchases. So
far it seems people are very keen on recommending various DVDs, while
very conservative on recommending books. But how does the behavior of
customers change as they get more involved into the recommendation
network? We would expect that most of the people are not heavily
involved, so their probability of buying is not high. In the extreme
case we expect to find people who buy almost everything they get
recommendations on.

There are two ways to measure the involvedness of a person in the
network: by the total number of incoming recommendations (on all
products) or the total number of different products they were
recommended. For every purchase of a book at time $t$, we count the
number of different books (DVDs, ...) the person received
recommendations for before time $t$. As in all previous experiments we
delete late recommendations, {\em i.e.} recommendations that arrived
after the first purchase of a product.

\begin{figure}[t]
\begin{center}
\begin{tabular}{cc}
  \includegraphics[width=0.45\textwidth]{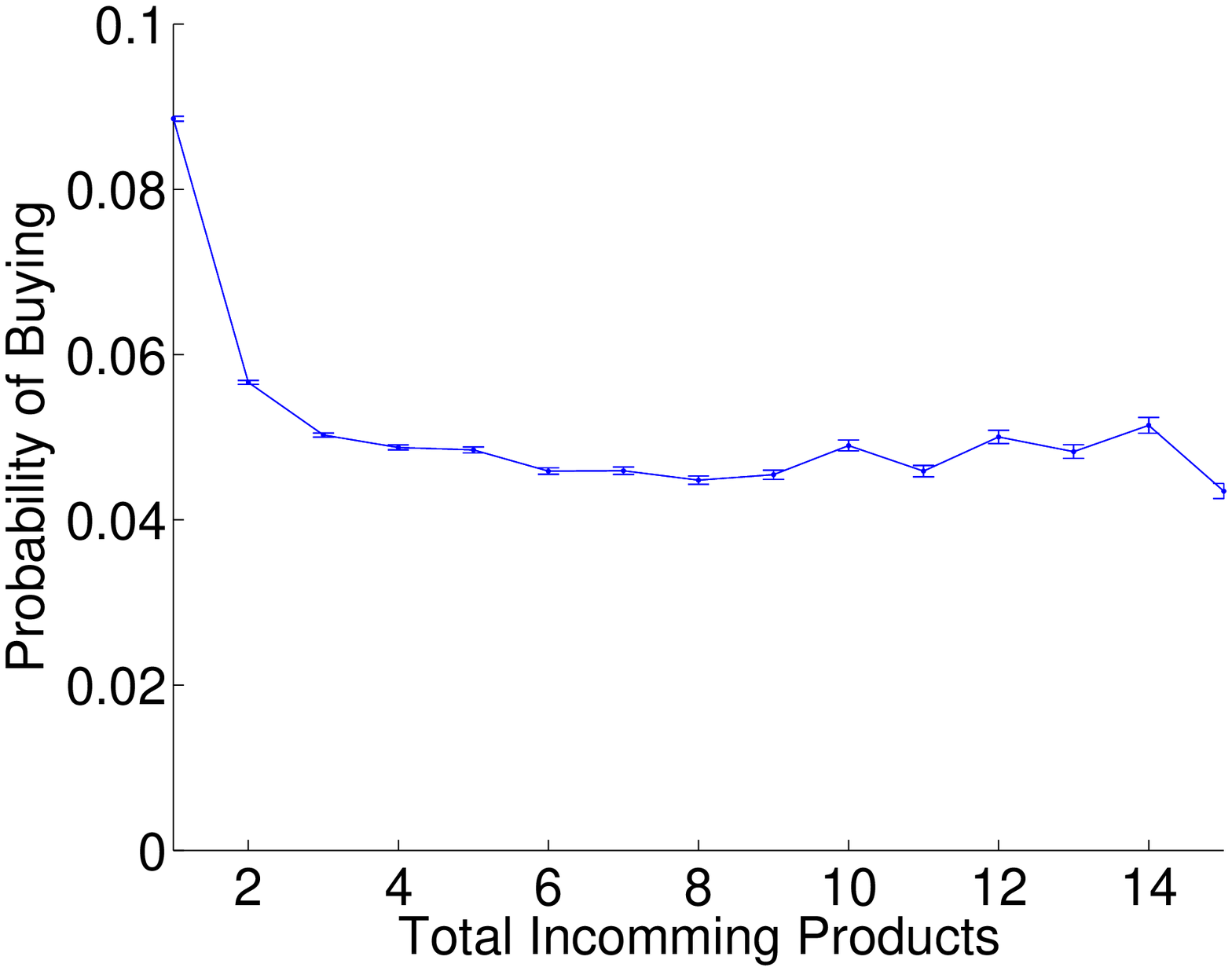} &
  \includegraphics[width=0.45\textwidth]{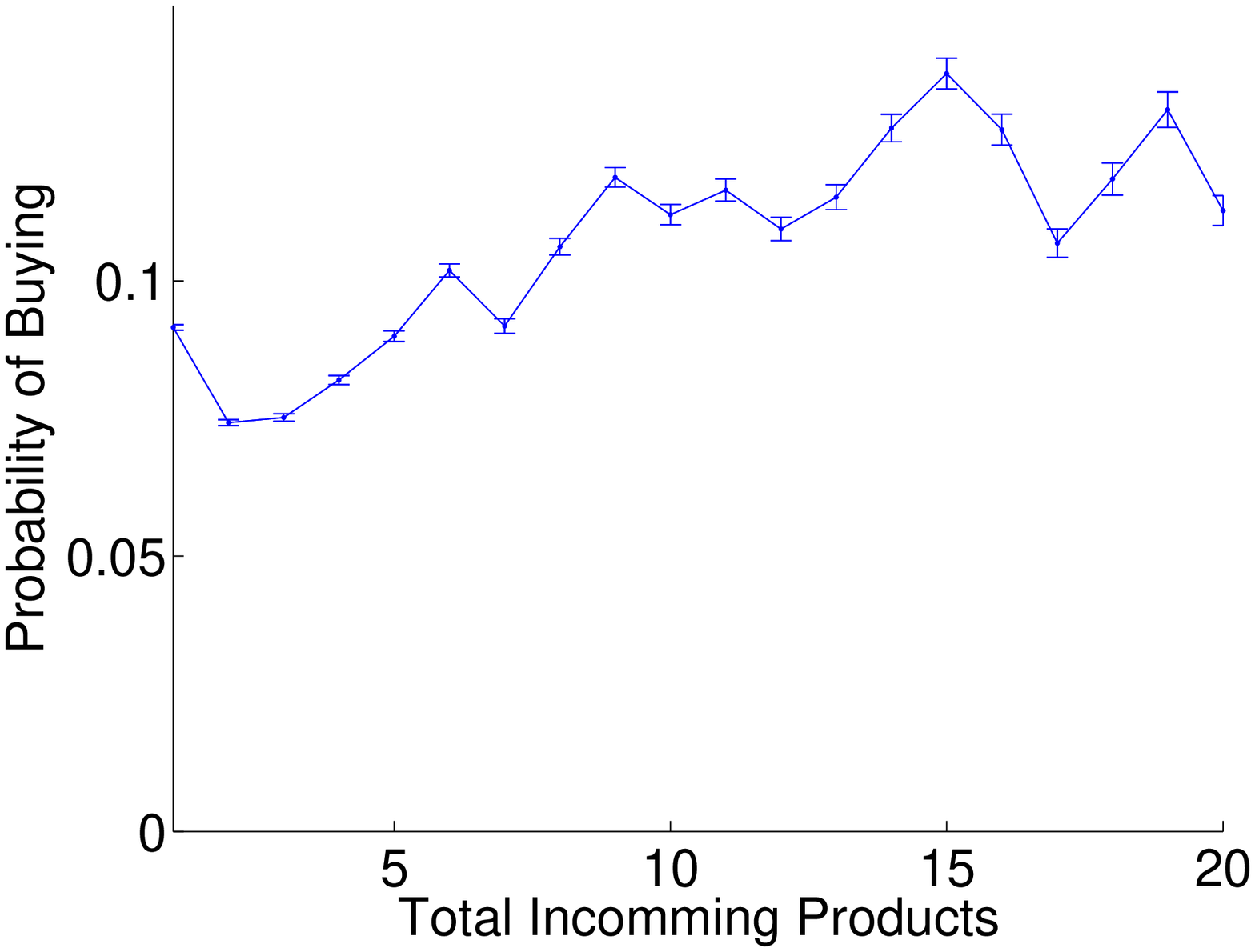} \\
  (a) Books & (b) DVD \\
  \includegraphics[width=0.45\textwidth]{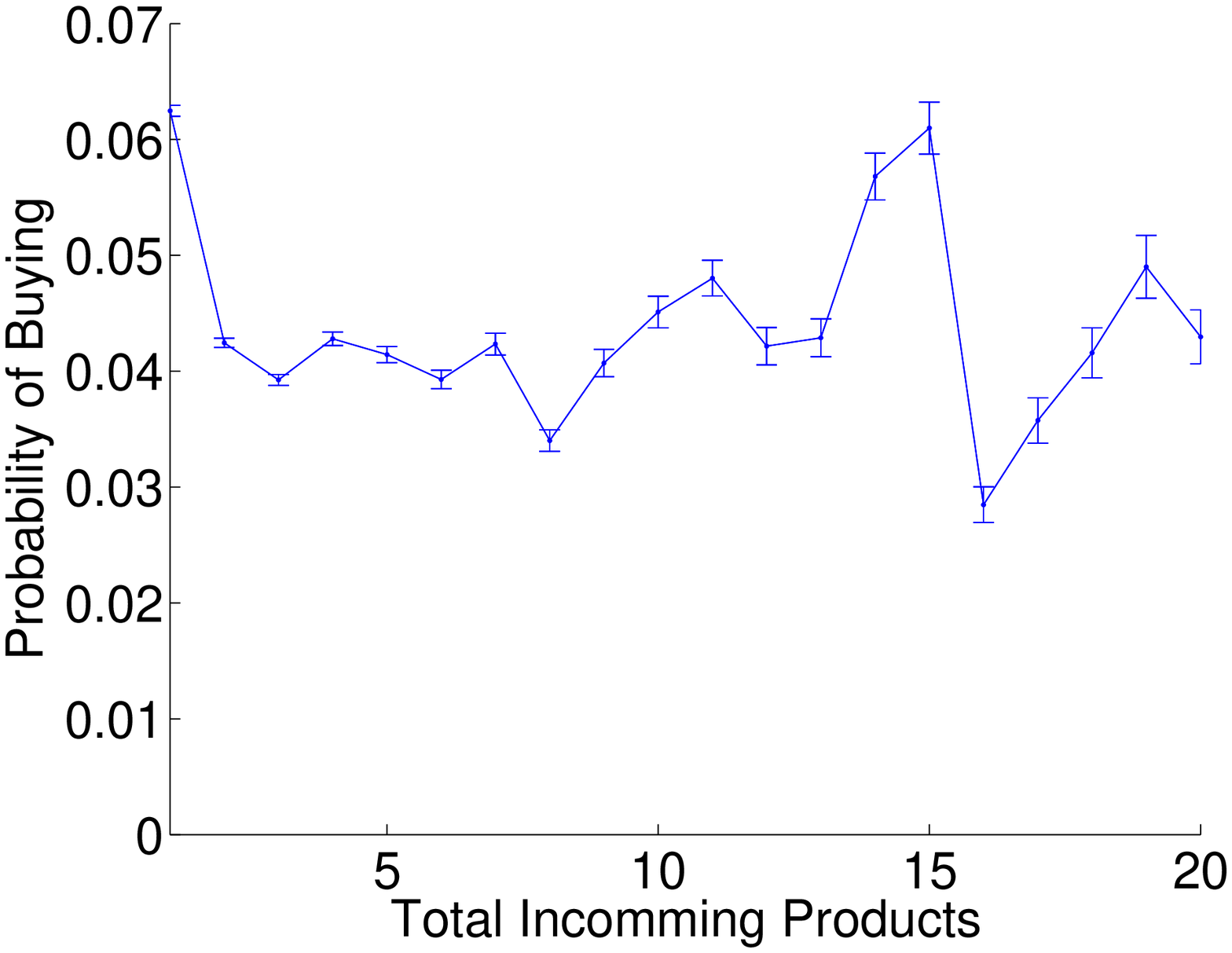} &
  \includegraphics[width=0.45\textwidth]{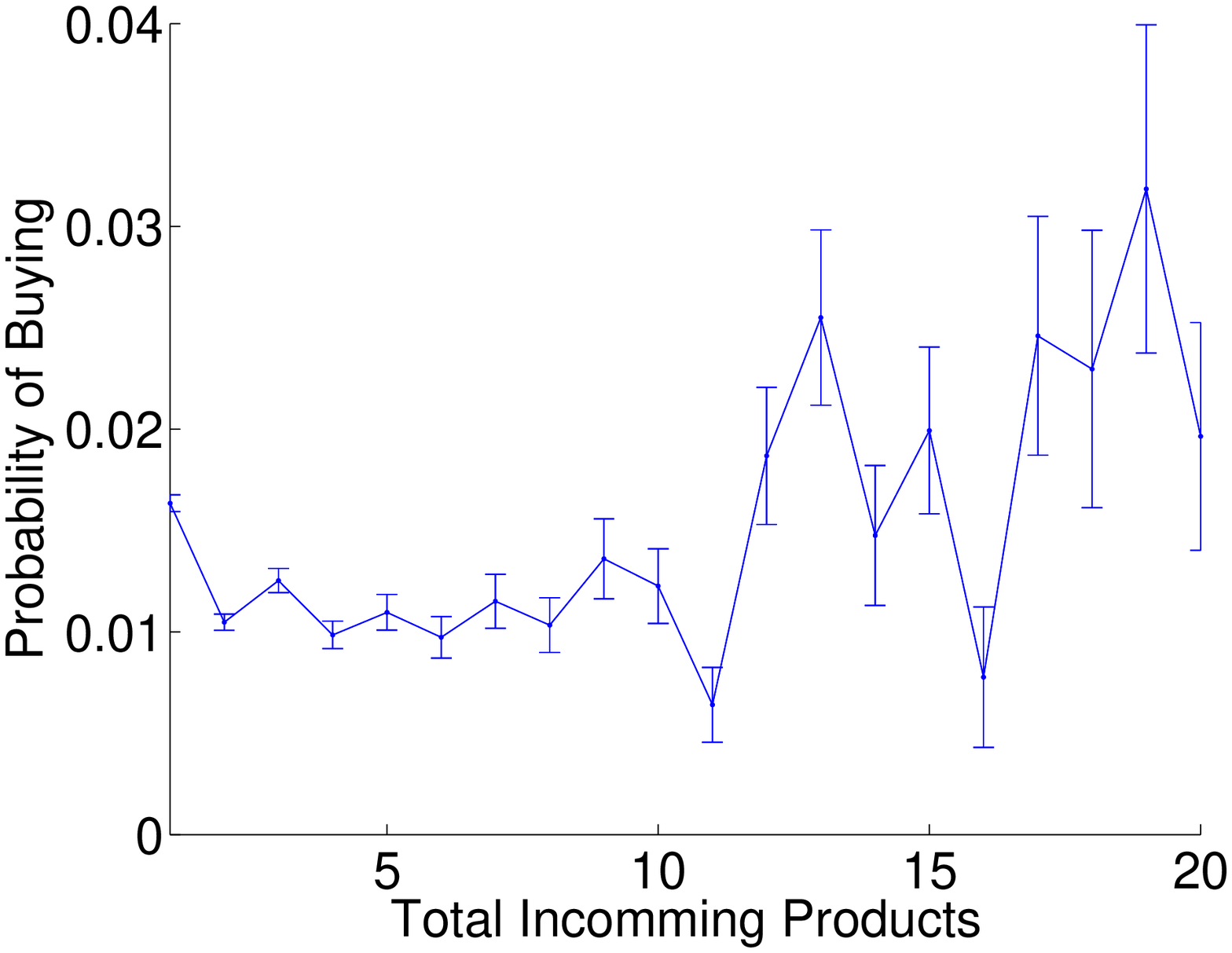} \\
  (c) Music & (d) Video \\
\end{tabular}
\caption{The probability of buying a product given a number of
different products a node got recommendations on.}
\label{fig:inAllBuyProb}
\end{center}
\end{figure}

We show the probability of buying as a function of the number of
different products recommended in Figure~\ref{fig:inAllBuyProb}.
Figure~\ref{fig:inAll1BuyProb} plots the same data but with the total
number of incoming recommendations on the x-axis. We calculate the error
bars as described in section~\ref{sec:errBar}. The number of
observations is large enough (error bars are sufficiently small) to draw
conclusions about the trends observed in the figures. For example, there
are more than $15,000$ observations (users) that had 15 incoming DVD
recommendations.

\begin{figure}[ht]
\begin{center}
\begin{tabular}{cc}
  \includegraphics[width=0.45\textwidth]{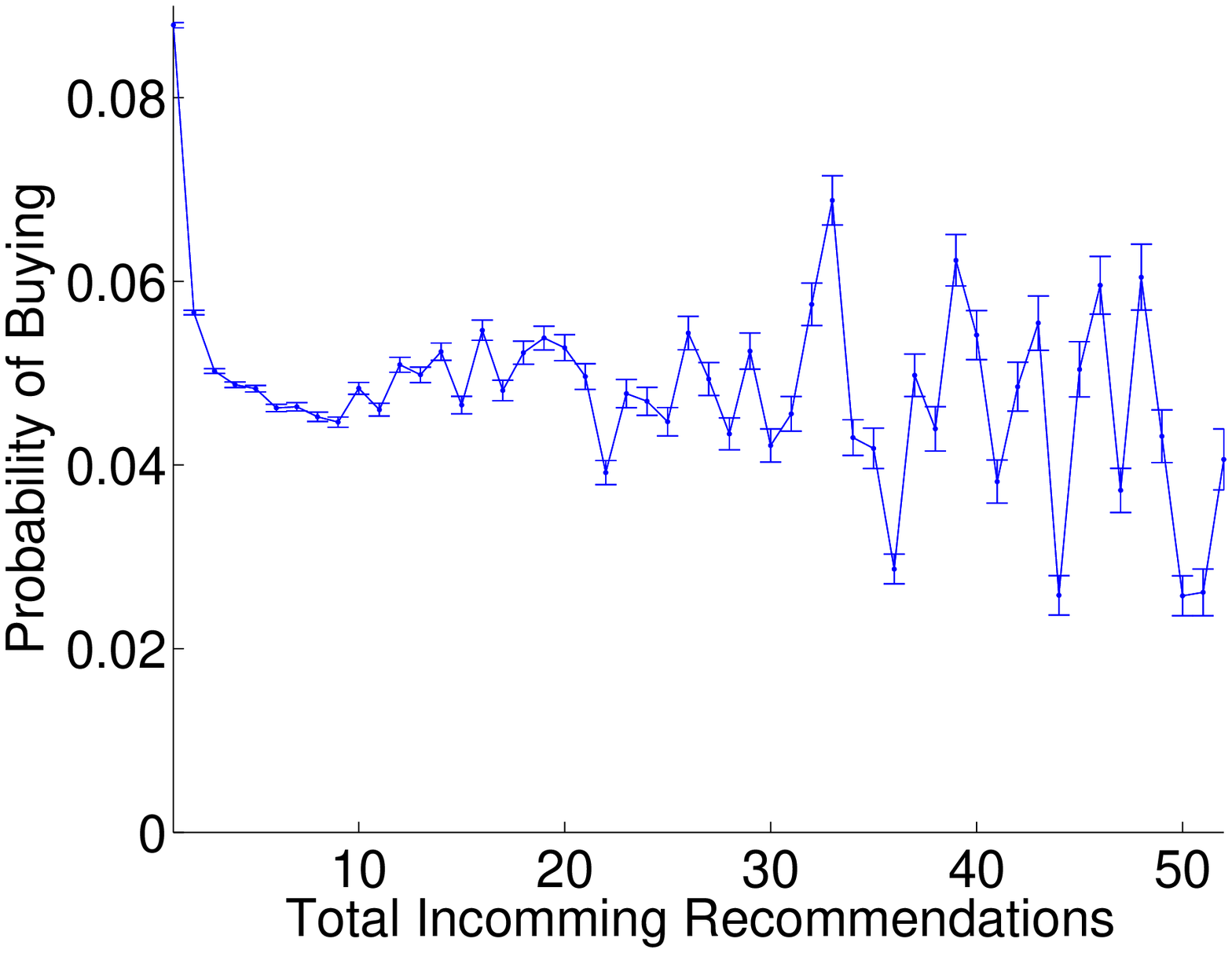} &
  \includegraphics[width=0.45\textwidth]{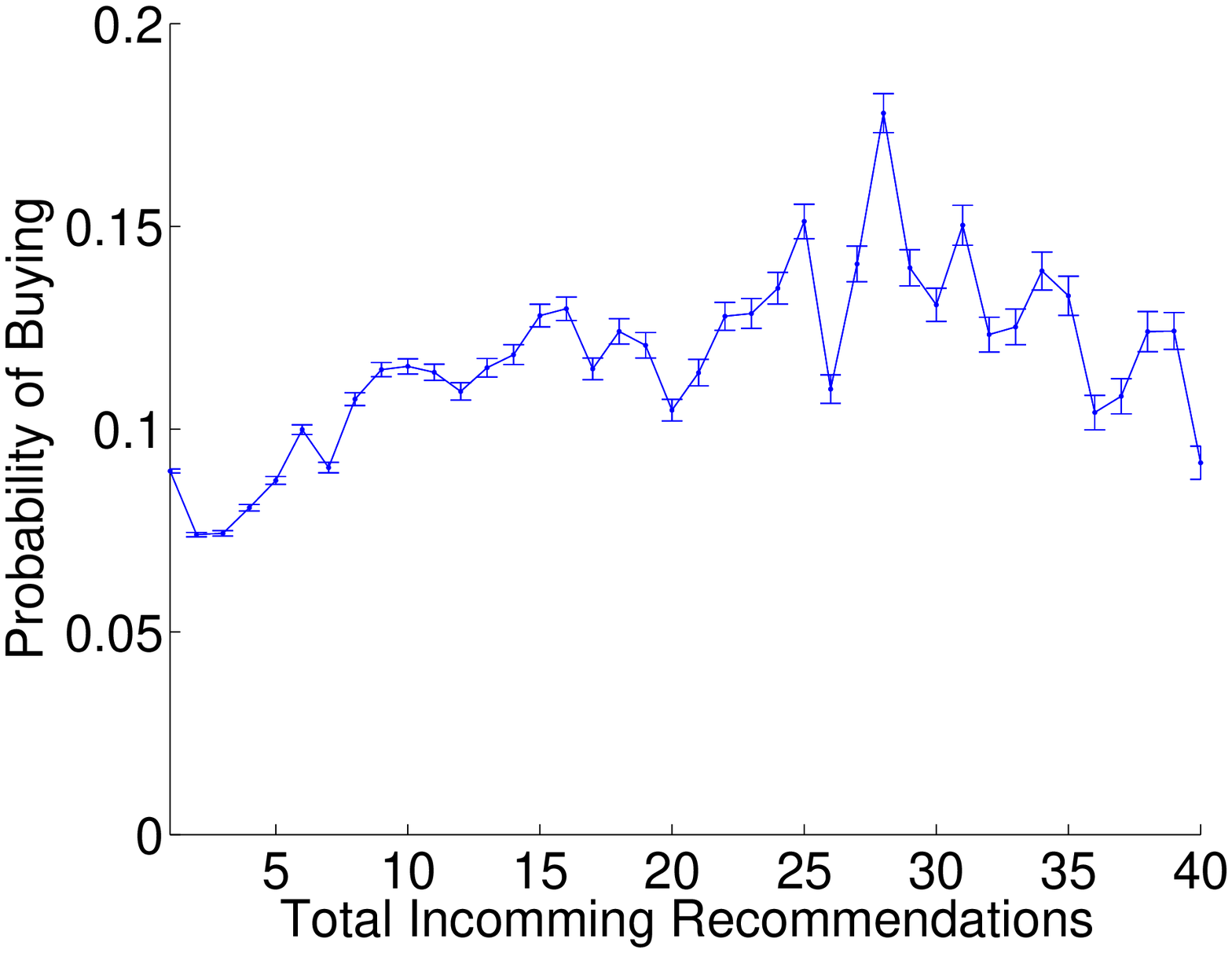} \\
  (a) Books & (b) DVD \\
  \includegraphics[width=0.45\textwidth]{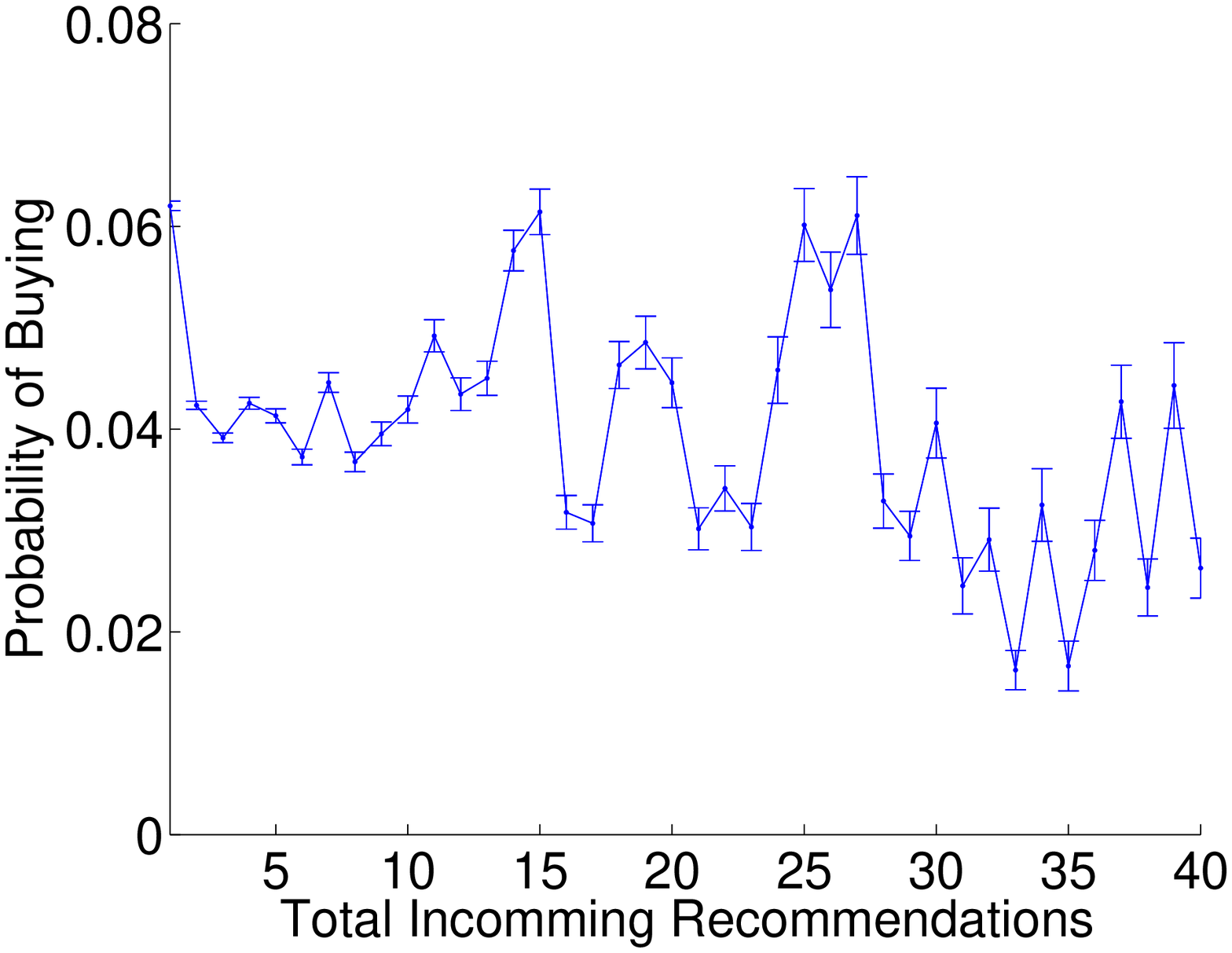} &
  \includegraphics[width=0.45\textwidth]{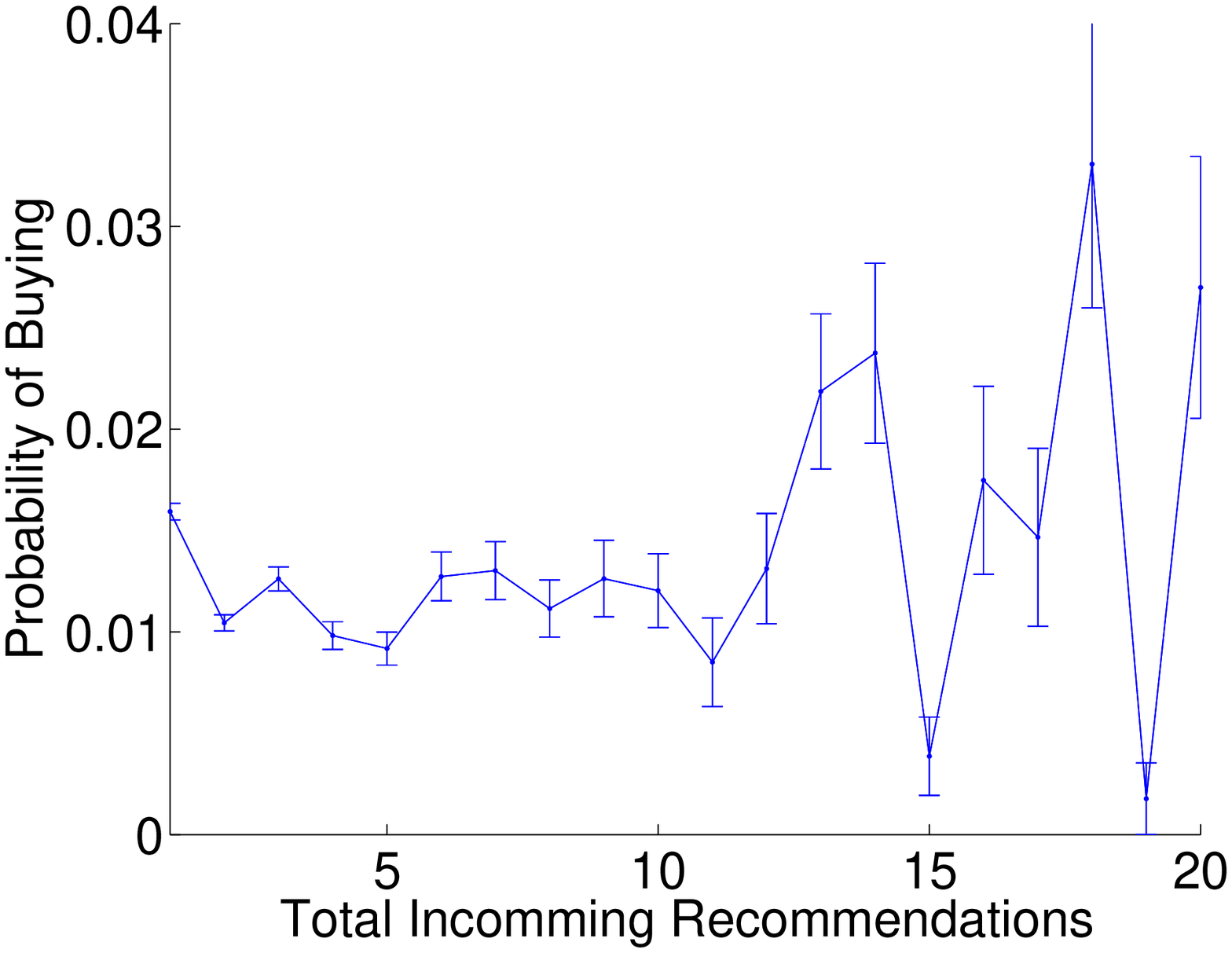} \\
  (c) Music & (d) Video \\
\end{tabular}
\caption{Probability of buying a product given a total number of
incoming recommendations on all products.} \label{fig:inAll1BuyProb}
\end{center}
\end{figure}

Notice that trends are quite similar regardless whether we measure how
involved is the user in the network by counting the number of products
recommended (figure~\ref{fig:inAllBuyProb}) or the number of incoming
recommendations (fig.~\ref{fig:inAll1BuyProb}).

We observe two distinct trends. For books and music
(figures~\ref{fig:inAllBuyProb} and~\ref{fig:inAll1BuyProb}, (a) and
(c)) the probability of buying is the highest when a person got
recommendations on just 1 item, as the number of incoming recommended
products increases to 2 or more the probability of buying quickly
decreases and then flattens.

Movies (DVDs and videos) exhibit different behavior
(figure~\ref{fig:inAllBuyProb} and~\ref{fig:inAll1BuyProb}, (b) and
(d)). A person is more likely to buy the more recommendations she gets.
For DVDs the peak is at around 15 incoming products, while for videos
there is no such peak -- the probability remains fairly level.
Interestingly for DVDs the distribution reaches its low at 2 and 3
items, while for videos it lies somewhere between 3 and 8 items. The
results suggest that books and music buyers tend to be conservative and
focused. On the other hand there are people who like to buy movies in
general. One could hypothesize that buying a book is a larger investment
of time and effort than buying a movie. One can finish a movie in an
evening, while reading a book requires more effort. There are also many
more book and music titles than movie titles.

The other difference between the book and music recommendations in
comparison to movies are the recommendation referral websites where
people could go to get recommendations. One could see these websites as
recommendation subscription services -- posting one's email on a list
results in a higher number of incoming recommendations. For movies,
people with a high number of incoming recommendations ``subscribed'' to
them and thus expected/wanted the recommendations. On the other hand
people with high numbers of incoming book or music recommendations did
not ``sign up'' for them, so they may perceive recommendations as spam
and thus the influence of recommendations drops.

Another evidence of the existence of recommendations referral websites
includes the DVD recommendation network degree distribution. The DVDs
follow a power law degree distribution with an exception of a peak at
out-degree 50. Other plots of DVD recommendation behavior also exhibited
abnormalities at around 50 recommendations. We believe these can be
attributed to the recommendation referral websites.

\section{Timing of recommendations and purchases}
\label{sec:lag}

The recommendation referral program encourages people to purchase as
soon as possible after they get a recommendation, since this maximizes
the probability of getting a discount. We study the time lag between the
recommendation and the purchase of different product groups, effectively
how long it takes a person to receive a recommendation, consider it, and
act on it.

\begin{figure}[t]
\begin{center}
\begin{tabular}{cc}
  \includegraphics[width=0.48\textwidth]{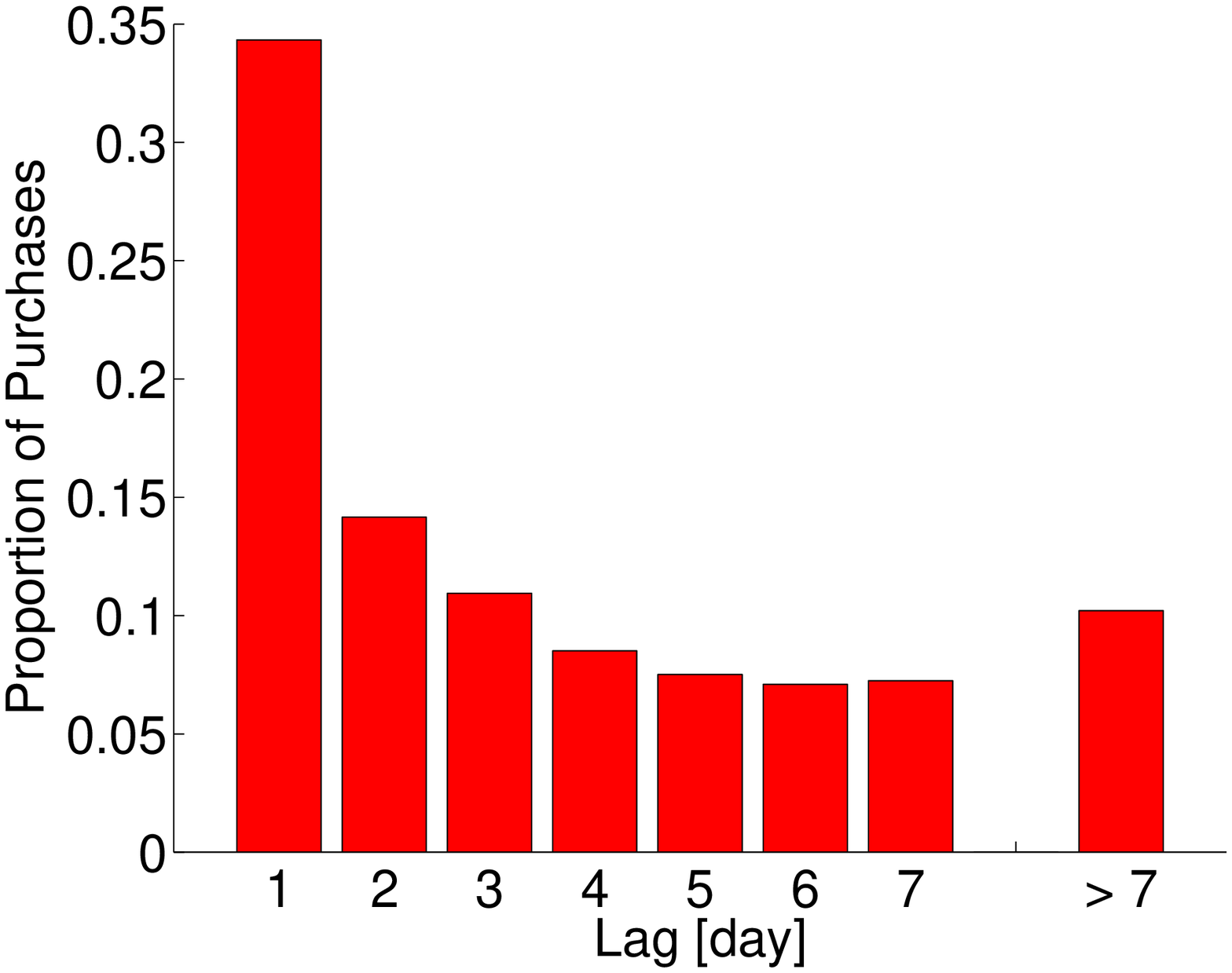} &
  \includegraphics[width=0.48\textwidth]{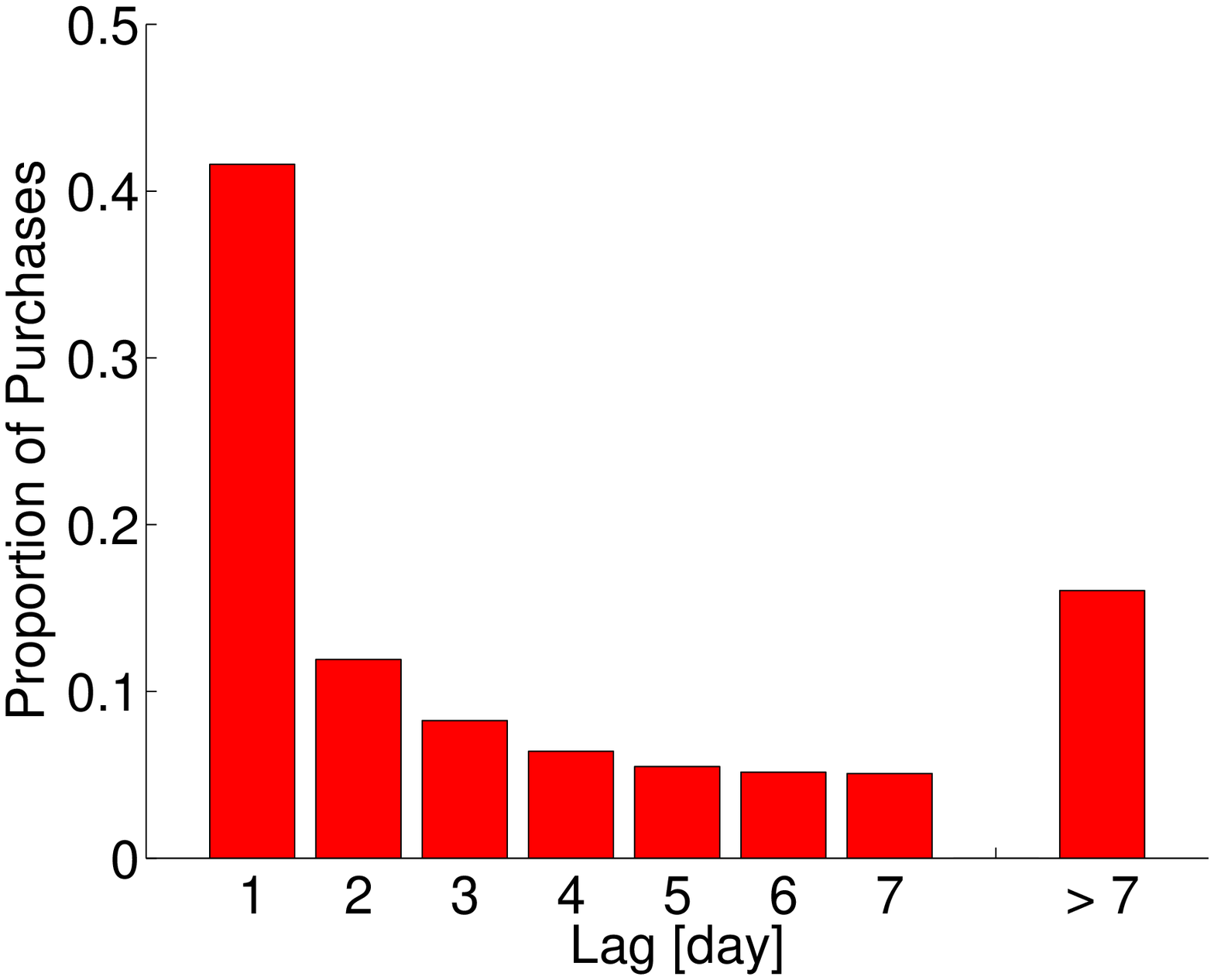} \\
  (a) Books & (b) DVD \\
\end{tabular}
\caption{The time between the recommendation and the actual
purchase. We use all purchases.} \label{fig:recBuyLagAll}
\end{center}
\end{figure}

We present the histograms of the ``thinking time'', i.e. the difference
between the time of purchase and the time the last recommendation was
received for the product prior to the purchase
(figure~\ref{fig:recBuyLagAll}). We use a bin size of 1 day. Around
35\%-40\% of book and DVD purchases occurred within a day after the last
recommendation was received. For DVDs 16\% purchases occur more than a
week after the last recommendation, while this drops to 10\% for books.
In contrast, if we consider the lag between the purchase and the
\emph{first} recommendation, only 23\% of DVD purchases are made within
a day, while the proportion stays the same for books. This reflects a
greater likelihood for a person to receive multiple recommendations for
a DVD than for a book. At the same time, DVD recommenders tend to send
out many more recommendations, only one of which can result in a
discount. Individuals then often miss their chance of a discount, which
is reflected in the high ratio (78\%) of recommended DVD purchases that
did not a get discount (see table~\ref{tab:networkSizes}, columns $b_b$
and $b_e$). In contrast, for books, only 21\% of purchases through
recommendations did not receive a discount.

\begin{figure}[t]
\begin{center}
\begin{tabular}{ccc}
  \includegraphics[width=0.30\textwidth]{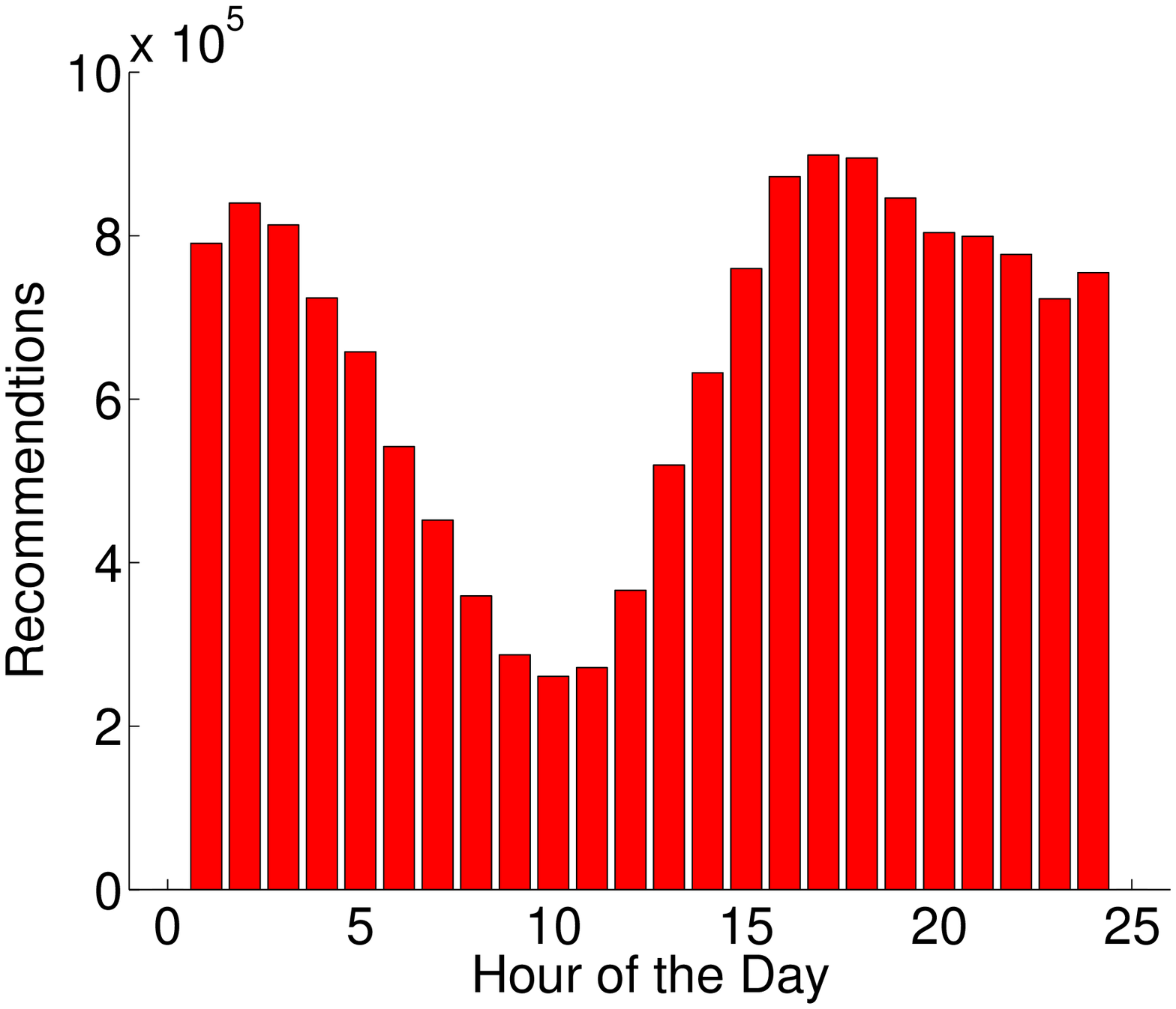} &
  \includegraphics[width=0.30\textwidth]{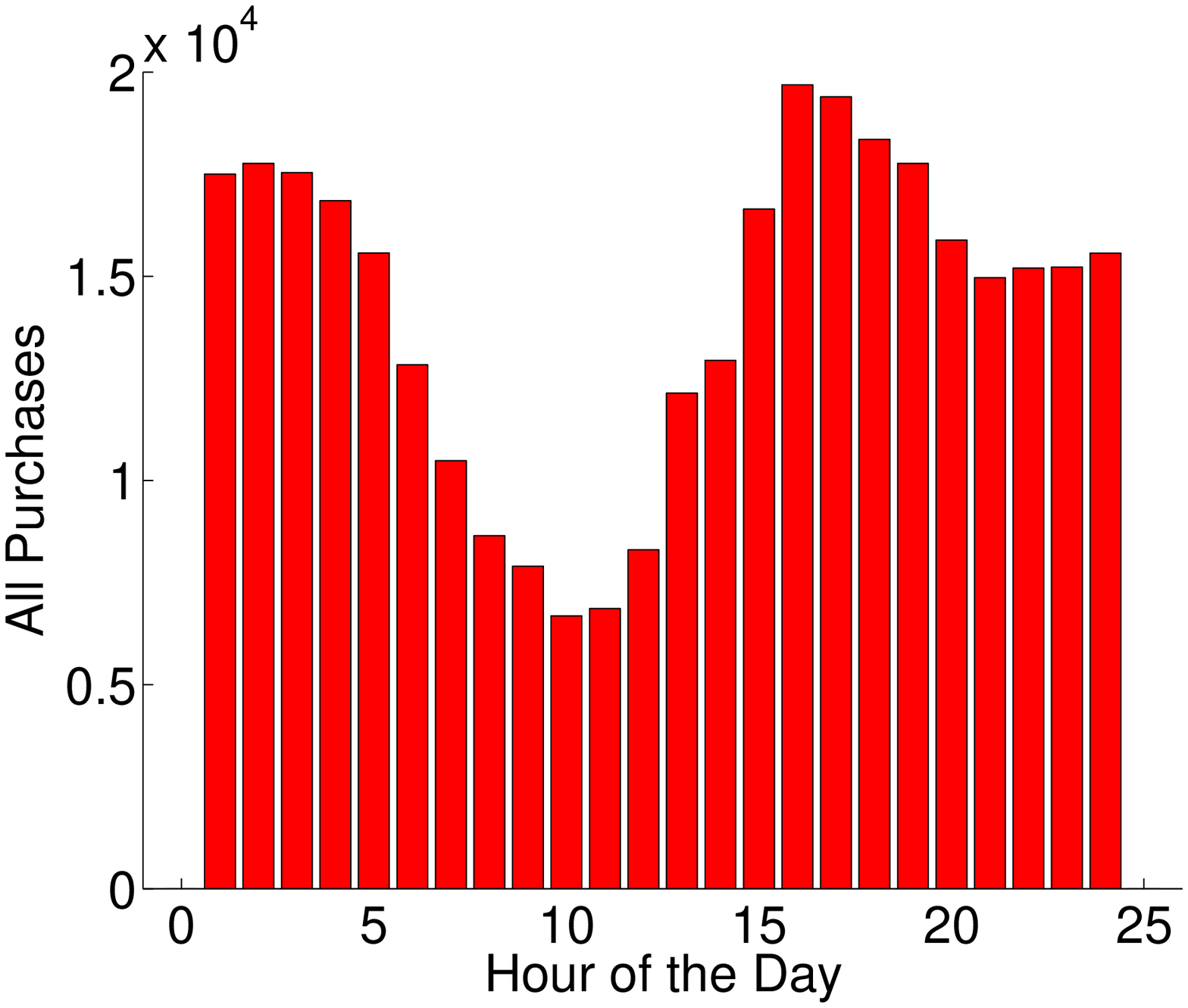} &
  \includegraphics[width=0.30\textwidth]{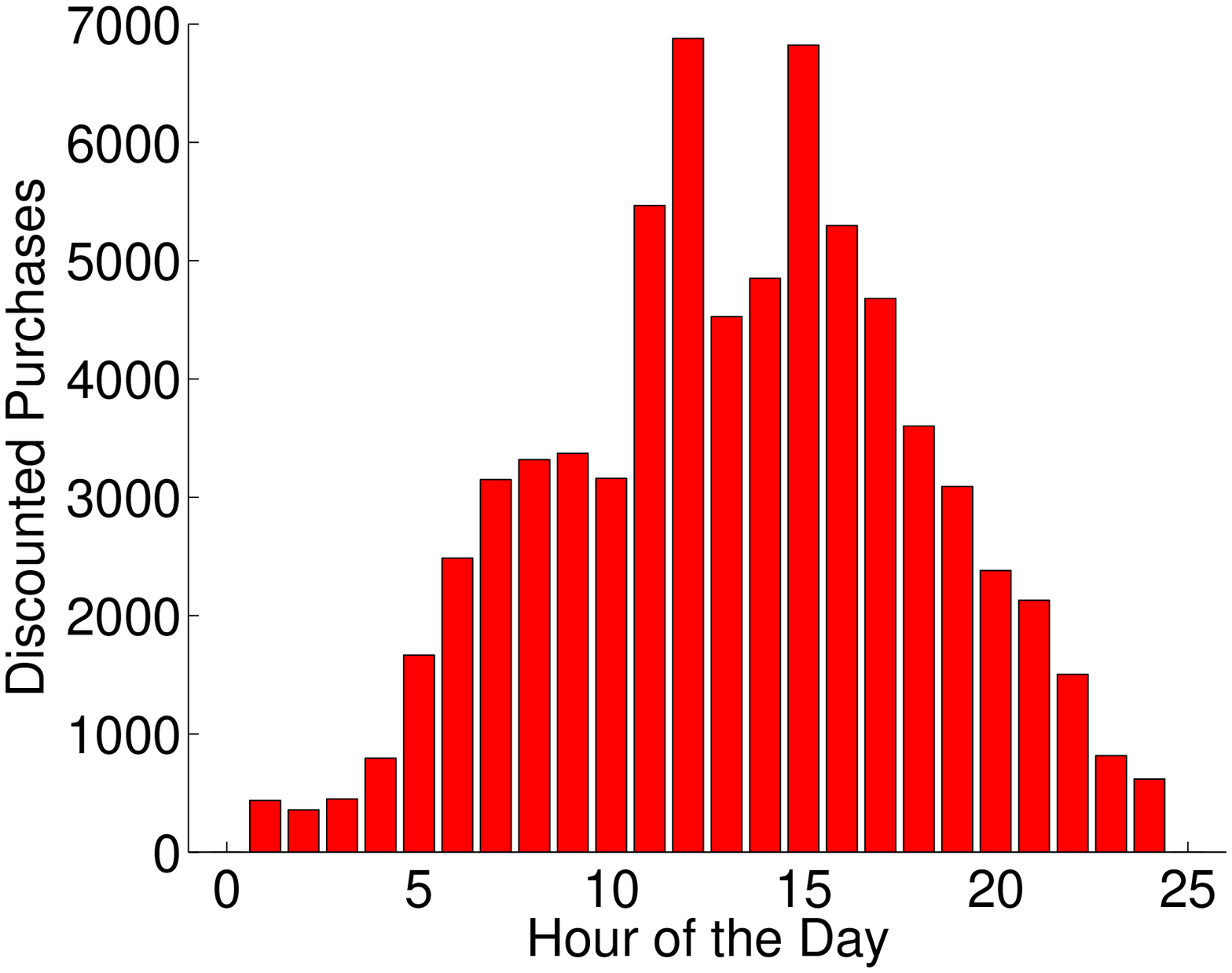} \\
  (a) Recommendations & (b) Purchases & (c) Purchases with Discount \\
\end{tabular}
\caption{Time of day for purchases and recommendations. (a) shows
the distribution of recommendations over the day, (b) shows all
purchases and (c) shows only purchases that resulted in a discount.}
\label{fig:recBuyTm}
\end{center}
\end{figure}

We also measure the variation in intensity by time of day for three
different activities in the recommendation system: recommendations
(figure~\ref{fig:recBuyTm}(a)), all purchases
 (figure~\ref{fig:recBuyTm}(b)), and finally just
the purchases which resulted in a discount
(figure~\ref{fig:recBuyTm}(c)). Each is given as a total count by hour
of day.

The recommendations and purchases follow the same pattern. The only
small difference is that purchases reach a sharper peak in the afternoon
(after 3pm Pacific Time, 6pm Eastern time). This means that the
willingness to recommend does not change with time, since about a
constant fraction of purchases also result in recommendations sent
(plots~\ref{fig:recBuyTm}(a) and (b) follow the same shape).

The purchases that resulted in a discount (fig.~\ref{fig:recBuyTm}(c))
look like a negative image of the first two figures. If recommendations
would have no effect then plot (c) should follow the same shape as (a)
and (b), since a fraction of people that buy would become first buyers,
i.e. the more recommendations sent, the more first buyers and thus
discounts. However, this does not seem to be the case. The number of
purchases with discount is the high when the number of purchases is
small. This means that most of discounted purchases happened in the
morning when the traffic (number of purchases/recommendations) on the
retailer's website was low. This makes sense since most of the
recommendations happened during the day, and if the person wanted to get
the discount by being the first one to purchase, she had the highest
chances when the traffic on the website was the lowest.

There are also other factors that come into play here. Assuming that
recommendations are sent to people's personal (non-work) email
addresses, then people probably check these email accounts for new email
less regularly while at work. So checking personal email while at work
and reacting to a recommendation would mean higher chances of getting a
discount. Second, there are also network effects, i.e. the more
recommendations sent, the higher chance of recommendation collision, the
lower chance of getting discount, since one competes with the larger set
of people.

\section{Recommendations and communities of interest}
\label{sec:communities}

Social networks are a product of the contexts that bring people
together. The context can be a shared interest in a particular topic or
kind of a book. Sometimes there are circumstances, such as a specific
job or religious affiliation, that would make people more likely to be
interested in the same type of book or DVD. We first apply a community
discovery algorithm to automatically detect communities of individuals
who exchange recommendations with one another and to identify the kinds
of products each community prefers. We then compare the effectiveness of
recommendations across book categories, showing that books on different
subjects have varying success rates.

\subsection{Communities and purchases}

In aggregating all recommendations between any two individuals in
Section~\ref{networkovertime} we showed that the network consists of one
large component, containing a little over 100,000 customers, and many
smaller components, the largest of which has 634 customers. However,
knowing that a hundred thousand customers are linked together in a large
network does not reveal whether a product in a particular category is
likely to diffuse through it. Consider for example a new science fiction
book one would like to market by word-of-mouth. If science fiction fans
are scattered throughout the network, with very few recommendations
shared between them, then recommendations about the new book are
unlikely to diffuse. If on the other hand one finds one or more science
fiction {\em communities}, where sci-fi fans are close together in the
network because they exchange recommendations with one another, then the
book recommendation has a chance of spreading by word-of-mouth.

In the following analysis, we use a community finding
algorithm~\cite{clauset04community} in order to discover the types of
products that link customers and so define a community. The algorithm
breaks up the component into parts, such that the modularity Q,

\begin{equation}
Q =  (\mathrm{number~of~edges~within~communities}) -
(\mathrm{expected~number~of~such~edges}),
\end{equation}

is maximized. In other words, the algorithm identifies communities such
that individuals within those communities tend to preferentially
exchange recommendations with one another.

The results of the community finding analysis, while primarily
 descriptive, illustrate both the presence of
communities whose members are linked by their common interests, and the
presence cross-cutting interests between communities. Applying the
algorithm to the largest component, we identify many small communities
and a few larger ones. The largest contains 21,000 nodes, 5,000 of whom
are senders of a relatively modest 335,000 recommendations. More
interesting than simply observing the size of communities is discovering
what interests bring them together. We identify those interests by
observing product categories where the number of recommendations within
the community is significantly higher than it is for the overall
customer population. Let $p_c$ be the proportion of all recommendations
that fall within a particular product category $c$. Then for a set of
individuals sending $x_g$ recommendations, we would expect by chance
that $x_g*p_c \pm \sqrt{x_g*p_c*(1-p_c)}$ would fall within category
$c$. We note the product categories for which the observed number of
recommendations in the community is many standard deviations higher than
expected. For example, compared to the background population, the
largest community is focused on a wide variety of books and music. In
contrast, the second largest community, involving 10,412 individuals
(4,205 of whom are sending over 3 million recommendations), is
predominantly focused on DVDs from many different genres, with no
particular emphasis on anime. The anime community itself emerges as a
highly unusual group of 1,874 users who exchanged over 3 million
recommendations.

\begin{table}[tb]
\centering
\begin{tabular}{r|r|l}
\# nodes & \# senders & topics \\ \hline
735 & 74 & books: American literature, poetry \\
710 & 179 & sci-fi books, TV series DVDs, alternative rock music \\
667 & 181 & music: dance, indie\\
653 & 121 & discounted DVDs \\
541 & 112 & books: art \& photography, web development, graphical
design, sci-fi \\
502 & 104 & books: sci-fi and other\\
388 & 77 & books: Christianity and Catholicism\\
309 & 81 & books: business and investing, computers, Harry Potter \\
192 & 30 & books: parenting, women's health, pregnancy\\
163 & 48 & books: comparative religion, Egypt's history, new age,
role playing games \\
\end{tabular}
\caption{A sample of the medium sized communities present in the largest
component} \label{tab:communitytable}
\end{table}

Perhaps the most interesting are the medium sized communities, some of
which are listed in Table~\ref{tab:communitytable}, having between 100
and 1000 members and often reflecting specific interests. Among the
hundred or so medium communities, we found, for example, several
communities focusing on Christianity. While some of the Christian
communities also shared an interest in children's books, broadway
musicals, and travel to Italy, others focused on prayer and bibles,
still others also enjoyed DVDs of the Simpsons TV series, and others
still took an interest in Catholicism, occult spirituality and kabbalah.

Communities were usually centered around a product group, such as books,
music, or DVDs, but almost all of them shared recommendations for all
types of products. The DVD communities ranged from bargain shoppers
purchasing discounted comedy and action DVDs to smaller anime or
independent movie communities, to a group of customers purchasing
predominantly children's movies. One community focused heavily on indie
music, and imported dance and club music. Another seemed to center
around intellectual pursuits, including reading books on sociology,
politics, artificial intelligence, mathematics, and media culture,
listening to classical music and watching neo-noir film. Several
communities centered around business and investment books and frequently
also recommended books on computing. One business and investment
community included fans of the Harry Potter fiction series, while
another enjoyed science fiction and adventure DVDs. One of communities
with the most particular interests recommended not only business and
investing books to one another, but also an unusual number of books on
terrorism, bacteriology, and military history. A community of what one
can presume are web designers recommended books to one another on art
and photography, web development, graphical design, and Ray Bradbury's
science fiction novels. Several sci-fi TV series such as Buffy the
Vampire Slayer and Star Trek appeared prominently in a few communities,
while Stephen King and Douglas Clegg featured in a community
recommending horror, sci-fi, and thrillers to one another. One community
focused predominantly on parenting, women's health and pregnancy, while
another recommended a variety of books but especially a collection of
cookie baking recipes.

Going back to components in the network that were disconnected from the
largest component, we find similar patterns of homophily, the tendency
of like to associate with like. Two of the components recommended
technical books about medicine, one focused on dance music, while some
others predominantly purchased books on business and investing. Given
more time, it is quite possible that one of the customers in one of
these disconnected components would have received a recommendation from
a customer within the largest component, and the two components would
have merged. For example, a disconnected component of medical students
purchasing medical textbooks might have sent or received a
recommendation from the medical community within the largest component.
However, the medical community may also become linked to other parts of
the network through a different interest of one of its members. At the
very least many communities, no matter their focus, will have
recommendations for children's books or movies, since children are a
focus for a great many people. The community finding algorithm on the
other hand is able to break up the larger social network to
automatically identify groups of individuals with a particular focus or
a set of related interests. Now that we have shown that communities of
customers recommend types of products reflecting their interests, we
will examine whether these different kinds of products tend to have
different success rates in their recommendations.

\subsection{Recommendation effectiveness by book category}

\label{sec:perCatStat} Some contexts result in social ties that are more
effective at conducting an action. For example, in small world
experiments, where participants attempt to reach a target individual
through their chain of acquaintances, profession trumped geography,
which in turn was more useful in locating a target than attributes such
as religion or hobbies~\cite{killworth78reverse,travers69smallworld}. In
the context of product recommendations, we can ask whether a
recommendation for a work of fiction, which may be made by any friend or
neighbor, is more or less influential than a recommendation for a
technical book, which may be made by a colleague at work or school.

\begin{table*}[!ht]
\centering
\begin{tabular}{l|rrrrrrrl}\hline
  category& $n_p$ & $n$ & $cc$ & $r_{p1}$ &
  $v_{av}$ & $c_{av}/$ & $p_m$ & $b*100$ \\
    & & & &  & & $r_{p1}$& &\\ \hline
  Books general & 370230 & 2,860,714 & 1.87  & 5.28 &  4.32 & 1.41 &
  14.95 &3.12 \\
  \hline
  \hline Fiction & \multicolumn{8}{l}{} \\
  \hline\hline
  Children & 46,451 & 390,283 & 2.82 & 6.44 & 4.52 & 1.12 & 8.76
  &2.06** \\
  Literature & 41,682 & 502,179 & 3.06  & 13.09 & 4.30 & 0.57 & 11.87
  &2.82* \\
  Mystery & 10,734 & 123,392 & 6.03 & 20.14 & 4.08 & 0.36 & 9.60&2.40**
  \\
  Science fiction & 10,008 & 175,168 & 6.17 & 19.90 & 4.15 & 0.64 &
  10.39 &2.34** \\
  Romance & 6,317 & 60,902 & 5.65 & 12.81 &  4.17 & 0.52 & 6.99 &
  1.78**
  \\
  Teens & 5,857 & 81,260 & 5.72 & 20.52& 4.36 & 0.41 & 9.56 & 1.94** \\
  Comics & 3,565 & 46,564 & 11.70 & 4.76 & 4.36 & 2.03 & 10.47 &2.30*
  \\
  Horror & 2,773 & 48,321 & 9.35 & 21.26  & 4.16 & 0.44 & 9.60 & 1.81**
  \\
  \hline \hline Personal & \multicolumn{8}{l}{} \\
  \hline\hline
  Religion & 43,423 & 441,263 & 1.89  & 3.87  & 4.45 & 1.73 &9.99& 3.13
  \\
  Health/Body & 33,751 & 572,704 & 1.54  & 4.34  & 4.41 & 2.39 & 13.96
  &
  3.04 \\
  History & 28,458 & 28,3406 & 2.74 & 4.34 & 4.30 & 1.27 & 18.00 &2.84
  \\
  Home/Garden & 19,024 & 180,009 & 2.91 & 1.78  & 4.31 & 3.48 & 15.37
  &2.26** \\
  Entertainment & 18,724 & 258,142 & 3.65  & 3.48 &  4.29 & 2.26 &
  13.97
  & 2.66* \\
  Arts/Photo & 17,153 & 179,074 & 3.49 & 1.56 &  4.42 & 3.85 & 20.95 &
  2.87 \\
  Travel & 12,670 & 113,939 & 3.91 & 2.74  & 4.26 & 1.87 &13.27 &
  2.39**
  \\
  Sports & 10,183 & 120,103 & 1.74  & 3.36  & 4.34 & 1.99 & 13.97 &
  2.26** \\
  Parenting & 8,324 & 182,792 & 0.73  & 4.71  & 4.42 & 2.57 & 11.87 &
  2.81 \\
  Cooking & 7,655 & 146,522 & 3.02  & 3.14  & 4.45 & 3.49 & 13.97&2.38*
  \\
  Outdoors & 6,413 & 59,764 & 2.23  & 1.93  & 4.42 & 2.50 & 15.00 &
  3.05
  \\
  \hline \hline Professional & \multicolumn{8}{l}{} \\
  \hline\hline
  Professional  & 41,794 & 459,889 & 1.72  & 1.91  & 4.30 & 3.22 &
  32.50
  &4.54** \\
  Business  & 29,002 & 476,542 & 1.55  & 3.61  & 4.22 & 2.94 &
  20.99&3.62** \\
  Science & 25,697 & 271,391 & 2.64 & 2.41  & 4.30 & 2.42 & 28.00
  &3.90** \\
  Computers  & 18,941 & 375,712 & 2.22  & 4.51  & 3.98 & 3.10 &
  34.95&3.61** \\
  Medicine &  16,047 & 175,520 & 1.08  & 1.41  & 4.40 & 4.19 &
  39.95&5.68** \\
  Engineering & 10,312 & 107,255 & 1.30 & 1.43  & 4.14 & 3.85 & 59.95 &
  4.10** \\
  Law & 5,176 & 53,182 & 2.64 & 1.89  & 4.25 & 2.67 & 24.95 & 3.66* \\
  \hline
  \hline Other & \multicolumn{8}{l}{} \\
  \hline\hline
  Nonfiction & 55,868 & 560,552 & 2.03 & 3.13  & 4.29 & 1.89 & 18.95&
  3.28** \\
  Reference & 26,834 & 371,959 & 1.94 & 2.49  & 4.19 & 3.04 & 17.47 &
  3.21 \\
  Biographies  & 18,233 & 277,356 & 2.80  & 7.65  & 4.34 & 0.90 & 14.00
  &2.96 \\
  \end{tabular}
  \caption{Statistics by book category: $n_p$:number of products in
  category, $n$ number of customers, $cc$ percentage of customers in
  the largest connected component, $r_{p1}$ avg. \# reviews in 2001 --
  2003, $r_{p2}$ avg. \# reviews 1st 6 months 2005, $v_{av}$ average
  star rating, $c_{av}$ average number of people recommending product,
  $c_{av}/r_{p1}$ ratio of recommenders to reviewers, $p_{m}$ median
  price, $b$ ratio of the number of purchases resulting from a
  recommendation to the number of recommenders. The symbol ** denotes
  statistical significance at the 0.01 level, * at the 0.05 level.}
  \label{tab:bookcategories}
\end{table*}

\tabref{tab:bookcategories} shows recommendation trends for all top
level book categories by subject. For clarity, we group the results by 4
different category types: fiction, personal/leisure,
professional/technical, and nonfiction/other. Fiction encompasses
categories such as Sci-Fi and Romance, as well as children's and young
adult books. Personal/Leisure encompasses everything from gardening,
photography and cooking to health and religion.

First, we compare the relative number of recommendations to reviews
posted on the site (column $c_{av}/r_{p1}$ of
table~\ref{tab:bookcategories}). Surprisingly, we find that the number
of people making personal recommendations was only a few times greater
than the number of people posting a public review on the website. We
observe that fiction books have relatively few recommendations compared
to the number of reviews, while professional and technical books have
more recommendations than reviews. This could reflect several factors.
One is that people feel more confident reviewing fiction than technical
books. Another is that they hesitate to recommend a work of fiction
before reading it themselves, since the recommendation must be made at
the point of purchase. Yet another explanation is that the median price
of a work of fiction is lower than that of a technical book. This means
that the discount received for successfully recommending a mystery novel
or thriller is lower and hence people have less incentive to send
recommendations.

Next, we measure the per category efficacy of recommendations by
observing the ratio of the number of purchases occurring within a week
following a recommendation to the number of recommenders for each book
subject category (column $b$ of table~\ref{tab:bookcategories}). On
average, only 2\% of the recommenders of a book received a discount
because their recommendation was accepted, and another 1\% made a
recommendation that resulted in a purchase, but not a discount. We
observe marked differences in the response to recommendation for
different categories of books. Fiction in general is not very
effectively recommended, with only around 2\% of recommenders
succeeding. The efficacy was a bit higher (around 3\%) for non-fiction
books dealing with personal and leisure pursuits. Perhaps people
generally know what their friends' leisure interests are, or even have
gotten to know them through those shared interests. On the other hand
they may not know as much about each others' tastes in fiction.
Recommendation success is highest in the professional and technical
category. Medical books have nearly double the average rate of
recommendation acceptance. This could be in part attributed to the
higher median price of medical books and technical books in general. As
we will see in~\secref{sec:regression}, a higher product price increases
the chance that a recommendation will be accepted.

Recommendations are also more likely to be accepted for certain
religious categories: 4.3\% for Christian living and theology and 4.8\%
for Bibles. In contrast, books not tied to organized religions, such as
ones on the subject of new age (2.5\%) and occult (2.2\%) spirituality,
have lower recommendation effectiveness. These results raise the
interesting possibility that individuals have greater influence over one
another in an organized context, for example through a professional
contact or a religious one. There are exceptions of course. For example,
Japanese anime DVDs have a strong following in the US, and this is
reflected in their frequency and success in recommendations. Another
example is that of gardening. In general, recommendations for books
relating to gardening have only a modest chance of being accepted, which
agrees with the individual prerogative that accompanies this hobby. At
the same time, orchid cultivation can be a highly organized and social
activity, with frequent `shows' and online communities devoted entirely
to orchids. Perhaps because of this, the rate of acceptance of orchid
book recommendations is twice as high as those for books on vegetable or
tomato growing.

\section{Products and recommendations}
\label{sec:progReg}

We have examined the properties of the recommendation network in
relation to viral marketing. Now we focus on the products themselves and
their characteristics that determine the success of recommendations.

\subsection{How long is the long tail?}

Recently a `long tail' phenomenon has been observed, where a large
fraction of purchases are of relatively obscure items where each of them
sells in very low numbers but there are many of those items. On
Amazon.com, somewhere between 20 to 40 percent of unit sales fall
outside of its top 100,000 ranked
products~\cite{brynjolfsson03consumer}. Considering that a typical brick
and mortar store holds around 100,000 books, this presents a significant
share. A streaming-music service streams more tracks outside than inside
its top 10,000 tunes~\cite{economist05longtail}.

\begin{figure}[t]
\begin{center}
  \begin{tabular}{cc}
    \includegraphics[width=0.45\textwidth]{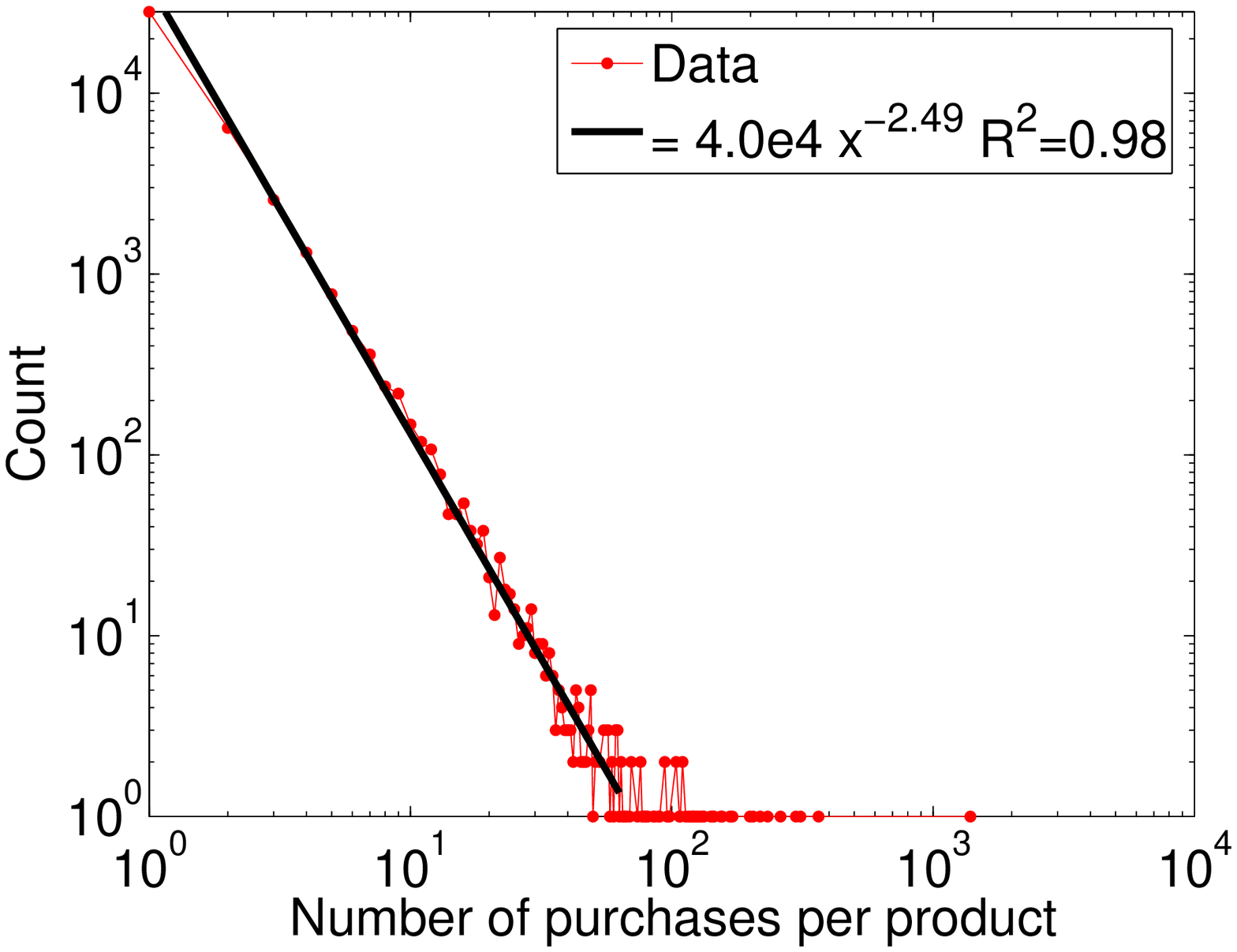} &
    \includegraphics[width=0.45\textwidth]{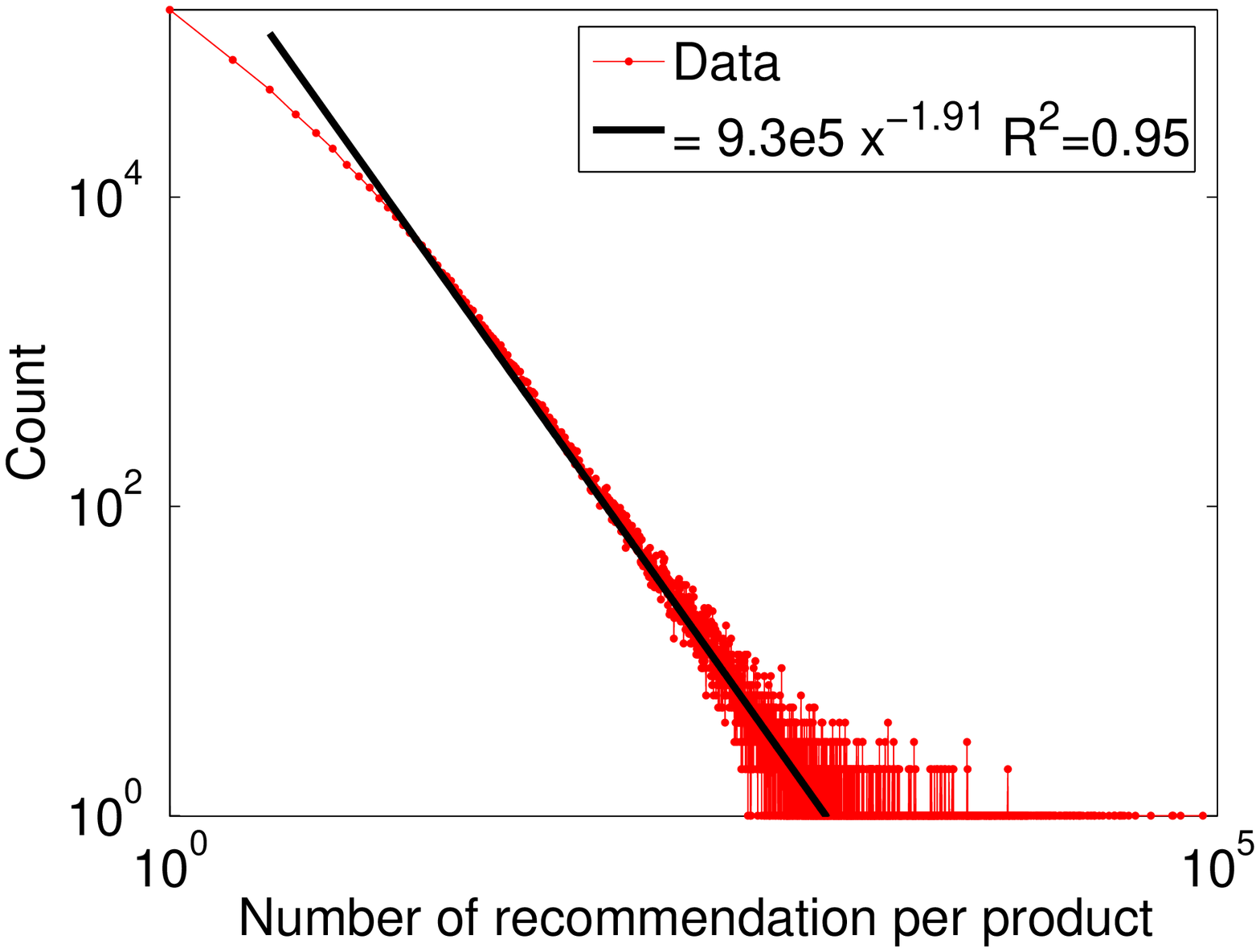} \\
    (a) Purchases & (b) Recommendations  \\
  \end{tabular}
  \caption{Distribution of number of purchases and recommendations of
  a product. (a) shows the number of purchases that resulted in a
  discount per product, and (b) shows the distribution of the number
  of recommendations per product.}
  \label{fig:ProdRecBuy}
\end{center}
\end{figure}

We performed a similar experiment using our data. Since we do not have
direct sales data we used the number of successful recommendations as a
proxy to the number of purchases. Figure~\ref{fig:ProdRecBuy} plots the
distribution of the number of purchases and the number of
recommendations per product. Notice that both the number of
recommendations and the number of purchases per product follow a
heavy-tailed distribution and that the distribution of recommendations
has a heavier tail.

Interestingly, figure~\ref{fig:ProdRecBuy}(a) shows that just the top
100 products account for 11.4\% of the all sales (purchases with
discount), and the top 1000 products amount to 27\% of total sales
through the recommendation system. On the other hand 67\% of the
products have only a single purchase and they account for 30\% of all
sales. This shows that a significant portion of sales come from products
that sell very few times. Recently there has been some debate about the
long tail~\cite{wallstreet,andersonbook}. Some argue that the presence
of the long tail indicates that niche products with low sales are
contributing significantly to overall sales online. We also find that
the tail is a bit longer than the usual 80-20 rule, with the top 20\% of
the products contributing to about half the sales. It is important to
note, however, that our observations do not reflect the total sales of
the products on the website, since they include only successful
recommendations that resulted in a discount. This incorporates both a
bias in the kind of product that is likely to be recommended, and in the
probability that a recommendation for that kind of product is accepted.

If we look at the distribution in the number of recommendations per
product, shown in Figure~\ref{fig:ProdRecBuy}(b), we observe an even
more skewed distribution. 30\% of the products have only a single
recommendation and the top 56,000 most recommended products (top 10\%)
account for 84\% of all recommendations. This is consistent with our
previous observations some types of products, e.g. anime DVDs, are more
heavily recommended than others.

\begin{figure}[t]
\begin{center}
  \begin{tabular}{cc}
    \includegraphics[width=0.45\textwidth]{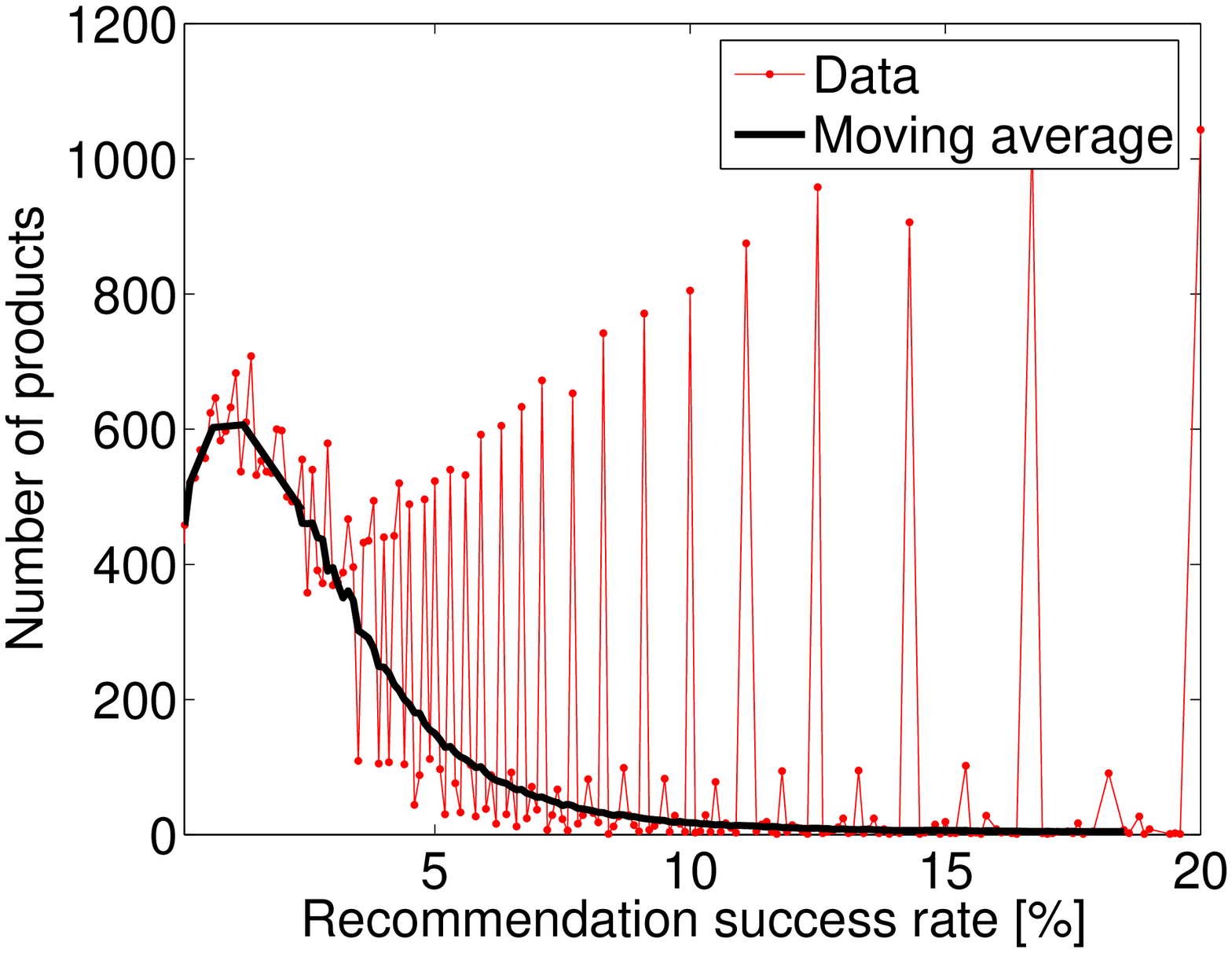} &
    \includegraphics[width=0.45\textwidth]{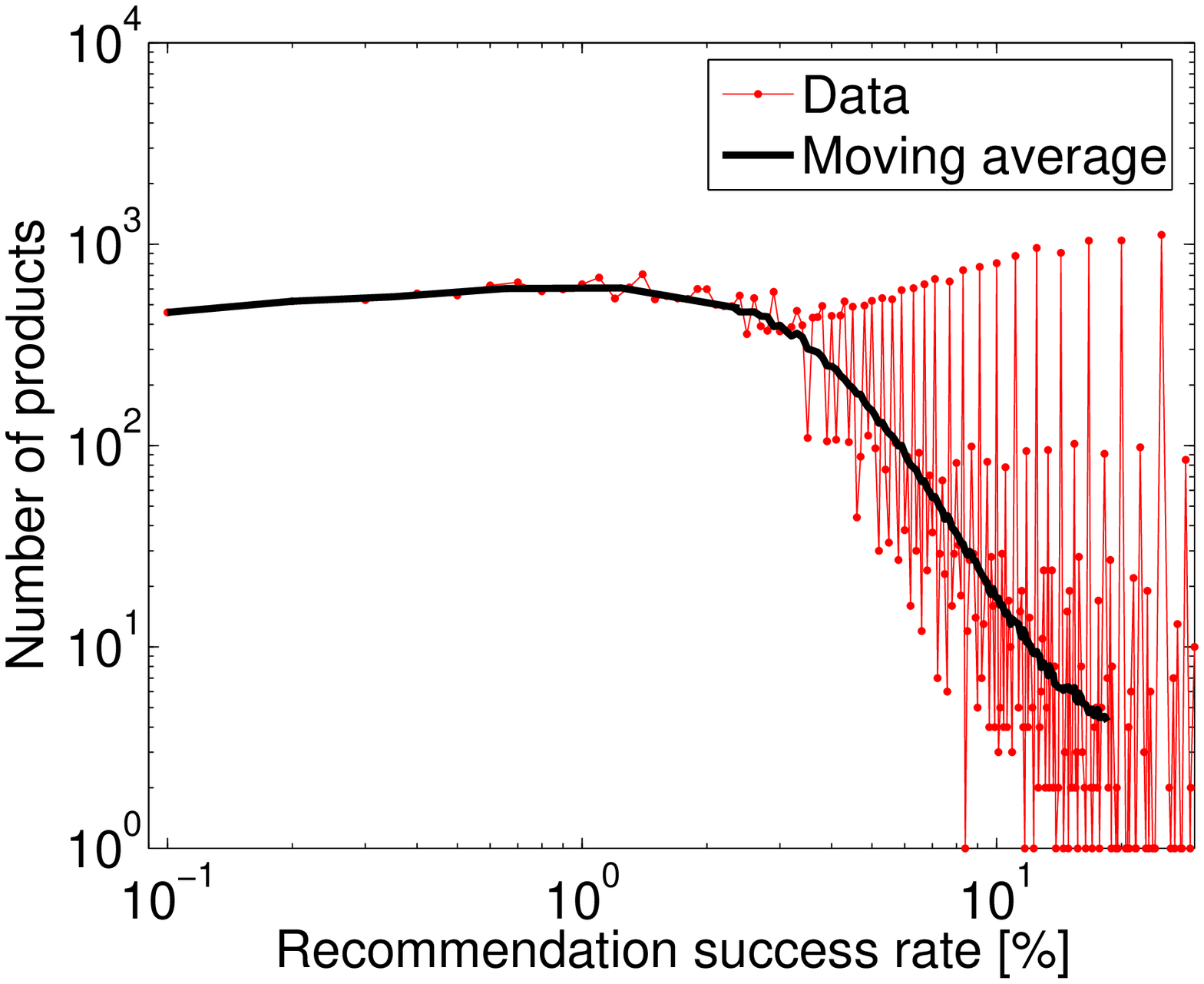}
    \\
    (a) Success rate (linear scale) & (b) Success rate (log scale)  \\
  \end{tabular}
  \caption{Distribution of product recommendation success rates. Both
  plots show the same data: (a) on a linear (lin-lin) scale, and (b)
  on a logarithmic (log-log) scale. The bold line presents the moving
  average smoothing.}
  \label{fig:ProdSuccRate}
\end{center}
\end{figure}

Next we examine the distribution of the product recommendation success
rate. Out of more than half a million products we took all the products
with at least a single purchase, of which there are 41,000 (7\%).
Figure~\ref{fig:ProdSuccRate} shows the success rate
(purchases/recommendations). Notice that the distribution is not heavy
tailed and has a mode at around 1.3\% recommendation success rate. 55\%
of the products have a success rate bellow 5\% and there are around 14\%
of the products that have a recommendation success rate higher than
20\%.

\subsection{Modeling the product recommendation success}
\label{sec:regression}

So far we have seen that some products generate many recommendations and
some have a better return than others on those recommendations, but one
question still remains: what determines the product's viral marketing
success? We present a model which characterizes product categories for
which recommendations are more likely to be accepted. We use a
regression of the following product attributes to correlate them with
recommendation success:

\begin{itemize}
  \item $n$: number of nodes in the social network (number of unique
      senders and receivers)
  \item $n_s$: number of senders of recommendations
  \item $n_r$: number of recipients of recommendations
  \item $r$: number of recommendations
  \item $e$: number of edges in the social network (number of unique
      (sender, receiver) pairs)
  \item $p$: price of the product
  \item $v$: number of reviews of the product
  \item $t$: average product rating
\end{itemize}

From the original set of the half-million products, we compute a success
rate s for the 8,192 DVDs and 50,631 books that had at least 10
recommendation senders and for which a price was given. In
section~\ref{sec:perCatStat} we defined recommendation success rate $s$
as the ratio of the total number purchases made through recommendations
and the number of senders of the recommendations. We decided to use this
kind of normalization, rather than normalizing by the total number of
recommendations sent, in order not to penalize communities where a few
individuals send out many recommendations
(figure~\ref{fig:recNetPlot}(b)). Note that in general $s$ could be
greater than 1, but in practice this happens extremely rarely (there are
only 107 products where $s>1$ which were discarded for the purposes of
this analysis).

Since the variables follow a heavy tailed distribution, we use the
following model:

\begin{equation}
  s = \exp(\sum_i {\beta_i\log(x_i)} + \epsilon_i)
\end{equation}

where $x_i$ are the product attributes (as described on previous page),
and $\epsilon_i$ is random error.

\begin{table}
\begin{center}
\begin{tabular}{c|c|c|c|c|c|c|c|c|c}
  \hline
  & $\log(s)$ & $\log(n)$ & $\log(n_s)$ & $\log(n_e)$ & $\log(r)$ & $\log(e)$ & $\log(p)$ & $\log(v)$ & $\log(t)$ \\
  \hline
  $\log(s)$ & 1 &  &  &  &  &  &  &  & \\
  $\log(n)$ & 0.275 & 1 &  &  &  &  &  &  & \\
  $\log(n_s)$ & 0.103 & 0.907 & 1 &  &  &  &  &  & \\
  $\log(n_r)$ & 0.310 & 0.994 & 0.864 & 1.000 &  &  &  &  & \\
  $\log(r)$ & 0.396 & 0.979 & 0.828 & 0.988 & 1 &  &  &  & \\
  $\log(e)$ & 0.392 & 0.981 & 0.831 & 0.990 & 0.999 & 1 &  &  & \\
  $\log(p)$ & 0.185 & 0.098 & 0.088 & 0.098 & 0.107 & 0.106 & 1 &  & \\
  $\log(v)$ & -0.050 & 0.465 & 0.490 & 0.449 & 0.421 & 0.423 & -0.053 & 1 & \\
  $\log(t)$ & -0.031 & 0.064 & 0.071 & 0.061 & 0.056 & 0.056 & -0.019 & 0.269 & 1\\
  \hline
\end{tabular}
\end{center}
\caption{Pairwise Correlation Matrix of the Books and DVD Product
Attributes. $\log(s)$: log recommendation success rate, $\log(n)$: log
number of nodes, $\log(n_s)$: log number of senders of recommendations,
$\log(n_r)$: log number of receivers, $\log(r)$: log number of
recommendations, $\log(e)$: log number of edges, $\log(p)$: log price,
$\log(v)$: log number of reviews, $\log(t)$: log average rating.}
\label{tab:correlations}
\end{table}

\begin{table}[tb]
\centering
  \begin{tabular}{c|c|c}
  \hline
  & Books & DVD \\
  Variable & Coefficient $\beta_i$ & Coefficient $\beta_i$\\
  \hline
  const & 1.317 (0.0038) ** & 0.929 (0.0100) ** \\
  $n$ & -0.579 (0.0060) ** & 0.171 (0.0124) ** \\
  $n_s$ & 0.144 (0.0018) ** & -0.070 (0.0023) ** \\
  $n_r$ & -0.006 (0.0064) & -0.360 (0.0104) ** \\
  $r$ & 0.062 (0.0084) ** & -0.002 (0.0083) \\
  $e$ & 0.383 (0.0106) ** & 0.251 (0.0088) ** \\
  $p$ & 0.013 (0.0003) ** & 0.007 (0.0016) ** \\
  $v$ & -0.003 (0.0001) ** & -0.003 (0.0006) ** \\
  $t$ & -0.001 (0.0006) * & 0.000 (0.0009) \\
  \hline
  $R^2$ & 0.30 & 0.81 \\
  \hline
  \end{tabular}
  \caption{Regression Using the Log of the Recommendation Success Rate
log(s), as the Dependent Variable for Books and DVDs separately. For
each coefficient we provide the standard error and the statistical
significance level (**:0.001, *:0.1). We fit separate models for books
and DVDs.} \label{tab:regressAll}
\end{table}

We fit the model using least squares and obtain the coefficients
$\beta_i$ shown in table~\ref{tab:regressAll}. With the exception of the
average rating, they are all significant, but just the number of
recommendations alone accounts for 15\% of the variance (taking all
eight variables into consideration yields an R2 of 0.30 for books and
0.81 for DVDs). We should also note that the variables in our model are
highly collinear, as can be seen from the pairwise correlation matrix
(table~\ref{tab:correlations}). For example, the number of
recommendations $r$ is highly negatively correlated with the dependent
variable ($\log(s)$) but in the regression model it exhibits positive
influence on the dependent variable. This is probably due to the fact
that the number of recommendations is naturally dependent on the number
of senders and number of recipients, but it is the high number of
recommendations relative to the number of senders that is of importance.

\begin{figure}[t]
\begin{center}
  \begin{tabular}{cc}
    \includegraphics[width=0.4\textwidth]{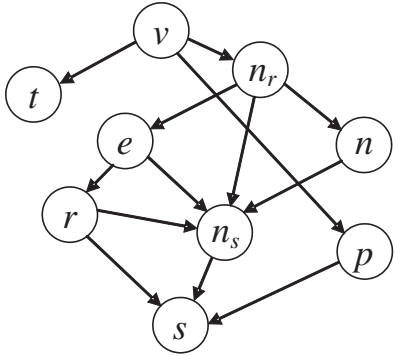}
  \end{tabular}
  \caption{A Bayesian network showing the dependencies between the
variables. $s$: recommendation success rate, $n$: number of nodes, $n_s$:
number of senders of recommendations, $n_r$: log number of receivers, $r$:
number of recommendations, $e$: number of edges, $p$: price, $v$: number of
reviews, $t$: average rating.}
  \label{fig:bayesNet}
\end{center}
\end{figure}

To illustrate the dependencies between the variables we train a Bayesian
dependency network~\cite{chickering03}, and show the learned structure
for the combined (Books and DVDs) data in figure~\ref{fig:bayesNet}. In
this a directed acyclic graph where nodes are variables, and directed
edges indicate that the distribution of a child depends on the values
taken in the parent variables.

Notice that the average rating ($t$) is not predictive of the
recommendation success rate ($s$). It is no surprise that the number of
recommendations $r$ is predictive of number of senders $n_s$. Similarly,
the number of edges $e$ is predictive of number of senders $n_s$.
Interestingly, price $p$ is only related to the number of reviews $v$.
Number of recommendations $r$, number of senders $n_s$ and price $p$,
are directly predictive of the recommendation success rate $s$.

Returning to our regression model, we find that the numbers of nodes and
receivers have negative coefficients, showing that successfully
recommended products are actually more likely to be not so widely
popular. The only attributes with positive coefficients are the number
of recommendations $r$, number of edges $e$, and price $p$. This shows
that more expensive and more recommended products have a higher success
rate. These recommendations should occur between a small number of
senders and receivers, which suggests a very dense recommendation
network where lots of recommendations are exchanged between a small
community of people. These insights could be of use to marketers ---
personal recommendations are most effective in small, densely connected
communities enjoying expensive products.

\section{Discussion and Conclusion}
\label{sec:conclusion}

Although the retailer may have hoped to boost its revenues through viral
marketing, the additional purchases that resulted from recommendations
are just a drop in the bucket of sales that occur through the website.
Nevertheless, we were able to obtain a number of interesting insights
into how viral marketing works that challenge common assumptions made in
epidemic and rumor propagation modeling.

Firstly, it is frequently assumed in epidemic models (e.g., SIRS type of
models) that individuals have equal probability of being infected every
time they interact~\cite{anderson92infectious,Bailey1975Diseases}.
Contrary to this we observe that the probability of infection decreases
with repeated interaction. Marketers should take heed that providing
excessive incentives for customers to recommend products could backfire
by weakening the credibility of the very same links they are trying to
take advantage of.

Traditional epidemic and innovation diffusion models also often assume
that individuals either have a constant probability of `converting'
every time they interact with an infected
individual~\cite{goldenberg01}, or that they convert once the fraction
of their contacts who are infected exceeds a
threshold~\cite{Granovetter:1978}. In both cases, an increasing number
of infected contacts results in an increased likelihood of infection.
Instead, we find that the probability of purchasing a product increases
with the number of recommendations received, but quickly saturates to a
constant and relatively low probability. This means individuals are
often impervious to the recommendations of their friends, and resist
buying items that they do not want.

In network-based epidemic models, extremely highly connected individuals
play a very important role. For example, in needle sharing and sexual
contact networks these nodes become the ``super-spreaders'' by infecting
a large number of people. But these models assume that a high degree
node has as much of a probability of infecting each of its neighbors as
a low degree node does. In contrast, we find that there are limits to
how influential high degree nodes are in the recommendation network. As
a person sends out more and more recommendations past a certain number
for a product, the success per recommendation declines. This would seem
to indicate that individuals have influence over a few of their friends,
but not everybody they know.

We also presented a simple stochastic model that allows for the presence
of relatively large cascades for a few products, but reflects well the
general tendency of recommendation chains to terminate after just a
short number of steps. Aggregating such cascades over all the products,
we obtain a highly disconnected network, where the largest component
grows over time by aggregating typically very small but occasionally
fairly large components. We observed that the most popular categories of
items recommended within communities in the largest component reflect
differing interests between these communities. We presented a model
which shows that these smaller and more tightly knit groups tend to be
more conducive to viral marketing.

We saw that the characteristics of product reviews and effectiveness of
recommendations vary by category and price, with more successful
recommendations being made on technical or religious books, which
presumably are placed in the social context of a school, workplace or
place of worship. A small fraction of the products accounts for a large
proportion of the recommendations. Although not quite as extreme in
proportions, the number of successful recommendations also varies widely
by product. Still, a sizeable portion of successful recommendations were
for a product with only one such sale - hinting at a long tail
phenomenon.

Since viral marketing was found to be in general not as epidemic as one
might have hoped, marketers hoping to develop normative strategies for
word-of-mouth advertising should analyze the topology and interests of
the social network of their customers. Our study has provided a number
of new insights which we hope will have general applicability to
marketing strategies and to future models of viral information spread.

\section*{ACKNOWLEDGEMENTS}

We thank anonymous reviewers for their insightful comments. This work
was partially supported by the National Science Foundation under grants
   SENSOR-0329549 
   IIS-0326322 
   IIS-0534205. 
This work is also supported in part by the Pennsylvania Infrastructure
Technology Alliance (PITA). Additional funding was provided by a
generous gift from Hewlett-Packard. Jure Leskovec was partially
supported by a Microsoft Research Graduate Fellowship.

\bibliographystyle{alpha} 

\renewcommand{\thefigure}{A-\arabic{figure}}
\setcounter{figure}{0}  
\renewcommand{\thetable}{A-\arabic{table}}
\setcounter{table}{0}  
\clearpage

\section*{Appendix}
\label{sec:appendix}

\begin{figure}[ht]
\begin{center}
\begin{tabular}{ccc}
  \includegraphics[width=0.45\textwidth]{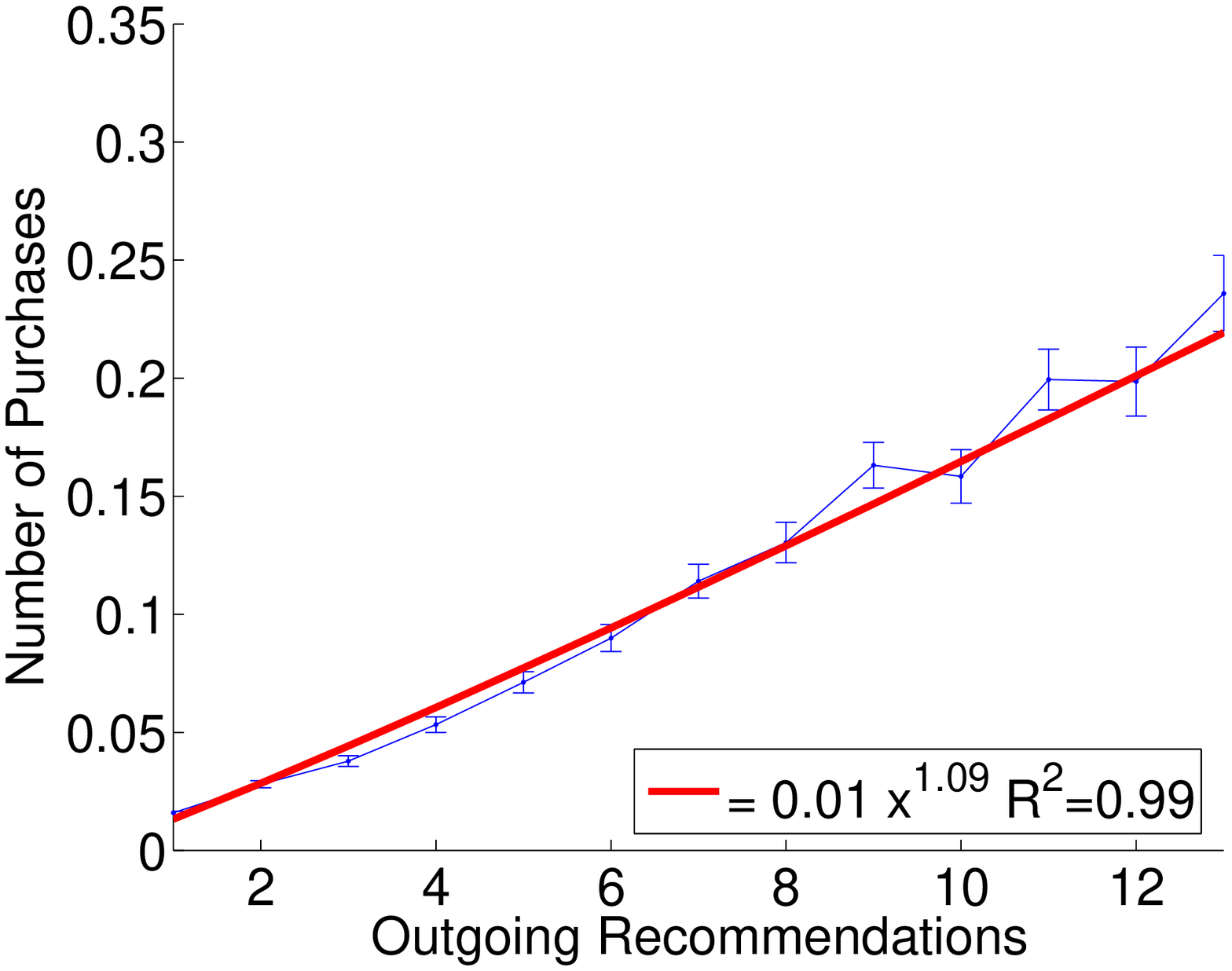} &
  \includegraphics[width=0.45\textwidth]{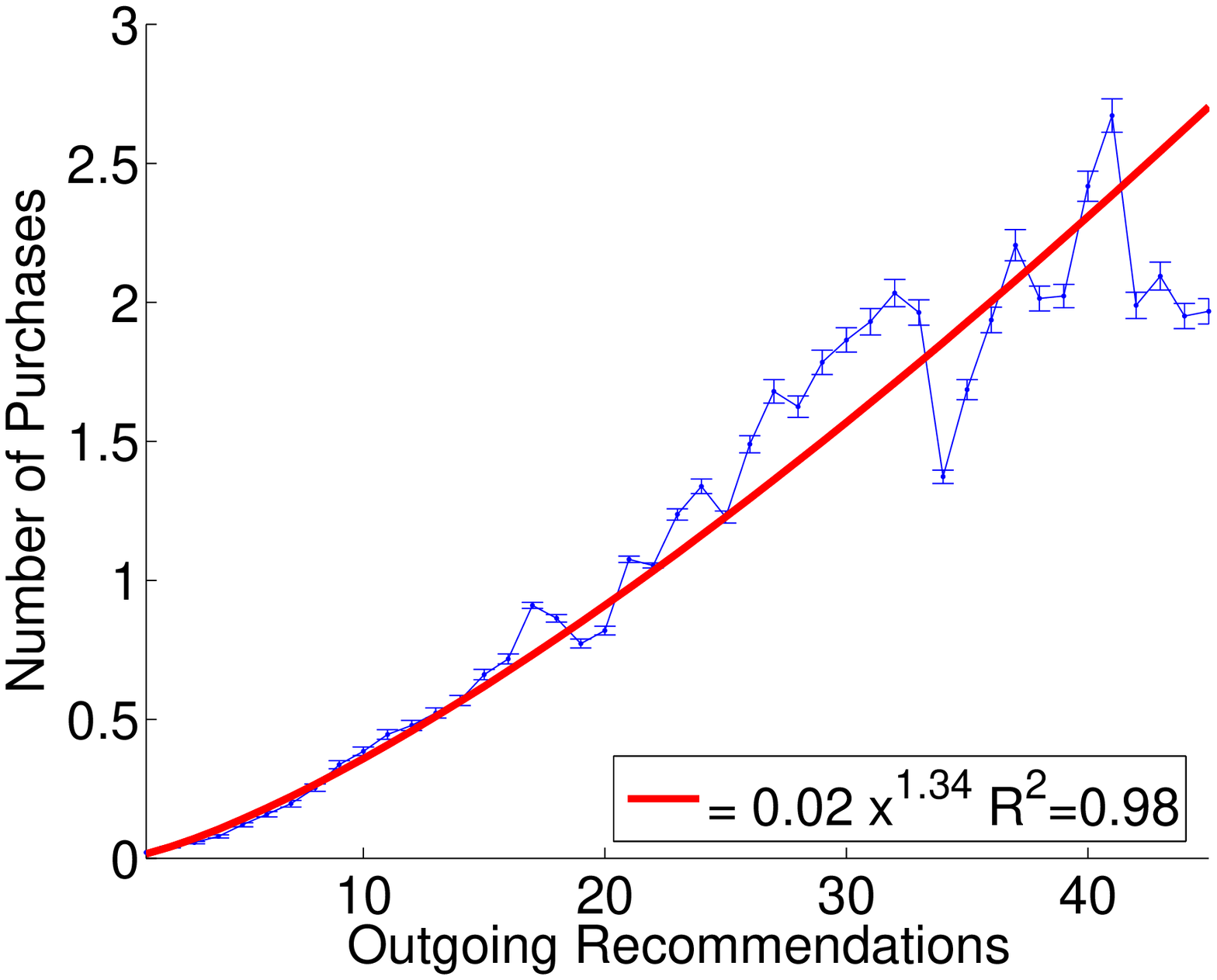} \\
  \includegraphics[width=0.45\textwidth]{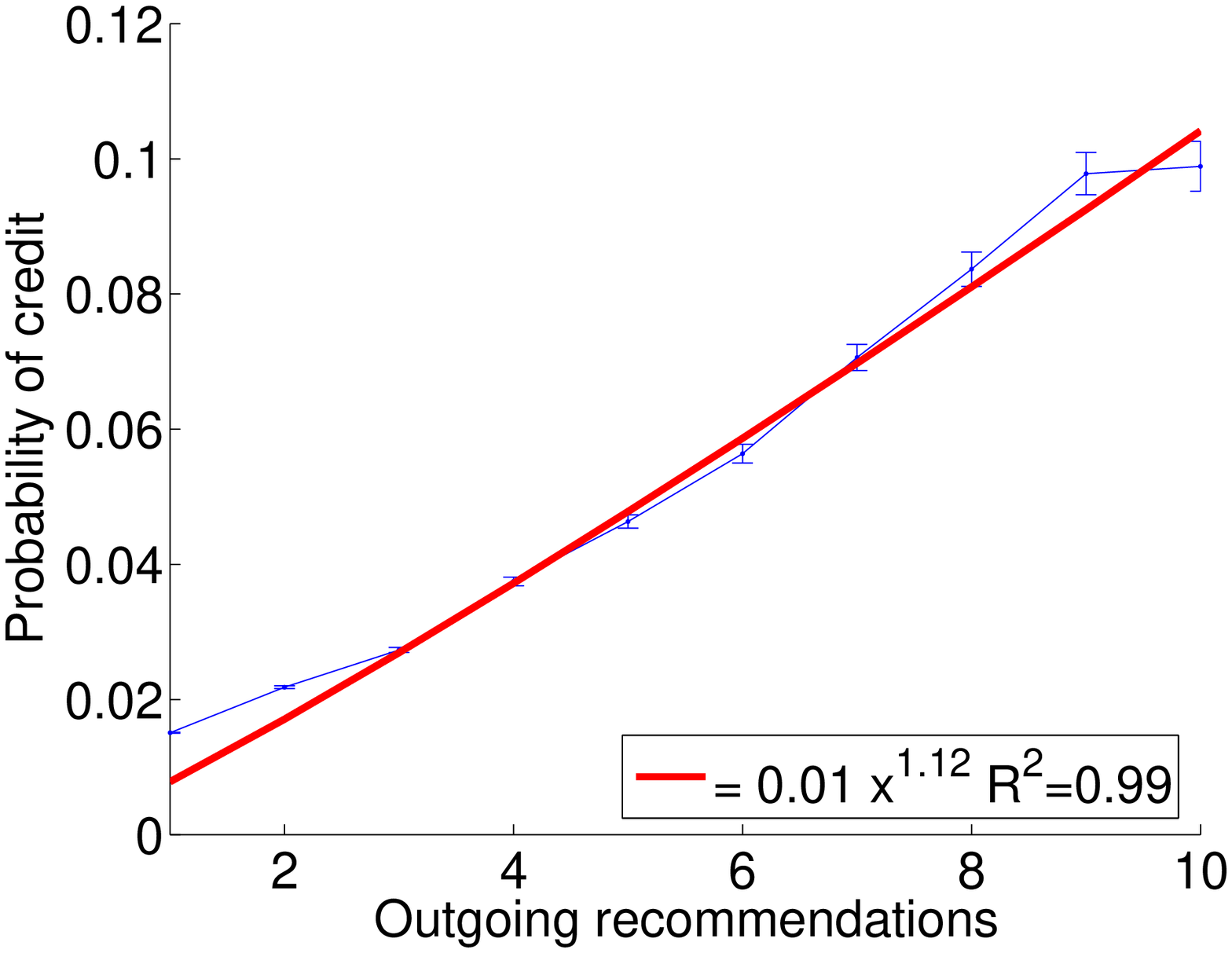} &
  \includegraphics[width=0.45\textwidth]{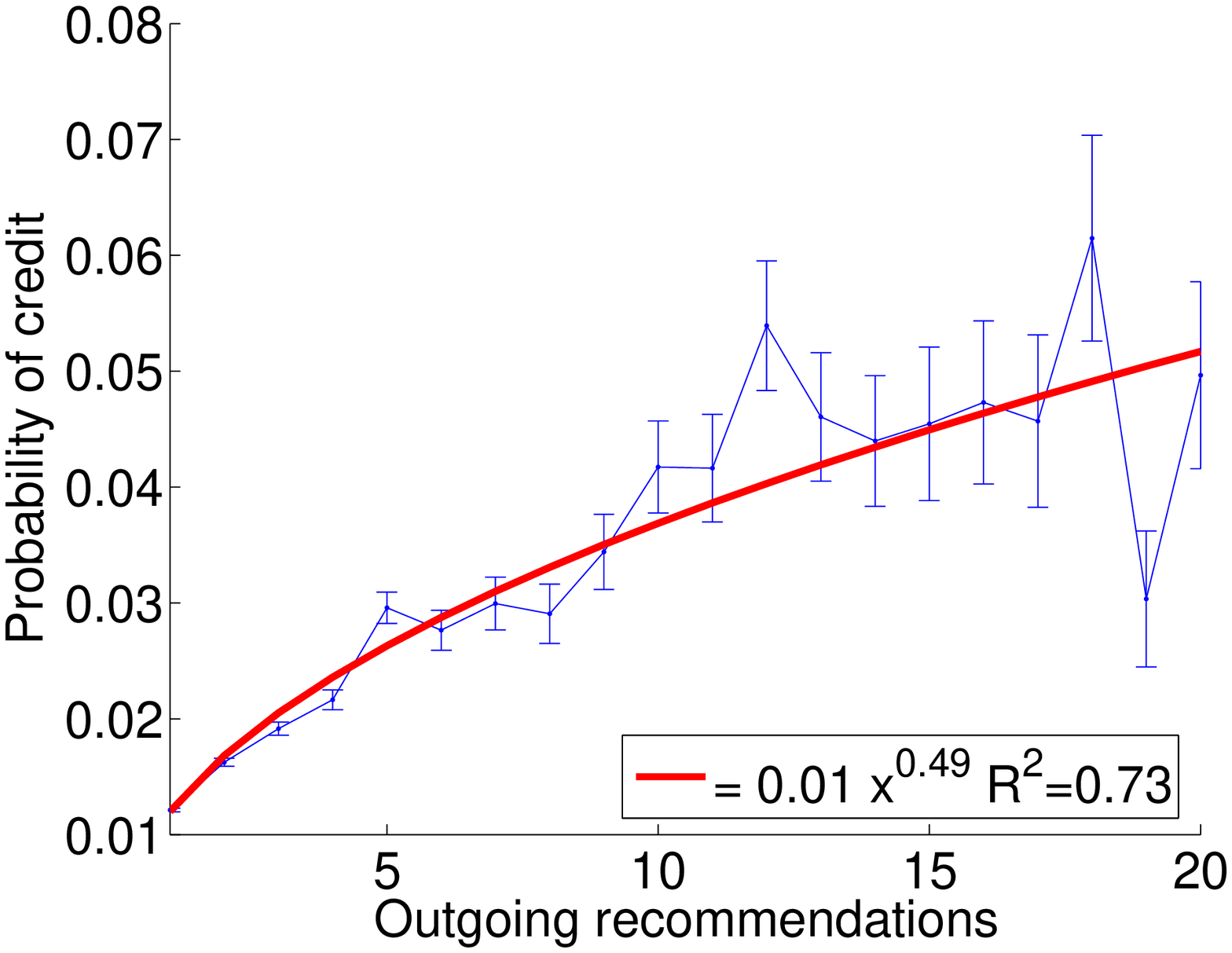} \\
  (a) Books & (b) DVD \\
\end{tabular}
\caption{Top row: Power fit to the non-linear part of the number of
resulting purchases given a number of outgoing recommendations.
Bottom row: Power fit to the probability of getting a credit given a
number of outgoing recommendations.} \label{fig:outRecBuyProbFit}
\end{center}
\end{figure}

\begin{figure}[ht]
\begin{center}
\begin{tabular}{cc}
  \includegraphics[width=0.45\textwidth]{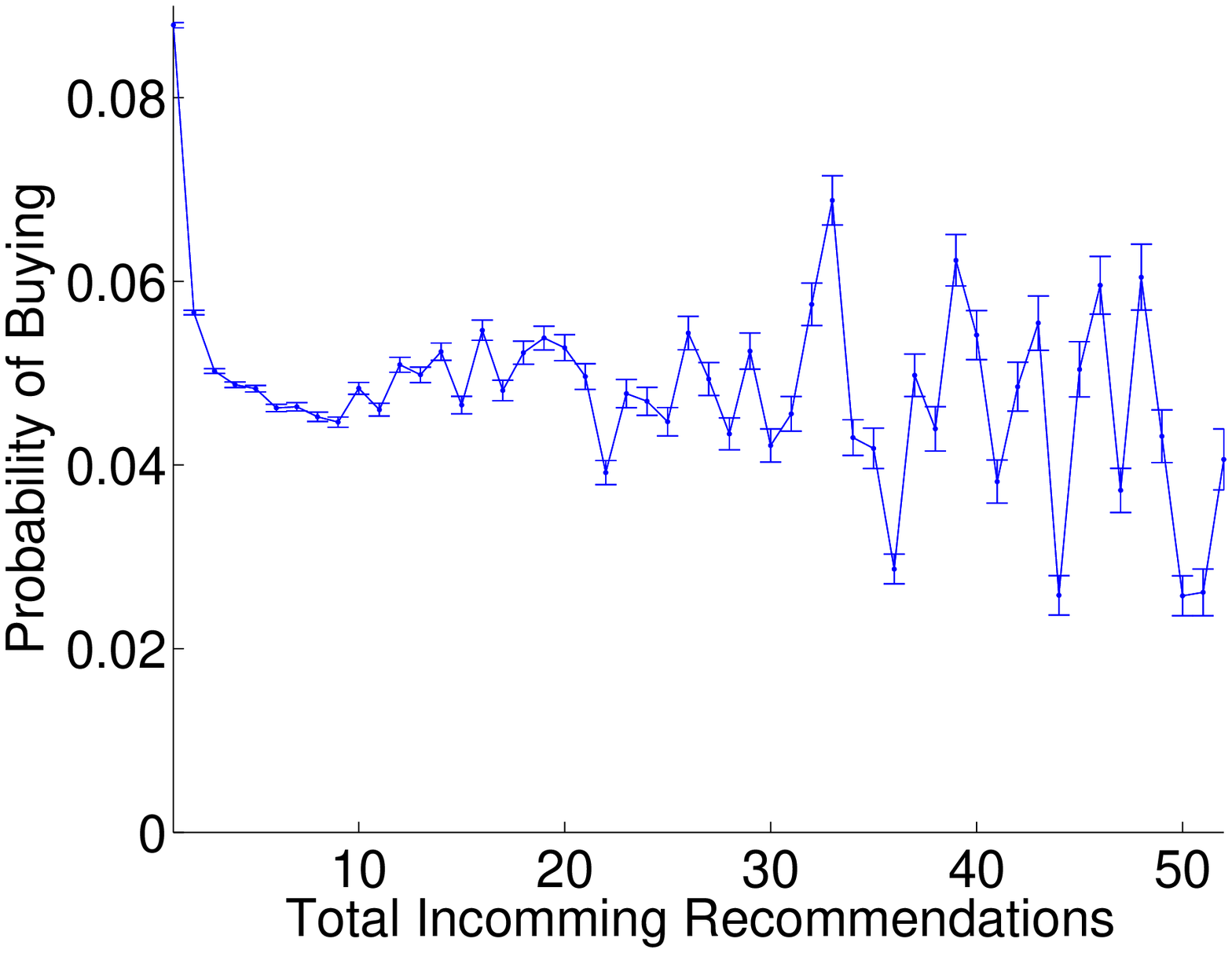} &
  \includegraphics[width=0.45\textwidth]{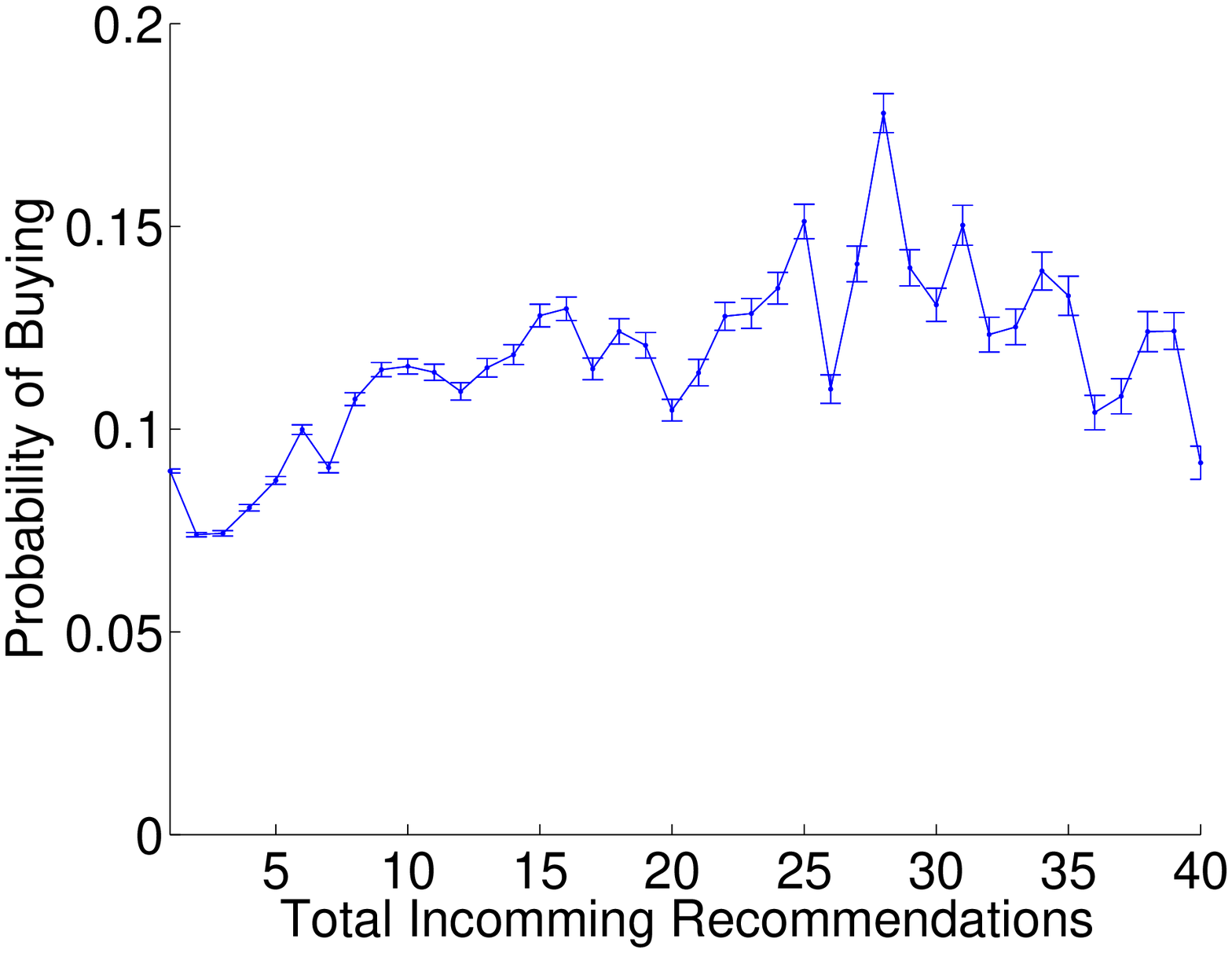} \\
  (a) Books & (b) DVD \\
  \includegraphics[width=0.45\textwidth]{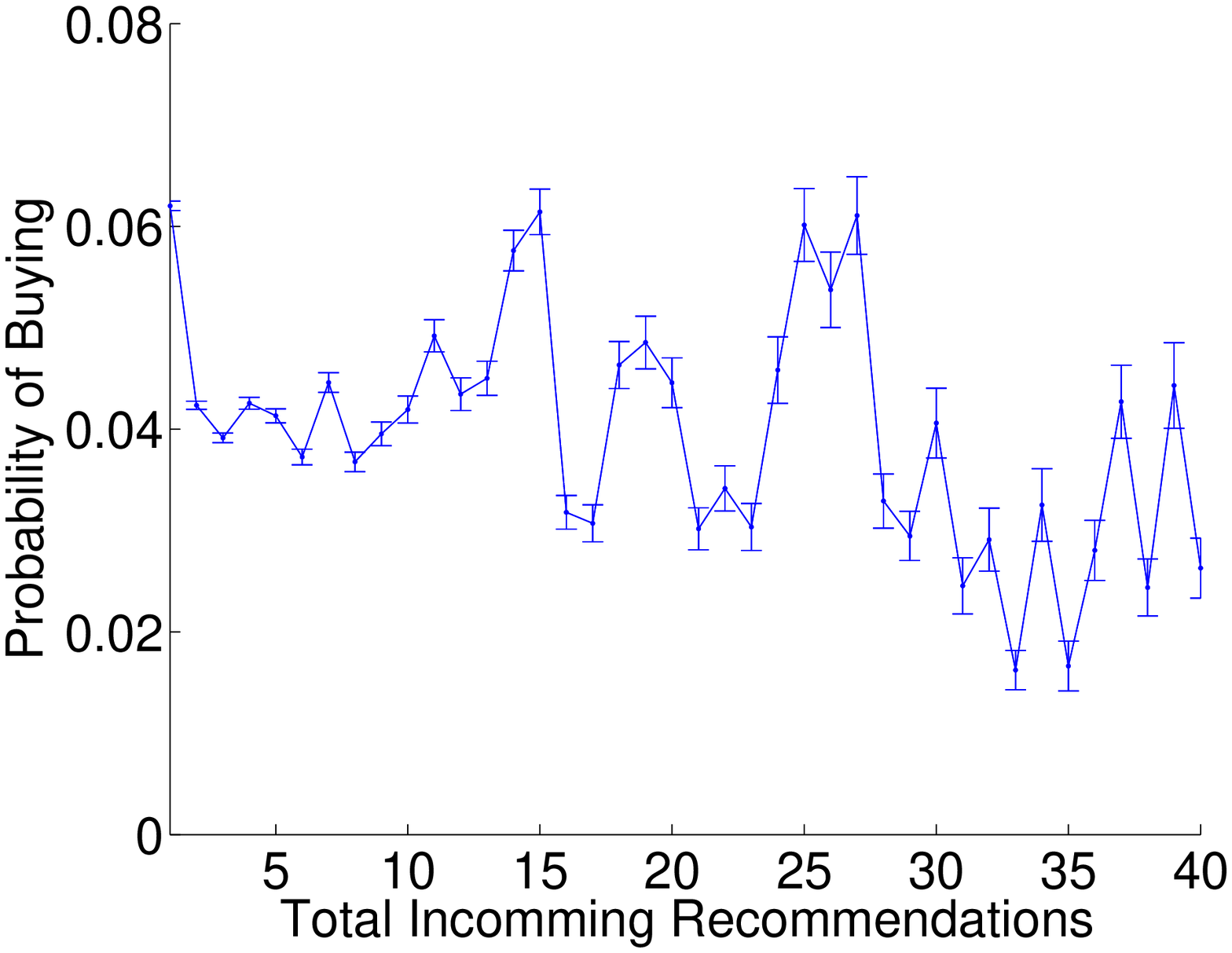} &
  \includegraphics[width=0.45\textwidth]{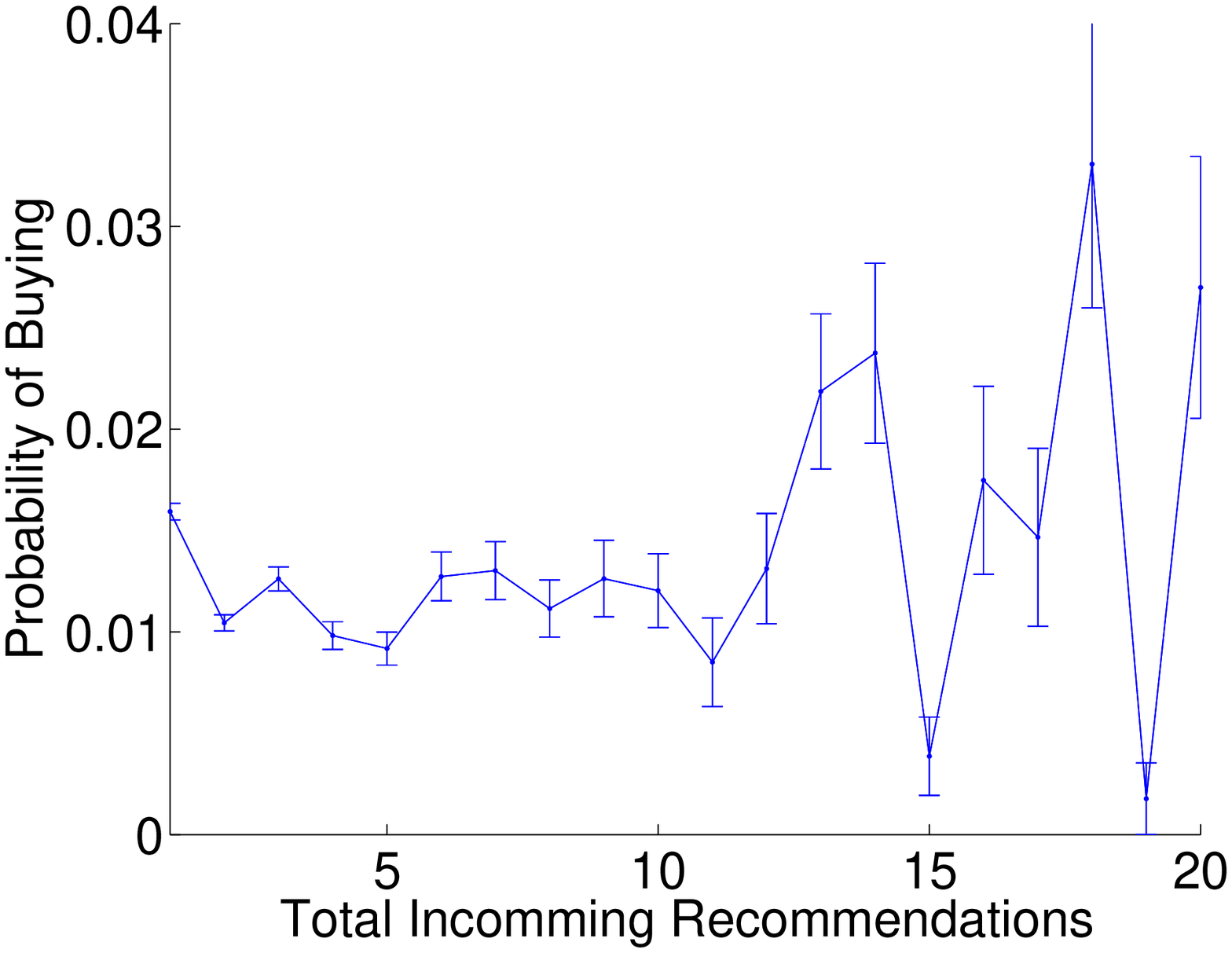} \\
  (c) Music & (d) Video \\
\end{tabular}
\caption{Probability of buying a product given a total number of
incoming recommendations on all products.} \label{fig:inAll1BuyProb}
\end{center}
\end{figure}

\begin{figure}[t]
\begin{center}
\begin{tabular}{cc}
  \includegraphics[width=0.45\textwidth]{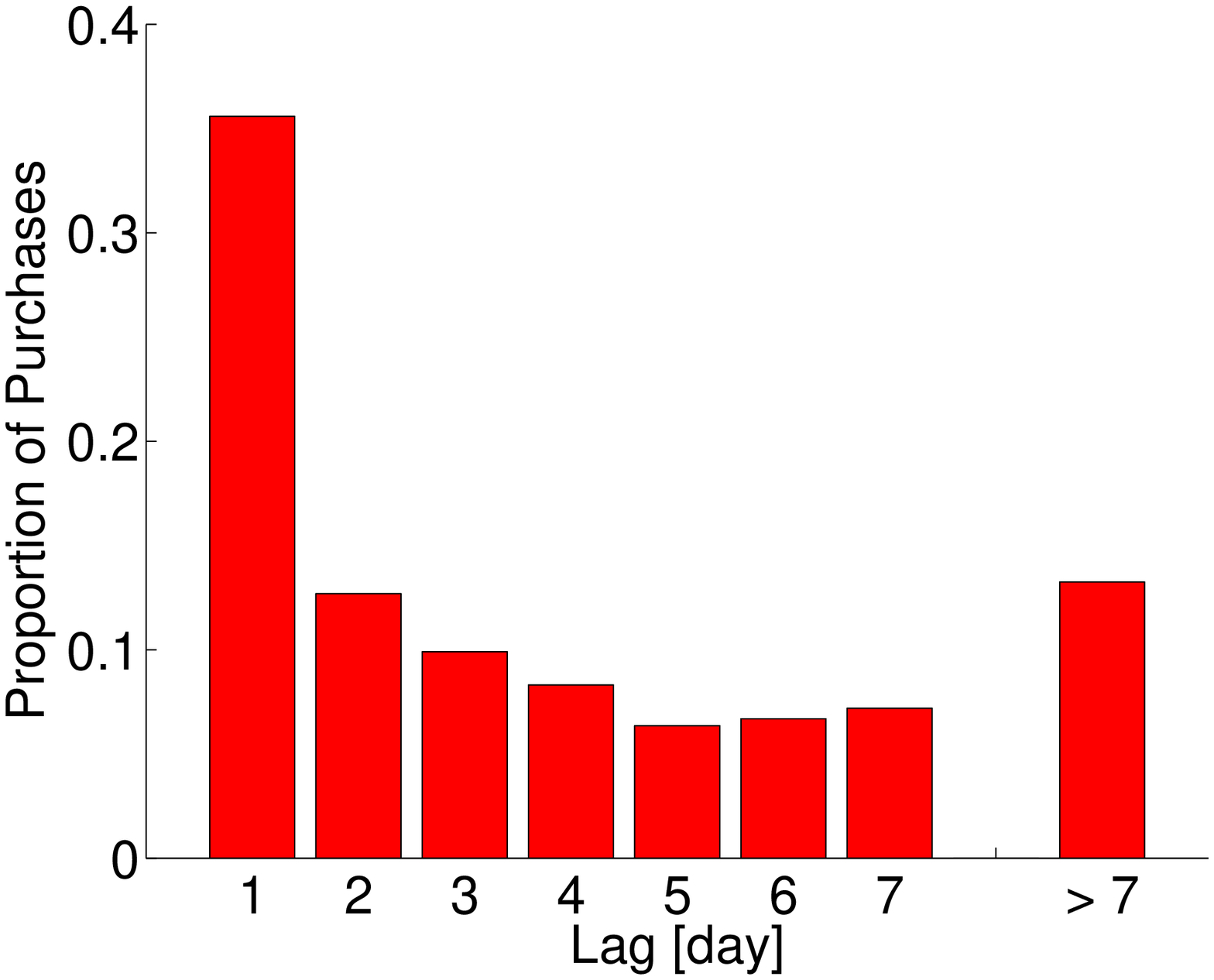} &
  \includegraphics[width=0.45\textwidth]{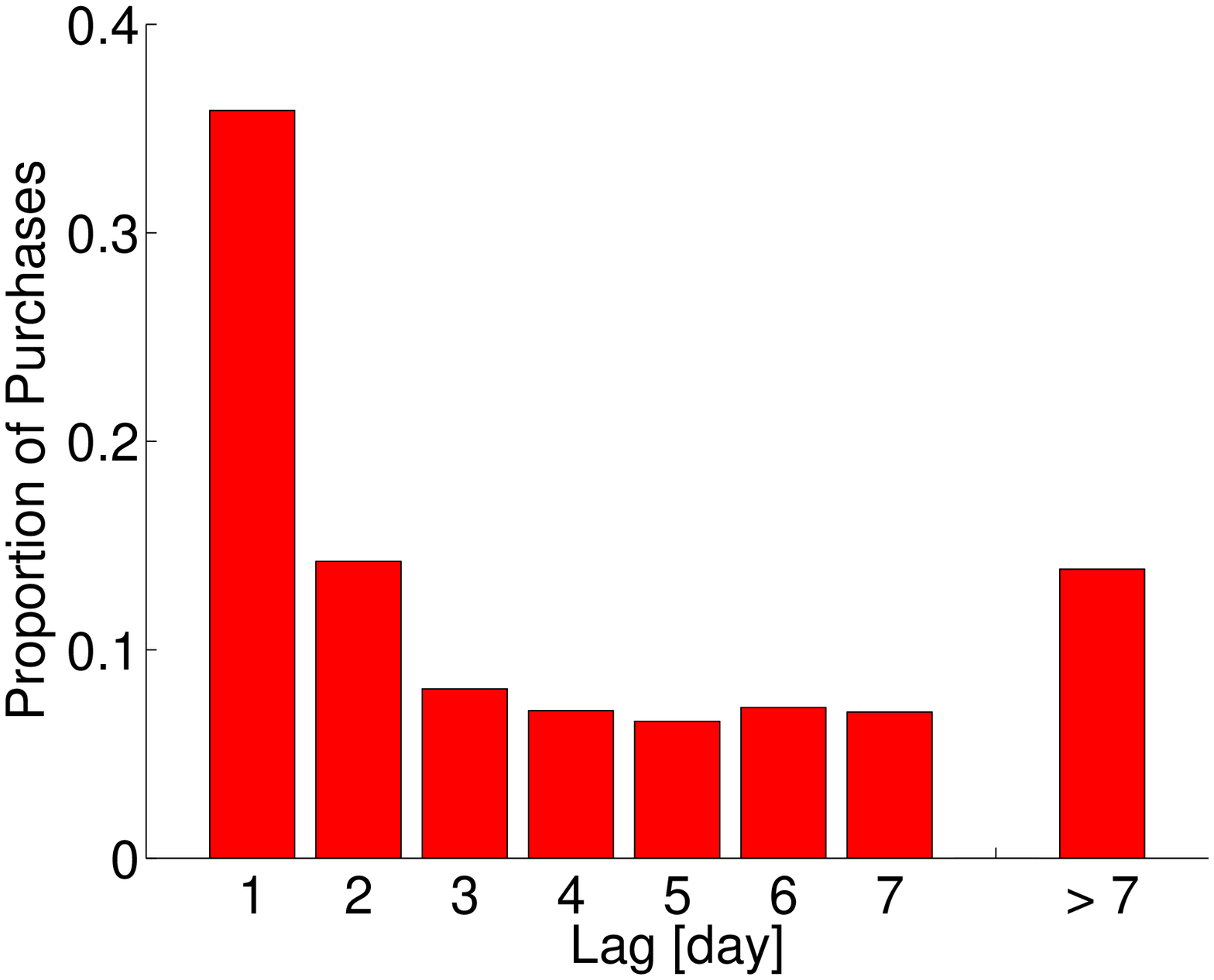} \\
  (a) Music & (b) Video \\
\end{tabular}
\caption{The time between the first recommendation and the purchase.
The histograms show how long does it take to accumulate sufficient
number of recommendations to trigger a purchase.
Figure~\ref{fig:recBuyLagAll} plots the same quantity for Books and DVD.
The bin size is 1 day. We use all purchases through recommendations.}
\label{fig:recBuyLagAllMn}
\end{center}
\end{figure}

\begin{figure}[t]
\begin{center}
\begin{tabular}{cc}
  \includegraphics[width=0.45\textwidth]{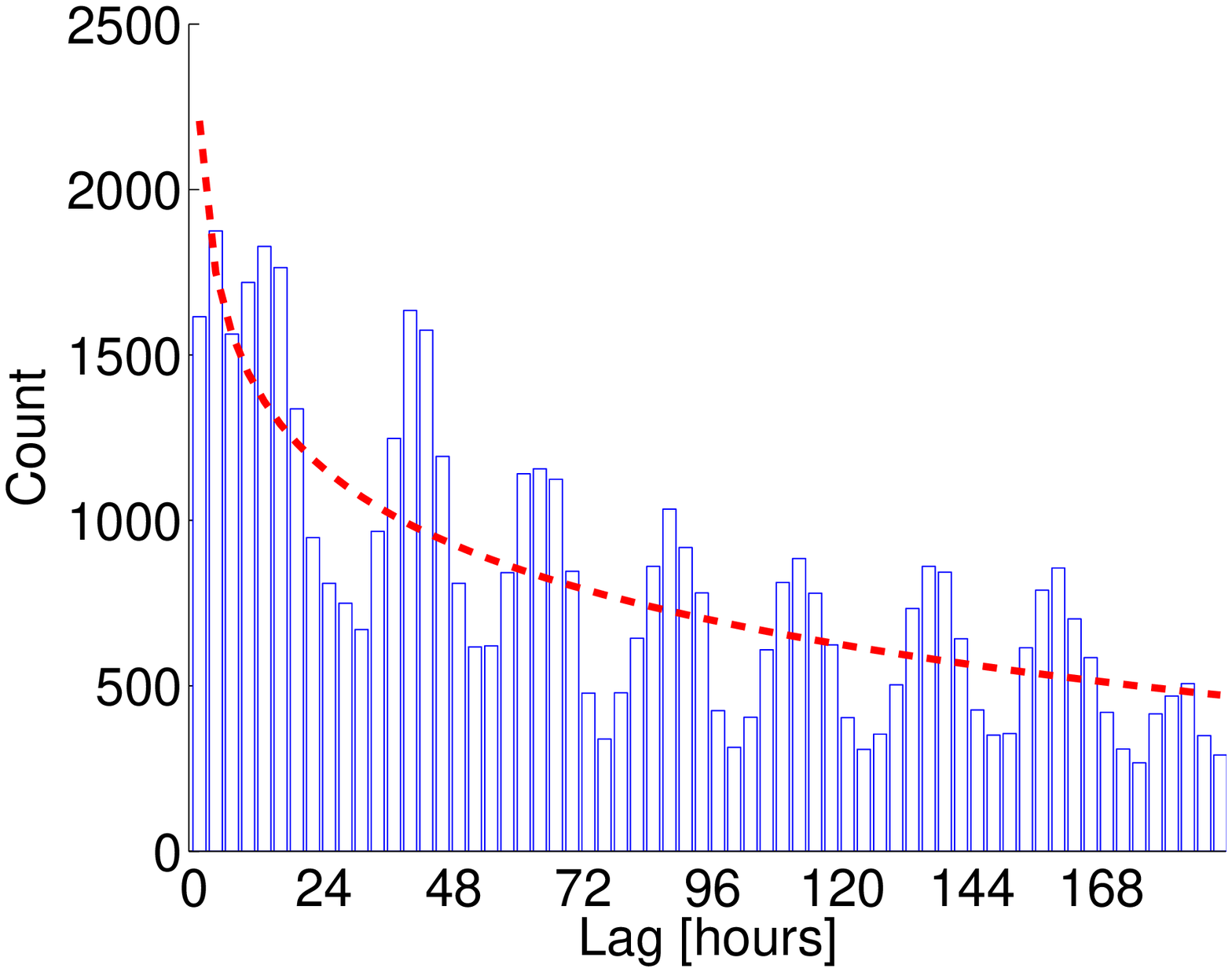} &
  \includegraphics[width=0.45\textwidth]{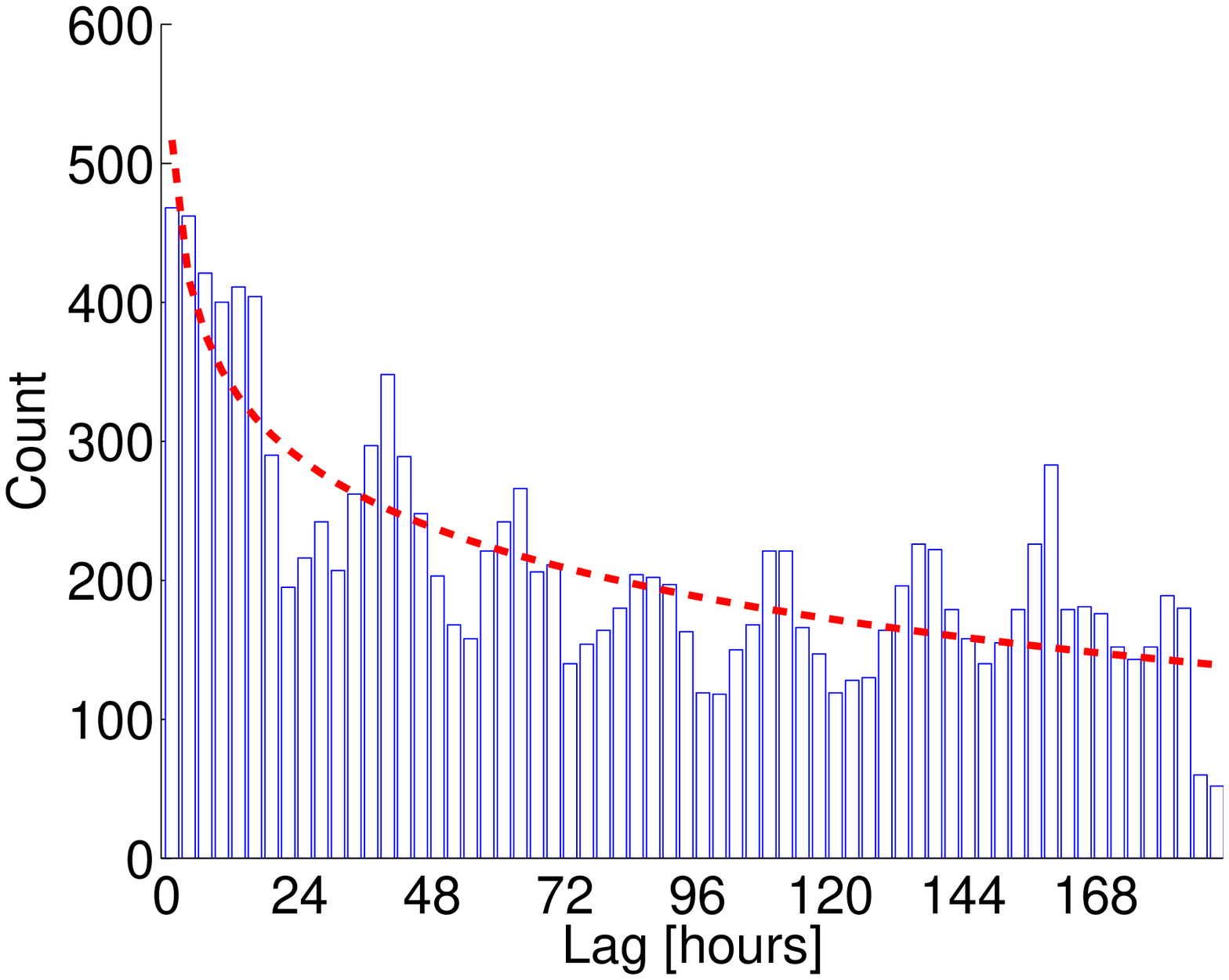} \\
  (a) Books & (b) DVD \\
  \includegraphics[width=0.45\textwidth]{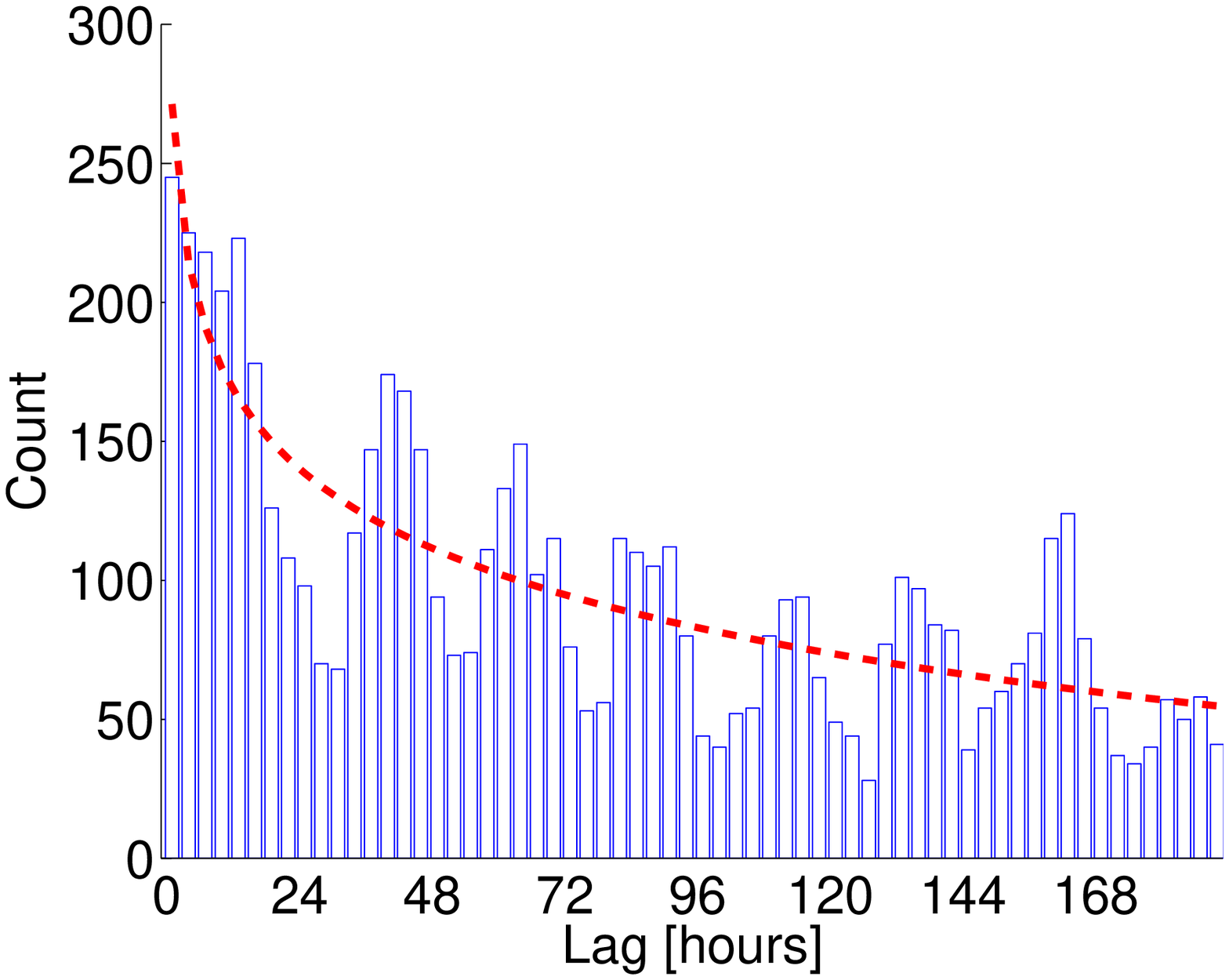} &
  \includegraphics[width=0.45\textwidth]{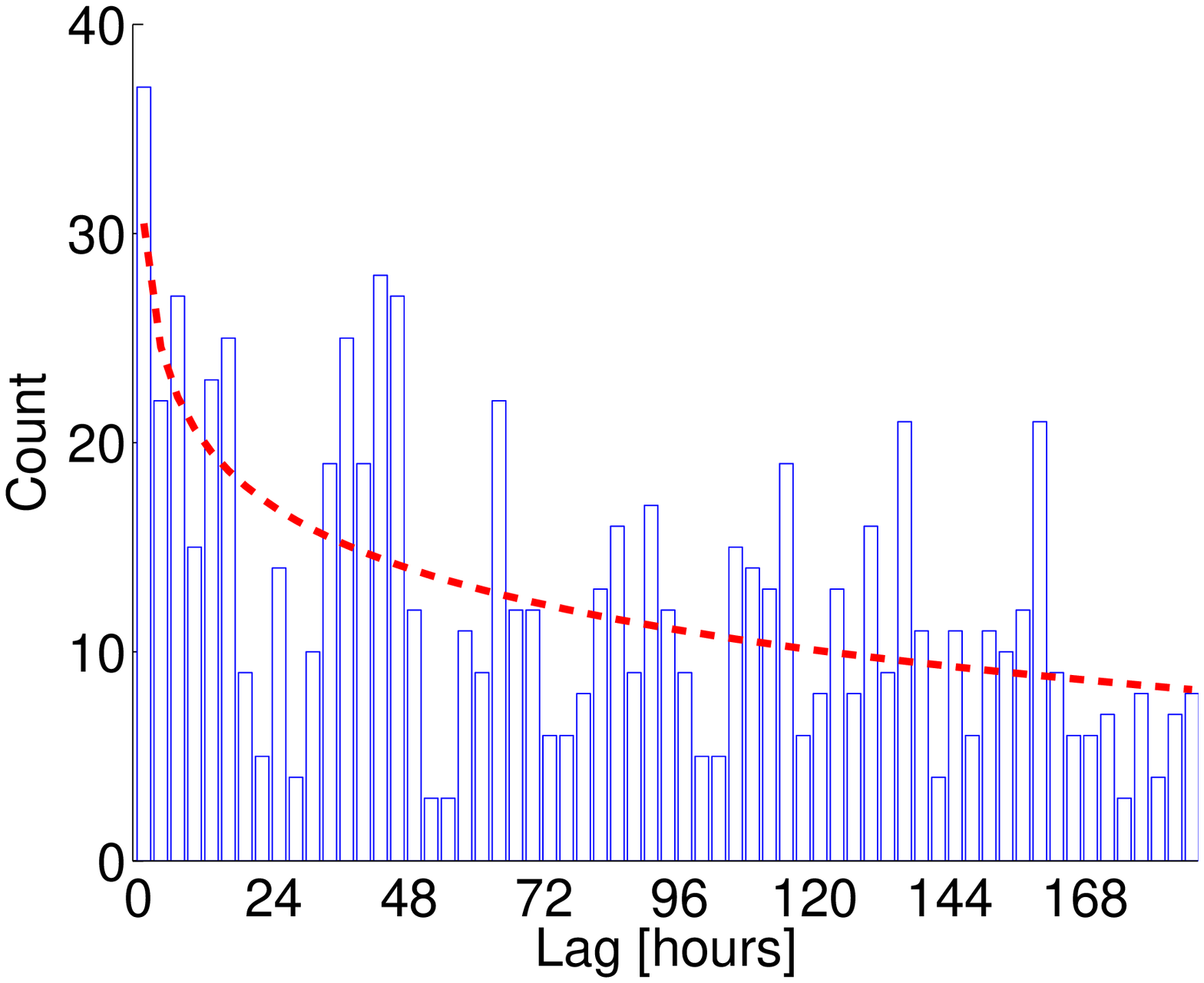} \\
  (c) Music & (d) Video \\
\end{tabular}
\caption{The time between the recommendation and the purchase taking
only the recommendations that resulted in a 10\% discount. The bin
size is 3 hours. The dashed line presents a logarithmic fit.}
\label{fig:recBuyLagBB}
\end{center}
\end{figure}

\begin{table}[tb]
\centering
\begin{tabular}{l|rrrrrrrrl}\hline
category& $n_p$ & $n$ & $cc$ & $r_{p1}$ &
$r_{p1}/r_{p2}$ & $v_{av}$ & $c_{av}/r_{p1}$ & $p_m$& $b$ \\
\hline
Anime and Manga & 1301 & 46941 & 18.92 & 14.40 & 17.17 & 4.19 & 2.96 & 26.96 & 28.44** \\
Classics & 266 & 24922 & 25.59 & 9.68 & 6.66 & 4.18 & 4.16 & 22.49 & 11.22** \\
Animation & 237 & 80092 & 11.99 & 41.90 & 19.17 & 4.03 & 3.88 & 22.49 & 10.43 \\
Science Fiction \& Fantasy & 1410 & 317420 & 6.61 & 59.18 & 16.66 & 3.85 & 2.51 & 17.99 & 9.62 \\
Art House \& International & 3185 & 276142 & 7.37 & 24.35 & 12.97 & 3.95 & 2.22 & 22.46 & 9.43* \\
Television & 1133 & 195948 & 8.17 & 18.95 & 11.68 & 4.22 & 5.32 & 17.99 & 8.90 \\
Horror & 1125 & 79744 & 13.15 & 30.00 & 9.10 & 3.59 & 1.37 & 17.98 & 8.72 \\
Action and Adventure & 2058 & 248674 & 7.00 & 39.80 & 15.11 & 3.80 & 1.96 & 17.96 & 8.42** \\
Mystery and Suspense & 1683 & 151101 & 9.28 & 26.73 & 10.45 & 3.82 & 2.20 & 17.98 & 7.57 \\
Military and War & 379 & 69180 & 12.53 & 39.31 & 11.14 & 4.12 & 2.26 & 17.96 & 7.41 \\
Cult Movies & 324 & 94049 & 11.24 & 37.93 & 8.45 & 3.89 & 3.34 & 17.98 & 7.28 \\
Kids and Family & 1357 & 230300 & 6.70 & 30.96 & 12.81 & 4.12 & 3.35 & 17.98 & 6.75 \\
Drama & 3376 & 255544 & 7.12 & 25.14 & 11.02 & 3.98 & 2.10 & 17.98 & 6.72* \\
Comedy & 2455 & 312033 & 6.08 & 26.25 & 11.14 & 4.02 & 3.30 & 17.98 & 6.01** \\
Musicals \& Performing Arts & 1091 & 88665 & 10.24 & 17.07 & 11.11 & 4.09 & 2.34 & 22.48 & 4.93 \\
Westerns & 234 & 17612 & 24.40 & 11.76 & 7.30 & 3.94 & 2.72 & 13.48 & 4.71* \\
Sports & 484 & 23191 & 16.92 & 8.64 & 7.89 & 3.97 & 2.49 & 17.98 & 4.55* \\
Documentary & 1058 & 53538 & 15.24 & 6.12 & 9.08 & 3.95 & 3.70 & 17.99 & 4.24 \\
Educational & 89 & 5532 & 19.60 & 3.39 & 2.63 & 3.97 & 5.48 & 19.95 & 3.99 \\
Music Video and Concerts & 2222 & 91657 & 8.44 & 8.06 & 11.16 & 4.09 & 2.88 & 17.99 & 3.85 \\
Special Interests & 963 & 43225 & 10.42 & 5.83 & 7.45 & 3.99 & 3.43 & 18.74  & 2.62 \\
Fitness and Yoga & 223 & 17160 & 2.23 & 14.65 & 6.66 & 3.88 & 2.93 & 17.96 & 1.98 \\
African American Cinema & 81 & 10609 & 17.92 & 16.00 & 9.06 & 4.15 & 3.41 & 17.98 & 1.56 \\
\end{tabular}
\caption{Statistics by DVD genre. * denotes significance at the 0.05
level, ** at the 0.01 level}
 \label{tab:dvdcategories}
\end{table}

\begin{table}[tb]
\centering
\begin{tabular}{l|rrrrrrrrl}\hline
category& $n_p$ & $n$ & $cc$ & $r_{p1}$ &
$r_{p1}/r_{p2}$ & $v_{av}$ & $c_{av}/r_{p1}$ & $p_m $& $b*100$ \\
\hline
Anime and Manga & 962 & 5081 & 9.64 & 13.98 & 18.76 & 4.39 & 0.26 & 17.99 & 1.99* \\
Educational & 607 & 6569 & 1.64 & 1.97 & 10.75 & 4.17 & 3.01 & 19.95 & 1.59 \\
Fitness & 920 & 24627 & 0.43 & 8.41 & 12.09 & 4.09 & 1.92 & 14.95 & 1.48 \\
Animation & 171 & 9500 & 4.04 & 61.83 & 19.58 & 4.29 & 0.36 & 17.99 & 1.36 \\
Kids and Family & 4736 & 84608 & 1.13 & 14.26 & 12.11 & 4.29 & 0.85 & 12.98 & 1.16 \\
Special Interests & 3769 & 36862 & 1.45 & 3.19 & 12.73 & 4.14 & 1.65 & 19.95 & 1.09 \\
Mystery and Suspense & 1514 & 13459 & 9.90 & 30.09 & 9.83 & 4.01 & 0.14 & 14.95 & 1.01 \\
Art House \& International & 2459 & 24713 & 3.52 & 17.54 & 10.09 & 4.18 & 0.28 & 17.99 & 0.84 \\
Science Fiction and Fantasy & 1583 & 29565 & 2.54 & 51.92 & 13.76 & 4.01 & 0.18 & 13.99 & 0.83 \\
Documentary & 2936 & 18884 & 1.15 & 3.33 & 9.83 & 4.21 & 0.95 & 19.95 & 0.82 \\
Television & 3632 & 31475 & 0.95 & 5.13 & 12.11 & 4.33 & 1.01 & 14.95 & 0.71 \\
Music Video \& Concerts & 1595 & 14360 & 4.46 & 8.75 & 11.26 & 4.40 & 0.49 & 16.99 & 0.70 \\
Musicals \& Performing Arts & 1621 & 22539 & 3.13 & 13.22 & 9.39 & 4.20 & 0.51 & 19.95 & 0.69 \\
Sports & 1251 & 7987 & 0.49 & 4.07 & 9.83 & 4.15 & 0.91 & 16.99 & 0.69 \\
Comedy & 3645 & 55868 & 2.13 & 22.26 & 10.60 & 4.13 & 0.36 & 13.99 & 0.59 \\
Drama & 4837 & 52691 & 1.87 & 21.72 & 9.25 & 4.15 & 0.26 & 14.95 & 0.56 \\
Military and War & 829 & 10859 & 1.13 & 28.54 & 9.39 & 4.22 & 0.21 & 14.95 & 0.56 \\
Westerns & 487 & 3743 & 1.58 & 9.42 & 6.01 & 4.12 & 0.43 & 9.99 & 0.56 \\
Classics & 326 & 3029 & 0.56 & 8.73 & 8.15 & 4.12 & 0.51 & 14.94 & 0.49 \\
African American Cinema & 87 & 1861 & 0.64 & 15.53 & 7.59 & 4.10 & 0.61 & 9.99 & 0.49 \\
Horror & 935 & 6728 & 1.07 & 36.38 & 9.02 & 3.81 & 0.10 & 12.99 & 0.40 \\
Action and Adventure & 2390 & 25921 & 1.84 & 33.13 & 11.90 & 3.96 & 0.17 & 13.99 & 0.31 \\
Cult Movies & 401 & 5260 & 0.65 & 32.06 & 7.63 & 3.90 & 0.18 & 9.99 & 0.30 \\
\end{tabular}
\caption{Statistics for videos in VHS format by genre}
 \label{tab:vhscategories}
\end{table}

\begin{table}[tb]
\centering
\begin{tabular}{l|rrrrrrrrl}\hline
category& $n_p$ & $n$ & $cc$ & $r_{p1}$ & $r_{p1}/r_{p2}$ & $v_{av}$
& $c_{av}/r_{p1}$ & $p_m$ & $b$ \\
\hline
Broadway and Vocalists & 5423 & 104396 & 4.25 & 6.03 & 13.86 & 4.49 & 1.68 & 14.49 & 2.01 \\
Country & 5876 & 98069 & 4.67 & 5.50 & 18.45 & 4.56 & 1.76 & 13.99 & 1.87 \\
Rock & 10717 & 196852 & 4.10 & 11.00 & 10.18 & 4.40 & 0.99 & 14.99 & 1.87 \\
Alternative Rock & 13405 & 216324 & 5.12 & 13.20 & 11.24 & 4.41 & 0.81 & 13.99 & 1.87 \\
Soundtracks & 4491 & 133507 & 4.81 & 7.92 & 13.82 & 4.38 & 1.77 & 14.99 & 1.87 \\
Classical & 14223 & 116937 & 5.34 & 2.65 & 11.60 & 4.52 & 1.82 & 15.49 & 1.83 \\
Folk & 5244 & 87580 & 5.33 & 4.40 & 13.54 & 4.60 & 2.05 & 14.99 & 1.81 \\
Pop & 16764 & 322431 & 3.30 & 9.55 & 13.19 & 4.43 & 1.22 & 13.99 & 1.78 \\
Opera and Vocal & 5402 & 61643 & 6.08 & 3.32 & 12.90 & 4.48 & 1.69 & 15.99 & 1.73 \\
Miscellaneous & 5823 & 80243 & 5.71 & 3.54 & 12.31 & 4.35 & 1.90 & 13.98 & 1.62 \\
Blues & 2987 & 31199 & 6.62 & 2.76 & 11.53 & 4.59 & 1.89 & 14.99 & 1.54 \\
Hard Rock and Metal & 4787 & 63893 & 4.96 & 18.23 & 7.92 & 4.33 & 0.42 & 14.99 & 1.52 \\
Christian and Gospel & 2977 & 37554 & 2.02 & 5.41 & 16.75 & 4.67 & 1.20 & 14.99 & 1.51 \\
Jazz & 11868 & 113078 & 4.49 & 2.91 & 11.40 & 4.59 & 1.99 7 & 14.99 & 1.50 \\
Classic Rock & 5711 & 117255 & 4.74 & 13.62 & 6.78 & 4.29 & 0.87 & 13.99 & 1.50 \\
Children s Music & 1755 & 37015 & 4.89 & 3.96 & 12.52 & 4.53 & 2.94 & 12.32 & 1.47 \\
Dance and DJ & 11332 & 139787 & 5.16 & 7.14 & 14.64 & 4.38 & 1.05 & 14.99 & 1.42 \\
New Age & 4219 & 60951 & 5.90 & 3.92 & 13.79 & 4.54 & 1.95 & 14.99 & 1.42 \\
International & 13139 & 130499 & 5.02 & 3.54 & 9.52 & 4.57 & 1.51 & 14.99 & 1.32 \\
Latin Music & 4634 & 38725 & 5.06 & 2.57 & 16.76 & 4.60 & 1.75 & 13.99 & 1.30 \\
Rap and Hip Hop & 3996 & 60135 & 3.67 & 12.23 & 9.64 & 4.38 & 0.67 & 14.99 & 1.14 \\
R\&B & 5965 & 85380 & 2.78 & 8.49 & 12.90 & 4.48 & 0.89 & 13.98 & 1.13 \\
\end{tabular}
\caption{Statistics by Music Style}
 \label{tab:musiccategories}
\end{table}

\end{document}